\documentclass[aps,prb,superscriptaddress,reprint,floatfix,longbibliography,showpacs,amsfonts,amsmath,amssymb,longbibliography]{revtex4-2}

\usepackage[T1]{fontenc}
\usepackage{graphicx}% Include figure files
\usepackage[bookmarks=false,colorlinks]{hyperref}
\hypersetup{linkcolor=Plum,citecolor=RubineRed,filecolor=Plum,urlcolor=PineGreen}
\usepackage[usenames,dvipsnames]{color}
\usepackage{hyperref}
\usepackage{setspace}

\newcommand{\CGT}{CrGeTe$_3$}
\newcommand{\Tc}{$T_{\rm Curie}$}

\newcommand{\etal}{$et\ al.$}

\newcommand{\rhoAHE}{$\rho_{\rm AHE}$} 
\newcommand{\rhoXX}{$\rho_{xx}$} 
\newcommand{\rhoXY}{$\rho_{xy}$}

\begin{document}

\title{Pressure tuning of intrinsic and extrinsic sources to the anomalous Hall effect in \CGT}

\author{Gili Scharf}
\affiliation{Raymond and Beverly Sackler School of Physics and Astronomy, Tel-Aviv University, Tel Aviv, 69978, Israel}

\author{Daniel Guterding}
\affiliation{Technische Hochschule Brandenburg, Magdeburger Straße 50, 14770 Brandenburg an der Havel, Germany}

\author{Bar Hen}
\affiliation{Raymond and Beverly Sackler School of Physics and Astronomy, Tel-Aviv University, Tel Aviv, 69978, Israel}

\author{Paul M. Sarte}
\affiliation{Materials Department, University of California, Santa Barbara, California, 93106, USA}

\author{Brenden R. Ortiz}
\affiliation{Materials Department, University of California, Santa Barbara, California, 93106, USA}

\author{Gregory Kh. Rozenberg}
\affiliation{Raymond and Beverly Sackler School of Physics and Astronomy, Tel-Aviv University, Tel Aviv, 69978, Israel}

\author{Tobias Holder}
\affiliation{Raymond and Beverly Sackler School of Physics and Astronomy, Tel-Aviv University, Tel Aviv, 69978, Israel}

\author{Stephen D. Wilson}
\affiliation{Materials Department, University of California, Santa Barbara, California, 93106, USA}

\author{Harald O. Jeschke}
\affiliation{Research Institute for Interdisciplinary Science, Okayama University, Okayama 700-8530, Japan}

\author{Alon Ron}
\affiliation{Raymond and Beverly Sackler School of Physics and Astronomy, Tel-Aviv University, Tel Aviv, 69978, Israel}

\begin{abstract}
The integrated Berry curvature is a geometric property that has dramatic implications for material properties. This study investigates the integrated Berry curvature and other contributions to the anomalous Hall effect in {\CGT} as a function of pressure. The anomalous Hall effect is absent in the insulating phase of {\CGT} and evolves with pressure in a dome-like fashion as pressure is applied. The dome's edges are characterized by Fermi surface deformations, manifested as mixed electron and hole transport. We corroborate the presence of bipolar transport by ab-initio calculations which also predict a nonmonotonic behavior of the Berry curvature as a function of pressure. Quantitative discrepancies between our calculations and experimental results indicate that additional scattering mechanisms, which are also strongly tuned by pressure, contribute to the anomalous Hall effect in {\CGT}.
\end{abstract}

\maketitle

\section{Introduction}
The Berry phase is a geometric property of the electronic band structure of solids that has dramatic impact on material properties~\cite{RevModPhys.82.1959}. A Berry phase is accumulated when a system is subject to a cyclic adiabatic transformation in its parameter space, and it is determined by the integrated Berry curvature. As a band structure property, one may conjecture that dramatic changes to the Fermi surface will result in considerable changes to the Berry curvature and thus may result in variation of its integrated value. An extreme case of such a change would be the insulator-to-metal transition~\cite{bhoi2021nearly}, where in the insulating state, there is no Fermi surface, and in the metallic state, a Fermi surface forms, which may host a nonzero integrated Berry curvature.

A common manifestation of the Berry phase in electronic transport properties is the anomalous Hall effect (AHE). The AHE is an additional contribution to the transverse resistivity ($\rho_{xy}$) on top of the ordinary Hall effect. It occurs in materials where time-reversal symmetry is broken in the presence of spin-orbit interaction~\cite{nagaosa2010anomalous}. As such, it can be probed through measurements of $\rho_{xy}$~as a function of the magnetic field and serve as a superb probe for investigating the integrated Berry curvature of electrons in solids. We chose {\CGT}, a ferromagnetic insulator which undergoes an insulator-to-metal  transition upon the application of hydrostatic pressure, as a platform for investigating the evolution of the integrated Berry curvature when the Fermi surface is strongly deformed.

{\CGT} is a layered ferromagnetic insulator with a Curie temperature ({\Tc}) of $\sim67$\,K~\cite{carteaux1995crystallographic}, which has recently attracted a lot of attention. Inelastic neutron scattering suggests that {\CGT} is a topological magnonic insulator~\cite{zhu2021topological}. It has also been predicted to sustain ferromagnetism to the 2D limit~\cite{monolayercalculationCGT,xu2018interplay,zhuang2015computational}. Additionally, short-range fluctuations seem to play an important role above the transition temperature~\cite{chen2022anisotropic,lin2017tricritical}, in great similarity to the closely related compound CrSiTe$_3$~\cite{ron2019dimensional,ron2020ultrafast,williams2015magnetic}. 

Application of hydrostatic pressure changes {\Tc} of {\CGT}. Up to $\mathrm{4.5\ GPa}$, the Curie temperature decreases as pressure is applied~\cite{sakurai2021pressure,bhoi2021nearly}. {\CGT} undergoes an insulator-to-metal transition at $\mathrm{\sim6\, GPa}$~\cite{bhoi2021nearly}. At the onset of metallicity around $\mathrm{4.5\, GPa}$, a ferromagnetic double-exchange mechanism comes into play~\cite{EbadAllah2024}, which dramatically increases {\Tc}, rising from $\sim54$\,K to $\sim250$\,K at 9.1\,GPa~\cite{bhoi2021nearly}. A similar enhancement of magnetism and conductivity was recently also observed in amorphous {\CGT} samples created by Xe ion irradiation~\cite{Zhang2023}. The coexistence of time-reversal symmetry breaking with the insulator-to-metal  transition in a material with a high atomic number $Z$ element (Te) makes {\CGT} an ideal candidate to search for an evolution of the Berry curvature as the system is tuned through the insulator-to-metal  transition.

In this work, we demonstrate that the various contributions to the AHE in {\CGT} can be tuned by hydrostatic pressure. At certain pressures, the AHE is present also above $\mathrm{300\, K}$ suggesting a possible enhancement of {\Tc}. Our measurements of the AHE for a wide range of hydrostatic pressures at T=2\,K reveal a dome-like behavior with onset at the insulator-to-metal transition that is quenched towards higher pressures. The AHE dome coincides with the pressure range where Fermi surface deformations are observed through the ordinary Hall effect. Our ab-initio calculations corroborate the presence of electron and hole pockets in the Fermi surface both of which host nonzero integrated Berry curvature. Both deform continuously as pressure is applied. A quantitative comparison between our experimental results and our calculations indicates that additional scattering mechanisms, which are also strongly tuned by hydrostatic pressure, contribute to the AHE.

\section{Methods - experimental}
To create the {\CGT} crystals, Cr powder (99.95\%, alfa), Ge powder (99.9999\%), and Te lump (99.999+\%) were sealed in a fused silica ampule at an approximate ratio of 1:1:8 Cr:Ge:Te. Fluxes were heated to 900\textdegree C at a rate of 200\textdegree C/hr, soaked at 900\textdegree C for 24h, and then slowly cooled down to 550\textdegree C at a rate of 2\textdegree C/h. The resulting fluxes were centrifuged at 550\textdegree C to remove molten Te from the crystals, after which thin platelets of dimensions $1 \,\mathrm{mm} \times 2 \,\mathrm{mm} \times 0.1 \,\mathrm{mm}$ were mechanically isolated. 

The pressure was exerted on the samples using miniature diamond anvil cells (DACs)~\cite{Diamond_anvil_cell}, with diamond anvil culets of $\mathrm{300\, \mu m}$. A rhenium gasket was drilled, then filled and covered with a powder layer of 75\% Al$_2$O$_3$~and 25\% NaCl for electrical insulation. Two pressure cells were loaded with $\sim 5\, \mathrm{\mu m}$ thick {\CGT} flakes and placed on top of the insulating layer, which functions as a pressure-transmitting medium. A $\sim 5\, \mathrm{\mu m}$ thick Pt foil was cut into triangular pieces and placed in contact with the {\CGT} flakes, allowing electrical transport measurements at elevated pressures in the Van der Pauw geometry. As such, {\rhoXX} and $\rho_{xy}$ are inferred from the measured resistance up to a factor of order unity due to uncertainties in the sample geometry and thickness, which are inevitable inside a DAC (a more detailed explanation is offered in the Supplemental Material in section \hyperref[section S1]{S1}). In addition, Ruby fragments were placed between the Pt leads for pressure determination~\cite{PhysRevB.78.104102}. 

The samples were compressed in steps of $\mathrm{2-4\,GPa}$~and then cooled down from ambient temperature to $\mathrm{2\,K}$. Measurements from the two cells are shown in this manuscript.
Due to the uncertainties in the samples' geometry and thickness, to minimize possible errors in the analysis, we use resistivities throughout the manuscript instead of analyzing and discussing the results in terms of conductivity (for further explanation, see section \hyperref[section S1]{S1} in the Supplemental Material). In addition, in order to compare the results between the two samples, we use a single geometric factor of order unity to normalize the longitudinal and transverse resistivity measurements of sample 1 with respect to sample 2. 

\section{Methods - theoretical}
We perform full-relativistic density functional theory (DFT) calculations within the full potential local orbital (FPLO) method~\cite{Koepernik1999} and using the generalized gradient approximation (GGA)~\cite{Perdew1996} for the exchange-correlation functional. Experimental crystal structures under pressure were imported from Ref.~\cite{yu2019pressure} and interpolated smoothly, as explained in Ref.~\cite{Xu2023}. All calculations were performed in ferromagnetic spin configuration.

We calculated the electronic band structure with orbital weights, the Fermi surface in the $k_x$-$k_y$-plane (at $k_z=0$), and the Berry curvature $\Omega(k)$ of {\CGT} as a function of pressure using FPLO. The high-symmetry path for the electronic band structure is the same as in Ref.~\cite{Xu2023}. Fermi surface and Berry curvature in the plane were evaluated on a $200 \times 200$ $k$-point grid for visualization purposes.

The anomalous Hall conductivity $\sigma_{xy}$ is defined as the integral of the total Berry curvature $\Omega_{z}(\vec{k})$ over the entire Brillouin zone (BZ)~\cite{Wang2006}:
\begin{equation}
\sigma_{xy} = -\frac{e^2}{\hbar} \int_\text{BZ} \frac{d\vec{k}}{(2 \pi)^3} \, \Omega_z (\vec{k})
\label{eq:conductivityBZintegral}
\end{equation}
The total Berry curvature $\Omega_z (\vec{k})$, calculated using Wannier interpolation within FPLO, is defined as the sum over all bands $n$ of the band-resolved Berry curvature $\Omega_{n, z}(\vec{k})$ weighted by the respective occupation number $f_n(\vec k)$~\cite{Wang2006}:
\begin{equation}
\Omega_z (\vec k) = \sum\limits_n f_n(\vec{k}) \, \Omega_{n, z} (\vec{k})
\label{eq:totalBerrycurvature}
\end{equation}

The integral over the entire BZ in Eq.~\eqref{eq:conductivityBZintegral} is computationally challenging since the relevant contributions of the total Berry curvature $\Omega_z (\vec{k})$ can be concentrated in tiny regions of momentum-space. We decided to implement the integral over the Brillouin zone using the \texttt{vegas} adaptive Monte Carlo algorithm~\cite{Lepage1978, Lepage2021}. 

For each calculation of the anomalous Hall conductivity $\sigma_{xy}$ we performed ten independent Monte Carlo runs with $10^6$ evaluations of the integrand for training the adaptive part of the algorithm and subsequent $10^6$ evaluations for the actual calculation of the conductivity. The ten independent runs allow us to estimate the standard deviation of the obtained results, i.e.~the Monte Carlo uncertainty.

We validated our implementation against literature results~\cite{Wang2007} for bcc-Fe, fcc-Ni, and hcp-Co and reproduced those with sufficient accuracy. We note that potentially more efficient approaches using line integrals on the Fermi surface have been suggested~\cite{Wang2007}.

From Eq.~\eqref{eq:conductivityBZintegral}, it is clear that the total Berry curvature $\Omega_z (\vec{k})$ and the conductivity $\sigma_{xy}$ have opposite signs. We decided to visualize and discuss $-\Omega_z (\vec{k})$ in the rest of our study so that positive contributions to $-\Omega_z (\vec{k})$ imply a positive contribution to $\sigma_{xy}$. 

The total Berry curvature has dimension length squared. All total Berry curvatures in this work are given in units of squared Bohr radii, denoted as $a_0^2$.

\begin{figure}[htb]
\includegraphics[width=\columnwidth]{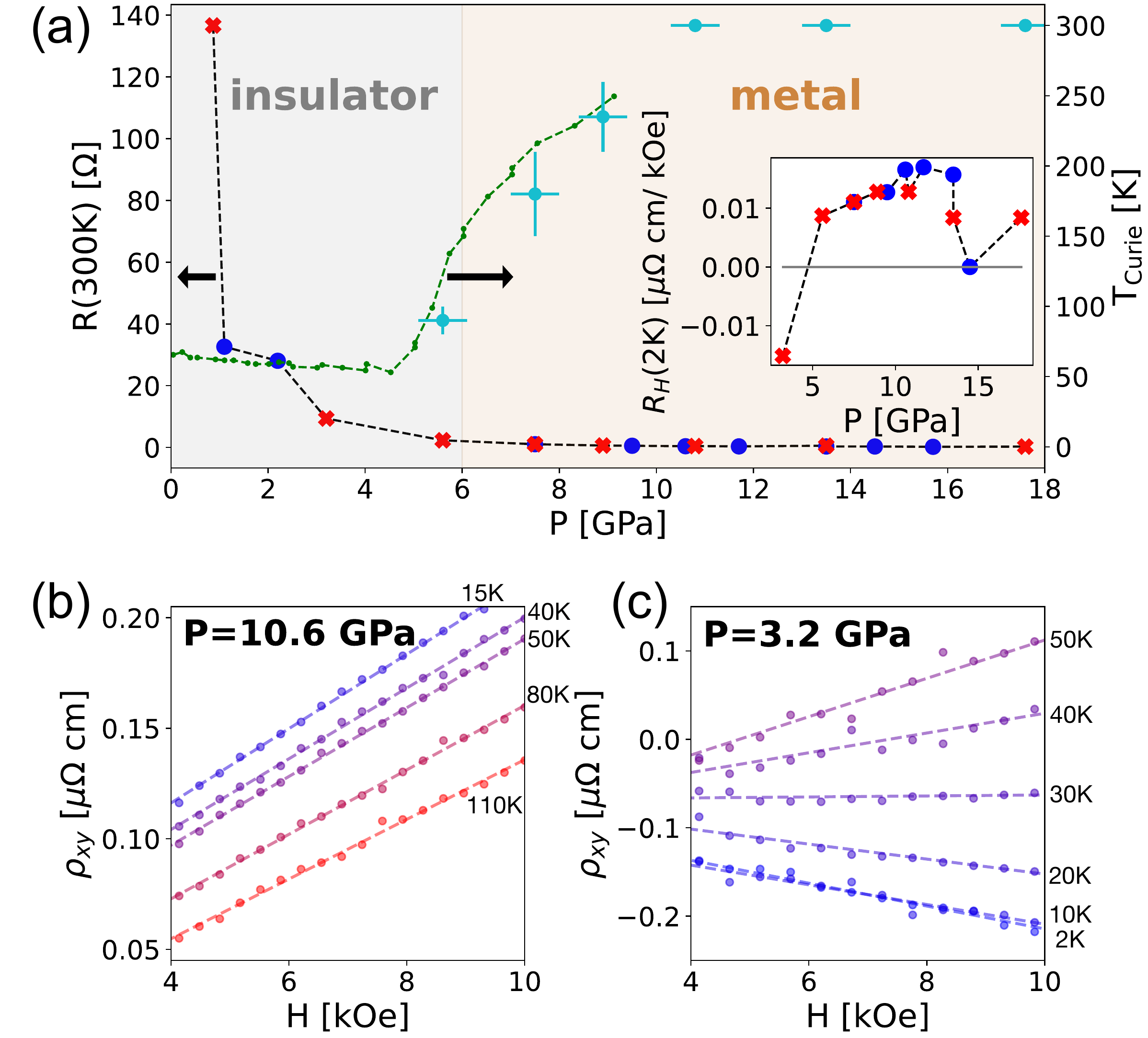} 
\caption{(a) Resistance $R$ (left axis) and Curie temperature {\Tc} (right axis) as function of pressure. Due to the insulator-to-metal transition, the sample resistance decreases significantly as pressure is applied. The blue and red points represent measurements from cells 1 and 2, respectively. The values are scaled by a single geometric factor of order unity, which is used throughout the manuscript for any longitudinal resistivity measurement. The cyan dots for the Curie temperature represent measurements from this work based on the AHE, which extend the pressure range covered in Ref.~\cite{bhoi2021nearly}, shown in green. The uncertainties in the value of {\Tc} are estimated as the interval between our sampling points, and the pressure uncertainties are estimated to be about 0.5\,GPa. The arrows signify that the value of 300\,K is a lower bound for the Curie temperature, as it is the highest temperature in which data was taken. The inset shows the Hall coefficient as a function of applied pressure at 2\,K extracted from a linear fit in the field range between 4\,kOe and 10\,kOe. The blue and the red points are measurements of $\rho_{xy}$~from the first and second cells, respectively. A single geometric factor of order unity was used to scale the values of $\rho_{xy}$ here and throughout the manuscript. Panels (b) and (c) show the antisymmetrized Hall measurements at pressures of 10.6\,GPa and 3.2\,GPa at different temperatures as a function of the magnetic field in the range which was used to calculate the Hall coefficient.}
\label{Figure 1}
\end{figure}

\begin{figure*}[htb]
\centering
\includegraphics[width=\textwidth]{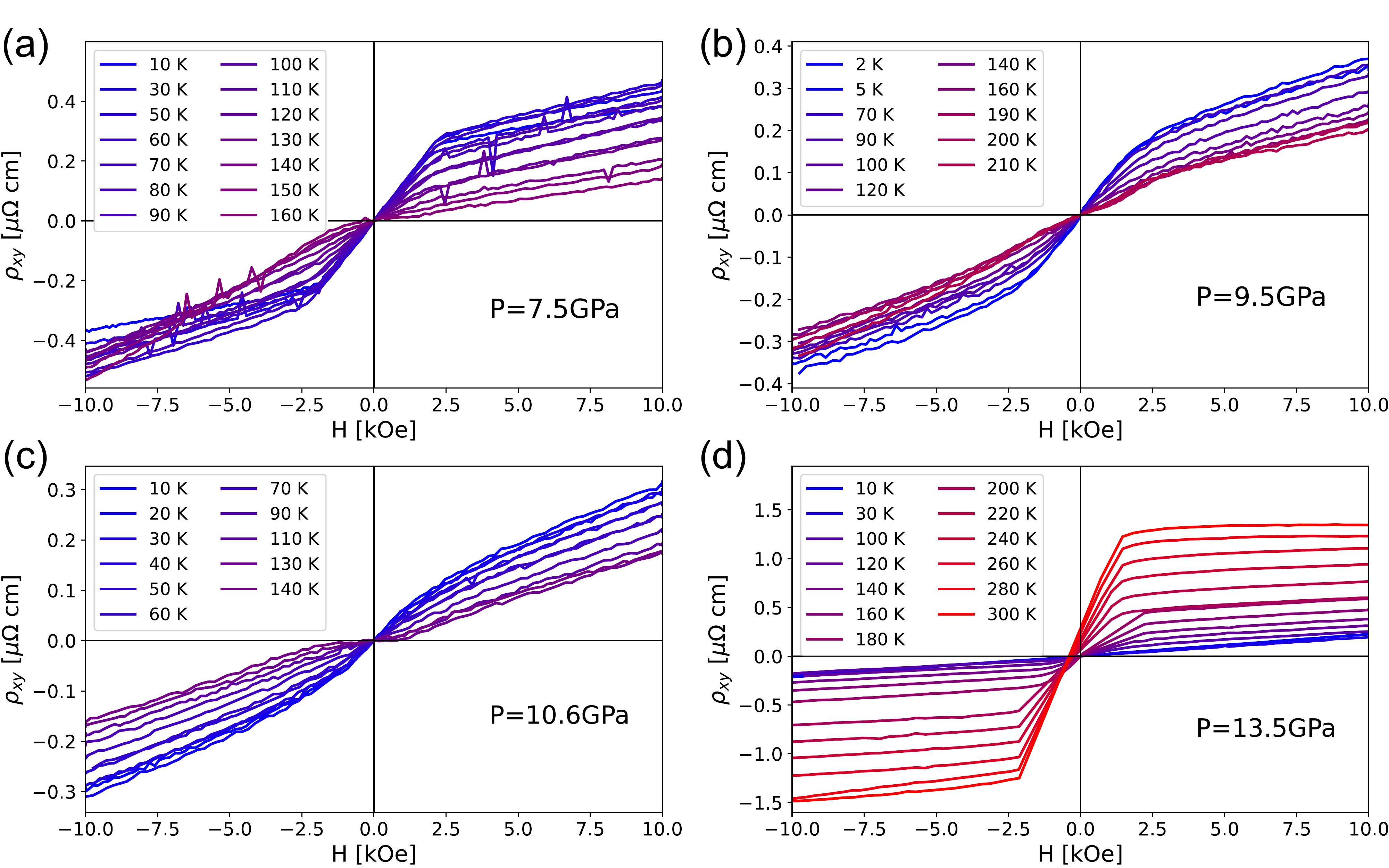}
\caption{\label{Figure 2}Measurements of the Hall effect in sample 1 at different temperatures and pressures of (a) 7.5\,GPa, (b) 9.5\,GPa, (c) 10.6\,GPa and (d) 13.5\,GPa. The steep slopes at low fields are due to the AHE. As can be seen in panel (a) for lower pressure, due to the non-perfect Van der Pauw geometry the raw data is not quite anti-symmetric as it should be. Therefore, {\rhoXX} contributions intermix in the measurements of {\rhoXY}. This effect is more prominent at lower pressures due to a higher longitudinal resistance, thus making the {\rhoXY} raw data less anti-symmetric.}
\end{figure*}

\section{Results and Discussion}
Fig.~\hyperref[Figure 1]{1(a)} shows that gradual application of pressure results in a significant drop to the sample resistance, which begins to saturate at pressures of $\sim6$\,GPa where an insulator-to-metal  transition occurs in agreement with Ref.~\cite{yu2019pressure,bhoi2021nearly} (see sections \hyperref[section S2]{S2} and \hyperref[section S3]{S3} in the Supplemental Materials for resistivity versus temperature plots). {\Tc} as a function of pressure from our AHE at high temperatures, is also shown in Fig.~\hyperref[Figure 1]{1(a)} and will be discussed later in this section. We note that the pressure at which we observe the insulator-to-metal  transition, and the dependence of Curie temperatures on pressure, are consistent with previous reports, indicating the similarity in sample quality and pressure conditions.

The inset of Fig.~\hyperref[Figure 1]{1(a)} shows the Hall coefficient as a function of pressure at $T=2$\,K, which exhibits a dome-like behavior as a function of pressure. In the metallic state (at $6<P<14.5$\,GPa), the positive sign of the Hall coefficient indicates that transport is hole-dominated at all temperatures, as demonstrated for 10.6\,GPa in Fig.~\hyperref[Figure 1]{1(b)}. The full data set is available in section \hyperref[section S4]{S4}. Fig.~\hyperref[Figure 1]{1(c)} shows that at the edges of the dome, both electrons and holes contribute to transport, as can be seen by the flattening and sign change of the Hall slope as a function of temperature. We note that a similar behavior also occurs at other pressures in the vicinity of 14\,GPa, as shown in Supplemental Materials, sections \hyperref[section S4]{S4} and \hyperref[raw data + AS]{S7}. These most likely originate from hole-like Te $5p$ and electron-like Cr $3d$ bands. It should be mentioned that measurements of the Hall effect were impracticable at pressures below $\mathrm{3.2\, GPa}$~due to the large longitudinal resistivity relative to the magnitude of the Hall effect (Supplemental Material section \hyperref[raw data + AS]{S7}).

Above 5.6\,GPa, when {\CGT} enters the metallic state, a significant AHE signal is observed. Fig.~\hyperref[Figure 2]{2(a)} shows a characteristic behavior of the AHE in the intermediate pressure regime. Here, the AHE is the strongest at low temperatures and monotonically weakens as the temperature increases. From this data, it is clear that the AHE persists to much higher temperatures than the Curie temperature at ambient pressure ($\sim 67$\,K). We note that the AHE can occur in paramagnetic materials~\cite{maryenko2017observation, PhysRevB.68.045327}, and therefore its presence at high temperatures does not necessarily prove enhancement of {\Tc}. However, since the trend observed by our measurements is a smooth continuation of the trend observed by magnetometry measurements~\cite{bhoi2021nearly} shown in Fig.~\hyperref[Figure 1]{1(a)}, we interpret the persistence of the AHE as an increase of {\Tc}. Fig.~\hyperref[Figure 2]{2(b)} shows measurements of the AHE at $13.5$\,GPa, characteristic of the high-pressure regime between $13.5$ and $17.6$\,GPa. At low temperatures (blue curves), the AHE is completely absent from measurements of $\rho_{xy}$, as can be seen by the absence of the steep low field AHE slope. As the temperature increases, the AHE gradually appears and is enhanced at elevated temperatures. We note that the AHE does not decay even at room temperature, continuing the trend observed in Ref.~\cite{bhoi2021nearly}, thus possibly indicating that {\Tc} in {\CGT} surpasses room temperature for this pressure range.

%Figure 2 was here

The qualitative differences in the evolution of the AHE with temperature suggest that different mechanisms are at play in different pressure regimes, i.e.~the intrinsic, skew-scattering, and side jump scattering mechanisms can change with pressure and temperature. To disentangle the contributions to the AHE, we separate the anomalous Hall resistivity {\rhoAHE} to intrinsic and extrinsic sources and follow Hou {\etal}~\cite{hou2015multivariable} who further distinguish between the static (temperature-independent) and dynamic (temperature-dependent) contributions to side jump and skew-scattering (i.e.~extrinsic) mechanisms. At low temperatures, quasiparticles are frozen, and there are no dynamic scattering events. Therefore, $\rho_{\rm AHE}(T=0)$ is a static effect, either intrinsic with geometric origins or extrinsic emanating from disorder. At higher temperatures, quasiparticles are thermally activated, and dynamic extrinsic scattering mechanisms may contribute to the AHE. We turn to interpreting our results in light of these distinctions in the three pressure regimes.

\begin{figure}[t]
\includegraphics[width=\columnwidth]{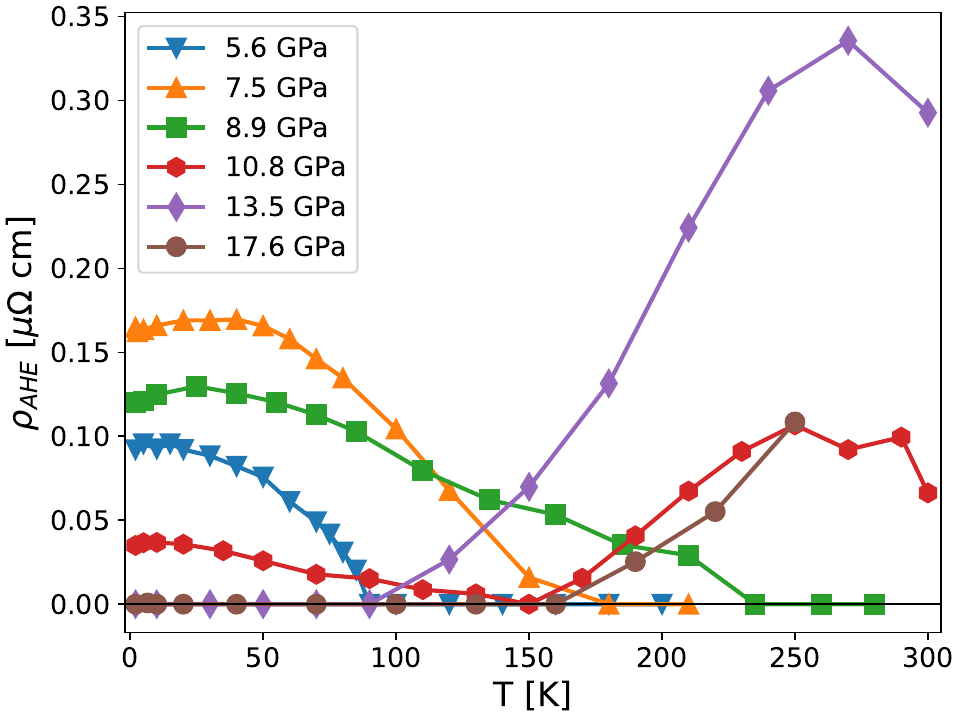}
\caption{\label{Figure 3}{\rhoAHE} as a function of temperature for various pressures, measured in sample 2. A similar plot for sample 1 is shown in section \hyperref[section S6]{S6} in the Supplemental Material. At the intermediate pressure regime, between the insulator-to-metal  transition and 13\,GPa, $\rho_{\rm AHE}\neq 0$ at low temperatures and decays smoothly as the temperature increases. In contrast, in the high-pressure regime, at low temperatures $\rho_{\rm AHE}=0$, and increases as the temperature increases.}
\end{figure}

{\rhoAHE}~is extracted using a procedure similar to Ref.~\cite{liu2018giant} detailed in the Supplemental Material, section \hyperref[section S5]{S5}. In the insulating state, the values of the longitudinal resistance ({\rhoXX}) intermixed in the measurements of {\rhoXY} are dominant for fields smaller than 0.2\,T (see section \hyperref[raw data + AS]{S7} in the Supplemental Material). Therefore, our measurements are insensitive to small anomalous Hall signals at this pressure range. Fig.~\ref{Figure 3} shows {\rhoAHE} as a function of temperature for various pressures above 5.6\,GPa. At the intermediate pressure regime, at low temperatures, $\rho_{\rm AHE}\neq 0$. This suggests that the pressure tuning into the metallic state activates either an intrinsic contribution to the AHE or a static extrinsic scattering mechanism. The signal decays as the temperature increases, and its disappearance marks the Curie temperature. In the high-pressure regime ($P>13.5$\,GPa), $\rho_{\rm AHE}= 0$ at low temperatures and smoothly increases as the temperature increases, persisting up to room temperature above which we could not heat our diamond anvil cells.

The absence of AHE indicates either a perfect cancellation of contributions from various scattering mechanisms, meaning that the sum of all the various contributions to the AHE is zero, or the complete nullification of all of them. Perfect cancellation typically occurs when two mechanisms contribute to {\rhoAHE} with opposite signs, which typically occurs at a specific temperature as seen, for example, in Ref.~\cite{fang2003anomalous, P.Pureur2004, PhysRevLett.96.037204, haham2011scaling}. In contrast, in {\CGT} at $P>13$\,GPa, $\rho_{\rm AHE} = 0$ for a wide temperature range (over 150\,K at 17.6\,GPa) rather than crossing zero at a particular temperature. Perfect accidental cancellation of various scattering mechanisms at such a wide temperature range is unlikely. Therefore we deduce that for $P>13$\,GPa, all scattering mechanisms are negligible at low temperatures, and the behavior shown in Fig.~\ref{Figure 3} is dominated by scattering off of thermally activated quasi-particles.

\begin{figure}[t]
\includegraphics[width=\columnwidth]{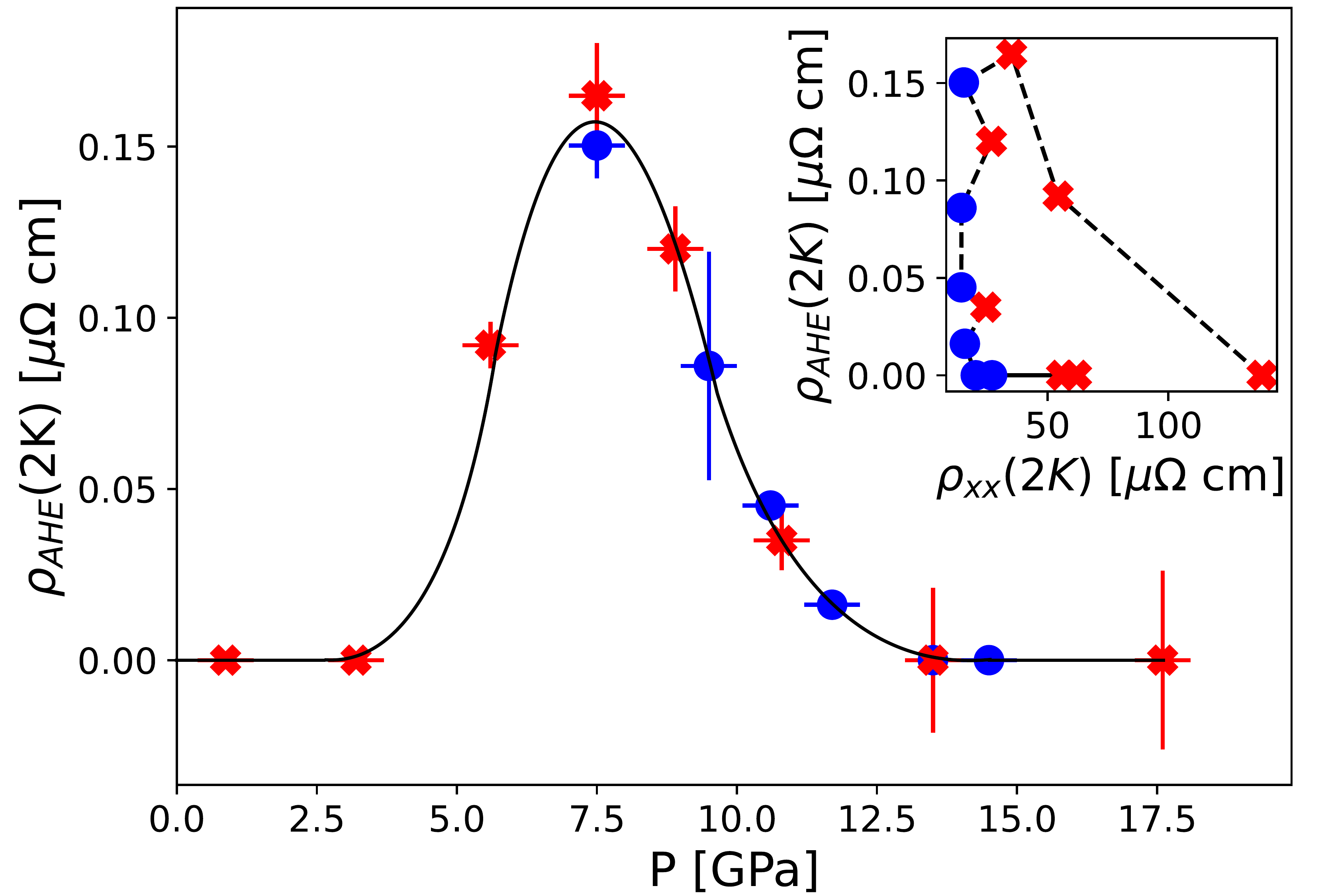}
\caption{\label{Figure 4}{\rhoAHE} measured at 2\,K as a function of applied pressure. The black line is a guide to the eye. The inset shows {\rhoAHE}, measured at 2\,K, as a function of the longitudinal resistivity \rhoXX, showing a hysteretic behavior that deviates from the parabolic relation in equation \ref{equation1}. The blue and the red points are from the first and second cells, respectively. Their resistivity values are scaled by a single geometric factor of order unity.} 
\end{figure}

In Fig.~\ref{Figure 4}, we plot {\rhoAHE} at 2\,K as a function of the pressure, which exhibits a dome-like shape starting at the insulator-to-metal  transition and finishing around 13\,GPa. To the best of our knowledge, this behavior has not been observed in the past. Typically, ferromagnets exhibit a monotonic behavior of the AHE as a function of pressure, as seen, for example, in $\mathrm{CeAlSi}$~\cite{PhysRevResearch.5.013068}
and $\mathrm{Co_3Sn_2S_2}$~\cite{liu2020pressure}, where in the former, the AHE is generated by skew scattering and in the latter by the intrinsic Berry phase. 
To understand this behavior, we look at the relation between {\rhoAHE} and {\rhoXX}, which at low temperatures, in the absence of dynamic scattering, simplifies to~\cite{hou2015multivariable}:
\begin{equation} \label{equation1}
    \rho_{AHE}=\alpha\rho_{xx}+\beta_0\rho_{xx}^2 \,,
\end{equation}
where $\alpha$~represents contributions from skew scattering, and $\beta_0$~is a mixture of intrinsic and static side jump mechanisms. In our experiment, we tune {\rhoXX} by changing the hydrostatic pressure P. The inset to Fig.~\ref{Figure 4} shows {\rhoAHE} as a function of \rhoXX$(P)$~at a constant temperature 2\,K, where a clear non-parabolic hysteretic behavior is observed. This means that the application of pressure changes not only {\rhoXX} but also $\alpha$~and $\beta_0$ and therefore transport measurements alone could not disentangle the contributions of the intrinsic and the static extrinsic mechanisms to the AHE. However, in {\CGT}, the AHE dome onsets and ends in regimes where mixed transport of electrons and holes is observed (Fig.~\hyperref[Figure 1]{1(a)} inset), indicative of Fermi surface deformations. In the next section, we compare our experimental results to ab-initio calculations of the Fermi surface, Berry curvature and the expected intrinsic AHE conductivity as a function of pressure to better understand the interplay between them.

\begin{figure*}[t]
\includegraphics[width=\textwidth]{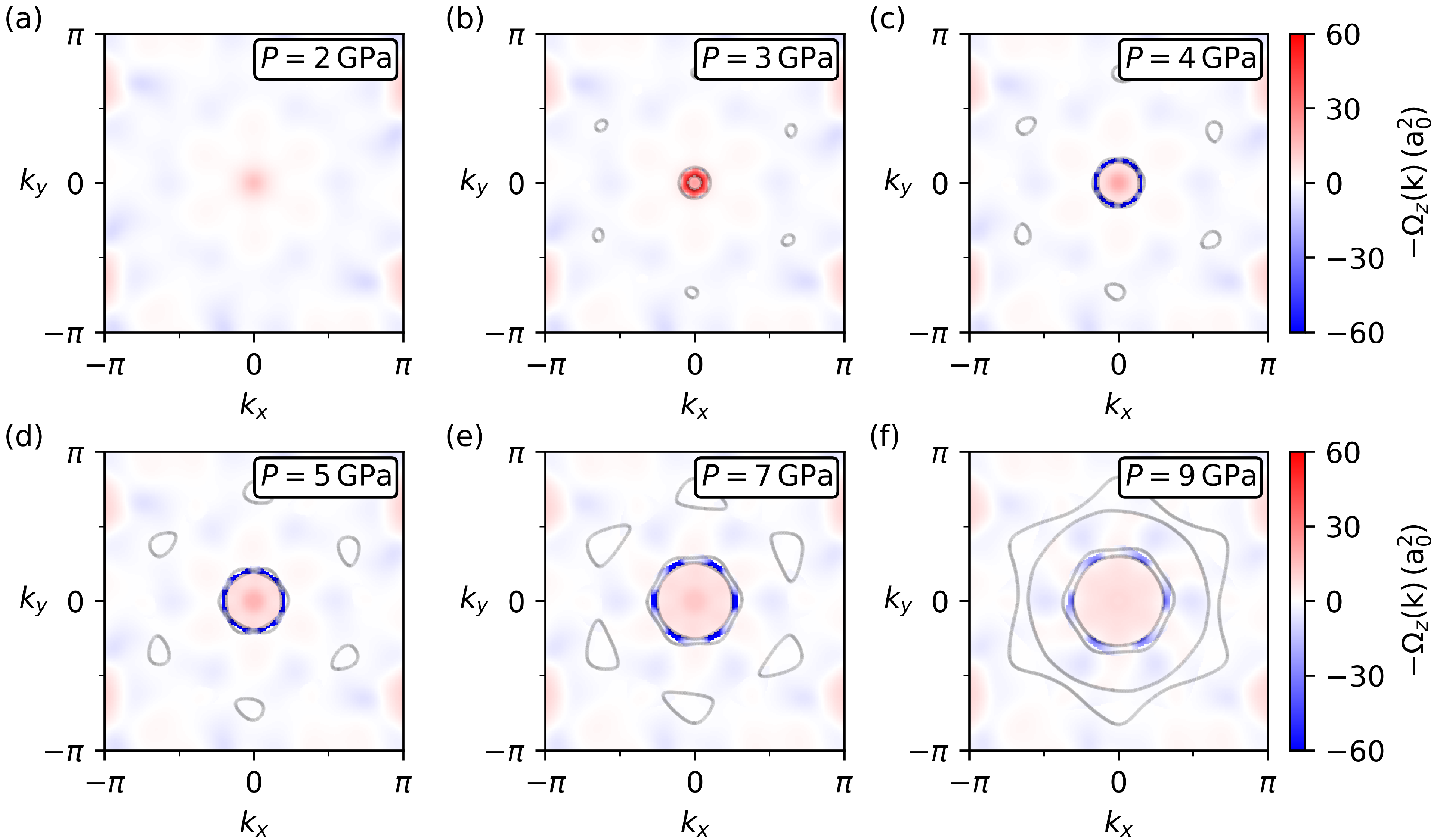}
\caption{Fermi surface (indicated by grey overlay) and total Berry curvature $-\Omega_z$ (indicated by the color scale) in the $k_x$-$k_y$-plane (at $k_z=0$) calculated from DFT in the ferromagnetic state for various pressures. (a) At $P=2$\,GPa small contributions to the total Berry curvature exist, even in the absence of a Fermi surface. (b) At $P=3$\,GPa two small hole pockets are present in the center of the BZ, while six electron pockets are located farther from the center. Strong contributions to the total Berry curvature originate from the space between the two hole pockets and from inside the inner hole pocket. Both contributions are positive. (c)-(e) At higher pressures the contribution between the two hole pockets becomes negative, while the contribution from inside the inner hole pocket remains positive. The pressure-tuning of the total Berry curvature is generated by expansion of the Fermi surfaces and quantitative changes in contributions with opposite sign. (f) At $P=9$\,GPa the electron pockets have merged to form a single electron cylinder. Both positive and negative contributions to the total Berry curvature have decreased in magnitude compared to the low-pressure cases. Note that some values of the total Berry curvature lie far beyond the scale used in this figure. These values were cut off to allow for an interpretable visualization.}
\label{fig:BerryFermiCuts}
\end{figure*}

\begin{figure*}[t]
\includegraphics[width=\textwidth]{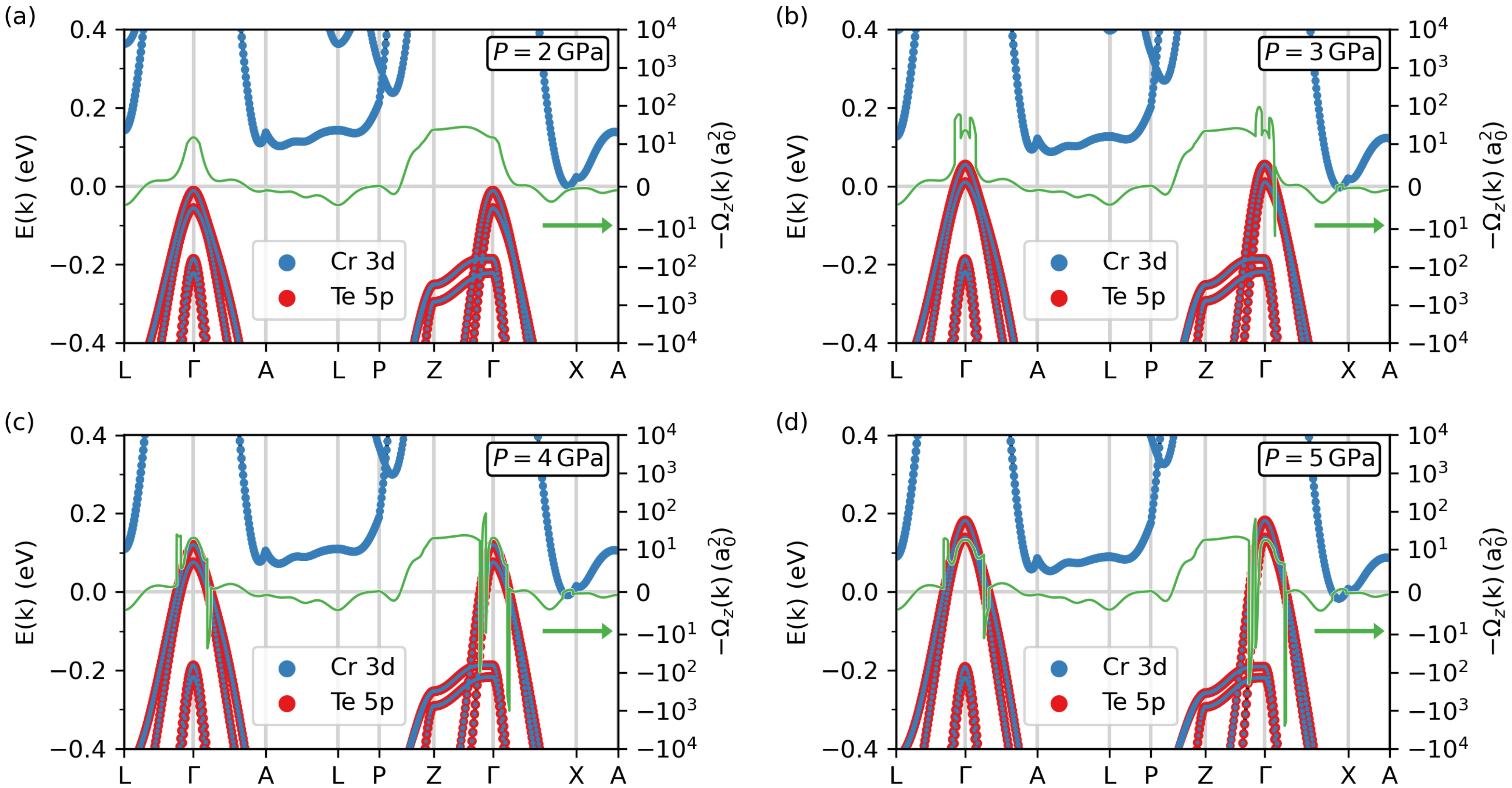}
\caption{Electronic band structure with orbital weights (left axis) and total Berry curvature $-\Omega_z$ (right axis) calculated from DFT in the ferromagnetic state for various pressures. (a) At $P=2$\,GPa there is no Fermi surface, but small contributions to the total Berry curvature are already present. (b) At $P=3$\,GPa hole pockets form around $\Gamma$, while electron pockets are present close to the X-point. The electron pockets carry almost exclusively Cr $3d$ weight and do not significantly affect the total Berry curvature. Major changes in total Berry curvature are induced by the hole pockets with mixed Cr $3d$ and Te $5p$ weights. (c)-(d) Further pressure modulates the total Berry curvature around the hole pockets in the center of the BZ, i.e.~close to $\Gamma$. Note that the total Berry curvature is plotted on a linear scale for values in the region $[-10, +10]$ and logarithmically otherwise, to make the extreme contributions visible.}
\label{fig:BerryBands}
\end{figure*}

\begin{figure}[t]
\includegraphics[width=\columnwidth]{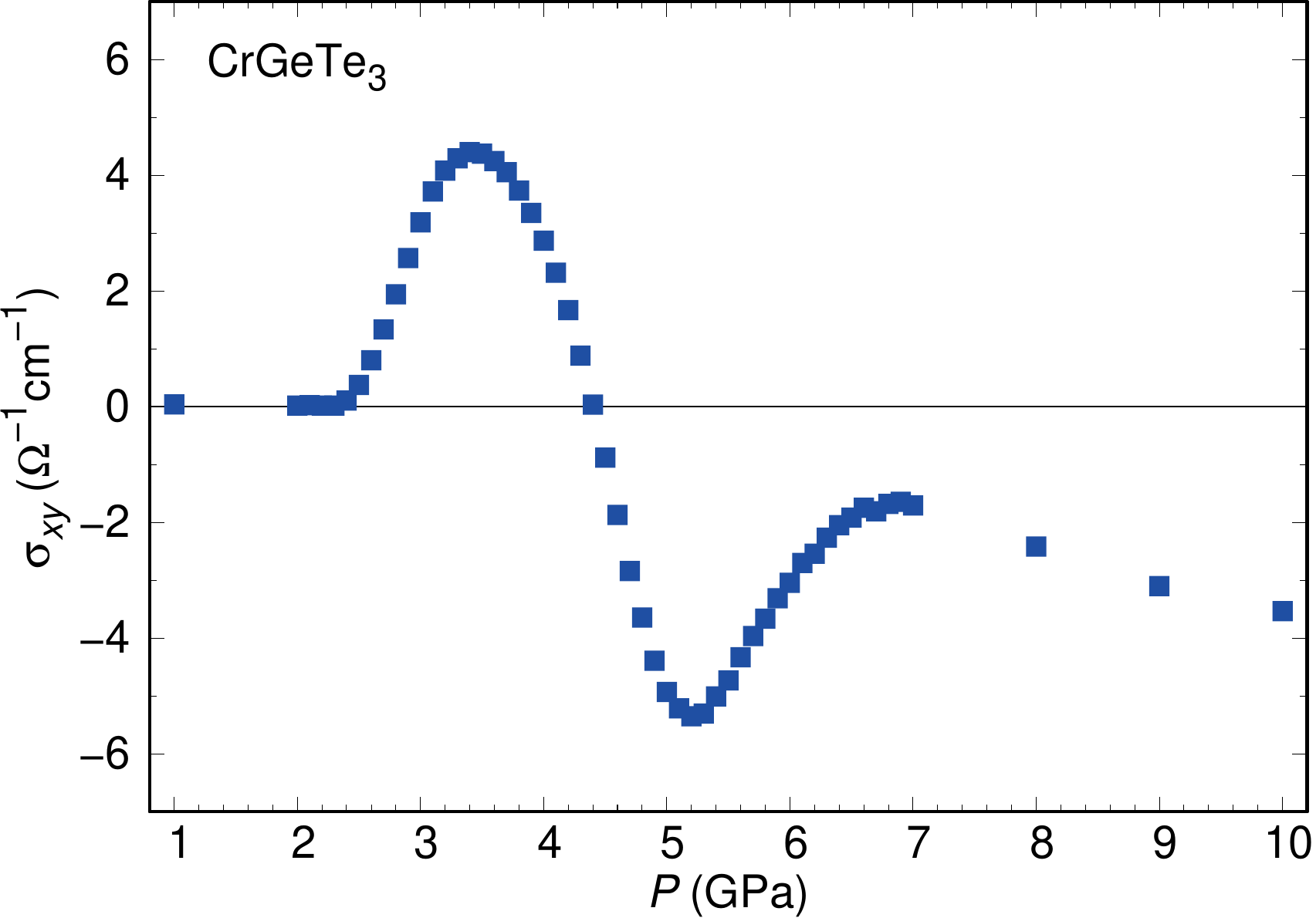}
\caption{Theoretical anomalous Hall conductivity in the xy-plane $\sigma_{xy}$ calculated from DFT in the ferromagnetic state. Pressure modulates the anomalous Hall conductivity and leads to a sign change at intermediate pressures. For each pressure value we performed ten independent adaptive Monte Carlo integrations. From these ten results we calculated the uncertainty as two standard deviations of the obtained conductivity values. For all pressures, the estimated uncertainty is smaller than the symbol size.}
\label{fig:TheoryAHEConductivity}
\end{figure}

\section{Theory - Results}
We first calculated Fermi surfaces and total Berry curvature of {\CGT} as a function of pressure. The results are shown in Fig.~\ref{fig:BerryFermiCuts}. In our DFT calculations, the insulator-to-metal transition occurs between $P=2$\,GPa and $P=3$\,GPa, so that all DFT results should be compared to experimental results at slightly higher pressures. In the insulating phase, {\CGT} already has small but nonzero total Berry curvature (see Fig.~\ref{fig:BerryFermiCuts}(a)). At the insulator-to-metal transition, both hole and electron pockets appear in the metallic phase. The respective bands do not overlap, i.e.~{\CGT} under pressure is a ferromagnetic semimetal. 

At $P=3$\,GPa (see Fig.~\ref{fig:BerryFermiCuts}(b)), two small hole pockets are present in the center of the BZ, while six electron pockets are located farther away from the $\Gamma$-point. Strong positive contributions to the total Berry curvature originate from the space between the two hole pockets and from inside the inner hole pocket. Large negative contributions are not yet present. At higher pressures, this changes as the contribution from the space between the two hole pockets becomes negative, while the contribution from inside the inner hole pocket remains positive (see Fig.~\ref{fig:BerryFermiCuts}(c)-(e)). The relative magnitude of positive and negative contributions changes as a function of pressure. At $P=9$\,GPa (see Fig.~\ref{fig:BerryFermiCuts}(f)), the electron pockets have merged to form a single electron cylinder. Both positive and negative contributions to the total Berry curvature are decreased in magnitude compared to the low-pressure cases. In summary, these figures show that the total Berry curvature in {\CGT} is substantially tuned by applying pressure.

To understand why electron pockets seemingly do not contribute to the total Berry curvature, we calculate the electronic band structure with orbital weights. As can be seen in Fig.~\ref{fig:BerryBands}, the hole bands close to the $\Gamma$-point carry both significant Cr $3d$ and Te $5p$ weight, while the electron bands that move below the Fermi level under pressure almost exclusively carry Cr $3d$ weight. Since spin-orbit coupling for Cr $3d$ is significantly weaker than for Te $5p$ states, a much smaller contribution of Cr $3d$ to the total Berry curvature can be expected. This differentiation in orbital weights explains why the appearance of electron pockets does not significantly affect the total Berry curvature.

Fig.~\ref{fig:BerryBands} also shows how the total Berry curvature on a high-symmetry path is affected by the application of pressure. The total Berry curvature in the absence of a Fermi surface is shown in Fig.~\ref{fig:BerryFermiCuts}(a). The pressure modulation occurs in tiny regions of momentum-space close to the Fermi surface (see Fig.~\ref{fig:BerryBands}(b)-(d)). These regions are so small that Fig.~\ref{fig:BerryBands}(b) shows a negative total Berry curvature close to $\Gamma$, which is not yet visible at the already high resolution of Fig.~\ref{fig:BerryFermiCuts}(b). This further illustrates the need for adaptive integration as implemented in our approach.

Finally, Fig.~\ref{fig:TheoryAHEConductivity} shows the calculated anomalous Hall conductivity $\sigma_{xy}$. We observe a strong modulation of the conductivity as a function of pressure. A maximum at around $P=3.5$\,GPa is followed by a sign change around $P=4.4$\,GPa and a minimum close to $P=5$\,GPa. The onset of conductivity between $P=2$\,GPa and $P=3$\,GPa is clearly related to the insulator-to-metal transition observed in our Fermi surfaces (see Fig.~\ref{fig:BerryFermiCuts}) and the electronic band structure (see Fig.~\ref{fig:BerryBands}).

When comparing these theoretical results for the conductivity to the experimental anomalous Hall resistivity, one must take into account that the longitudinal resistivity also enters when calculating the conductivity from experimental data:
\begin{equation}
\sigma_{xy} = -\frac{\rho_{xy}}{\rho_{xx}^2 + \rho_{xy}^2}
\label{eq:conductivityresisitivty}
\end{equation}
Since the experimental longitudinal resistivity is much larger than the transversal one (see Fig.~\ref{Figure 4}), we can expect that transverse resistivity and conductivity are roughly proportional to each other. Note, however, the additional minus sign in Eq.~\eqref{eq:conductivityresisitivty}. Although the experiment seems to observe a pressure modulation similar to the high-pressure region of our calculation (different sign due to Eq.~\eqref{eq:conductivityresisitivty}), the positive anomalous Hall conductivity (negative resistivity) of our calculations is not seen in experiment. Furthermore, even after correctly converting according to Eq.~\eqref{eq:conductivityresisitivty}, the absolute scales do not match. The theoretically calculated resistivity is several orders of magnitude smaller. In our DFT calculations we also rotated the spin quantization axis from the $k_z$-direction toward the $k_x$-direction and observed only a small but systematic effect on the Berry curvature (see Supplemental Material section \hyperref[section S8]{S8}). The giant effect of magnetic field rotation on the AHE in ferromagnetic Weyl semimetal EuB$_6$~\cite{Shen2023} seems absent in calculations for {\CGT}. This shows that the Berry curvature is not the sole contribution to the anomalous Hall conductivity in {\CGT} and that extrinsic mechanisms must also be at play. 

The two extrinsic mechanisms that are to be considered are skew scattering and side jumps. Calculation of the magnitude of skew scattering and side jumps alike include matrix elements that contain the Bloch functions before and after the scattering event~\cite{nagaosa2010anomalous}. Our ab initio calculations show that the band structure and Bloch wavefunctions change considerably due to the effects of hydrostatic pressure which suggests that together with changes to the Berry curvature the magnitude of the extrinsic mechanisms should change as well. However, in absence of exact knowledge of the scattering potential it is impossible to calculate the magnitude of these effects. However, assuming an effective mass of order unity, one can calculate the typical lifetime for carrier scattering from our Hall and {\rhoXX} data. For pressures above the metal insulator transitions one gets values of the order of 1 to 10 femtoseconds. This indicates that side jumps are most likely prominent in their contribution to the AHE in {\CGT}~\cite{nagaosa2010anomalous}.

\section{Conclusions}                  
In summary, we have measured and calculated the AHE conductivity in {\CGT} as a function of applied hydrostatic pressure. Both experiment and theory show that the anomalous Hall effect in {\CGT} can be substantially tuned by pressure. Comparison between experiment and theory suggests that the measured values are not purely intrinsic and that the majority of the signal originates from extrinsic skew scattering. Our calculations reveal significant changes to the band structure of {\CGT} as pressure is increased in the metallic phase. These are accompanied by changes to the Bloch wave functions which affect the matrix elements of skew scattering and side jump mechanisms.  

\begin{acknowledgments}
A.R. acknowledges support from the Zuckerman Foundation, and the Israel Science Foundation (Grant No. 1017/20).
S.D.W., B.R.O., and P.S. gratefully acknowledge support via the UC Santa Barbara NSF Quantum Foundry funded via the Q-AMASE-i program under award DMR-1906325.
G.Kh.R. acknowledges the Israel Science Foundation (Grants No. 1748/20).
T.H.\ acknowledges financial support by the European Research Council (ERC) under grant QuantumCUSP (Grant Agreement No. 101077020). 
H.O.J. acknowledges support through JSPS KAKENHI
Grant No. 24H01668.
G.S. thanks Shay Sandik, Itai Silber, and Gal Tuvia for the help with the cryogenic equipment. G.S. acknowledges support from the Israeli Clore fellowship. 
\end{acknowledgments}

\providecommand{\noopsort}[1]{}\providecommand{\singleletter}[1]{#1}%

\newpage

\onecolumngrid
\setcounter{secnumdepth}{0}  % Disables section numbering

\vspace{1cm}
\begin{center}
\begin{spacing}{1.25} 
    \textbf{\large Supplemental Material for: "Pressure tuning of intrinsic and extrinsic sources to the anomalous Hall effect in \CGT"}
\end{spacing}
\end{center}

\setcounter{equation}{0}
\setcounter{figure}{0}
\setcounter{table}{0}
\setcounter{page}{1}
\setcounter{section}{0}
\makeatletter
\renewcommand{\theequation}{S\arabic{equation}}
\renewcommand{\thefigure}{S\arabic{figure}}
\renewcommand{\thetable}{S\arabic{table}}
\renewcommand{\thesection}{\textbf{S\arabic{section} - }}

\renewcommand{\bibnumfmt}[1]{[S#1]}
\renewcommand{\citenumfont}[1]{S#1}

\vspace{-0.8cm}
\section{S1 - Data analysis using $\mathrm{\rho}$ rather than with $\mathrm{\sigma}$}\label{section S1}
\vspace{-1em}
Throughout the manuscript, we chose not to work with conductivity but rather with resistivity for the following reason: Calculations of $\mathrm{\rho_{xx}}$~and $\mathrm{\rho_{xy}}$~from resistance data incorporate division by multiplicative factors, $\mathrm{c_{xx}}$~and $\mathrm{c_{xy}}$~respectively, originating from geometry. In contrast to experiments under ambient conditions in which the sample geometry could be well defined and the positions of the leads determined to great accuracy, experiments in a DAC result in inevitable uncertainties in $\mathrm{c_{xx}}$~and $\mathrm{c_{xy}}$. These result in a linear multiplicative factor for $\mathrm{\rho_{xx}}$~and $\mathrm{\rho_{xy}}$. In contrast, the extraction of the conductivity values involves the inversion of the resistivity matrix, which results in the following expression for the transverse conductivity: $\mathrm{\sigma_{xy}=-\frac{\rho_{xy}}{\rho_{xx}^2+\rho_{xy}^2 }}$, when expressed as a function of the measured quantities as well as $\mathrm{c_{xx}}$~and $\mathrm{c_{xy}}$~the following form is obtained: $\mathrm{\sigma_{xy}=-\frac{\frac{R_{xy}}{c_{xy}}}{(\frac{R_{xx}}{c_{xx}})^2+(\frac{R_{xy}}{c_{xy}})^2 }}$. This nonlinear relation between $\mathrm{\sigma_{xy}}$ and the inevitable error in $\mathrm{c_{xx}}$~and $\mathrm{c_{xy}}$~will result in resistivity dependent errors in the estimation of any conductivity quantities from our data. 
\par

Throughout the manuscript the data in both samples was factored by a geometric factor of order unity which was extracted by comparing the resistivity and Hall coefficients at pressure points common to both samples (7.5\,GPa).

%\vspace{0.2em}

\section{S2 - The MIT - R(T) at different pressures}\label{section S2} 
\vspace{-1em}
The longitudinal resistance as a function of temperature, normalized at $\mathrm{9.5\, K}$, measured at different pressures in the first cell. The change in the behavior of the graph from decreasing to increasing as a function of the temperature in the low-temperature regime by application of pressure indicates a metal-insulator transition driven by the application of pressure on the \CGT.

\vspace{-0.7em}

\begin{figure}[ht]
  \centering
  \includegraphics[width=0.7\columnwidth]{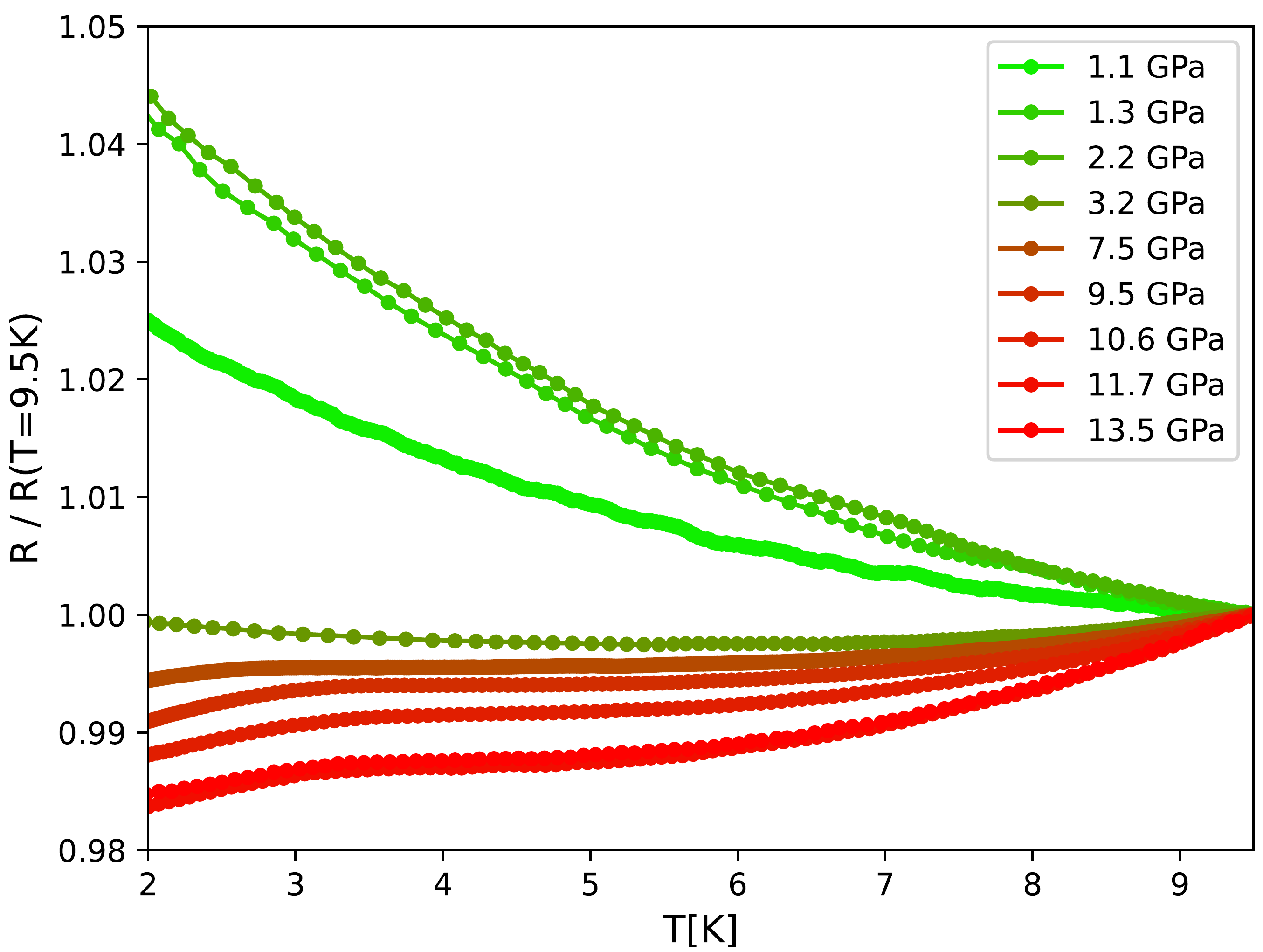}
  
  \caption{The longitudinal resistance as a function of temperature, normalized at $\mathrm{9.5\, K}$, measured at different pressures in the first cell. The change in the behavior of the graph from decreasing to increasing as a function of the temperature by application of pressure indicates a MIT driven by the application of pressure on the \CGT.}
  
  \vspace{-0.5em}
\end{figure}

\newpage

\section{S3 - \rhoXX~at different pressures and temperatures}\label{section S3}
\vspace{-1em}
In Figure \ref{rhoxx(T) graphs}, we present the longitudinal resistivity as a function of temperature in the metallic state, measured on both samples. As can be seen, they show very similar behavior as all of them are monotonic - increasing with the temperature and showing similar values. As such, going back to our measurements of the AHE as a function of the temperature (see Figures 2 and \ref{rhoAH(T) 1st sample}), the observed behaviors cannot be explained just by the scaling of \rhoAHE~with the \rhoXX. First, the scaling of \rhoXX~cannot explain the change in the behavior of the AHE between the intermediate pressure regime ($\mathrm{5.6<P<13}$\,GPa) and the high-pressure regime ($\mathrm{13\, Gpa<P}$). Second, it cannot explain why at $\mathrm{13.5\, GPa}$, the AHE is stronger than at higher pressures in the first sample and thus probably not also in the second cell. Finally, going back to the low-temperature behavior of the AHE as shown in Figure 4, the values of \rhoXX~at low temperatures (shown in log-scale in Figure \ref{rhoxx(P) at 2K}) cannot explain the dome-like behavior of the AHE.

\begin{figure}[ht]
  \centering
  \includegraphics[width=0.8\columnwidth]{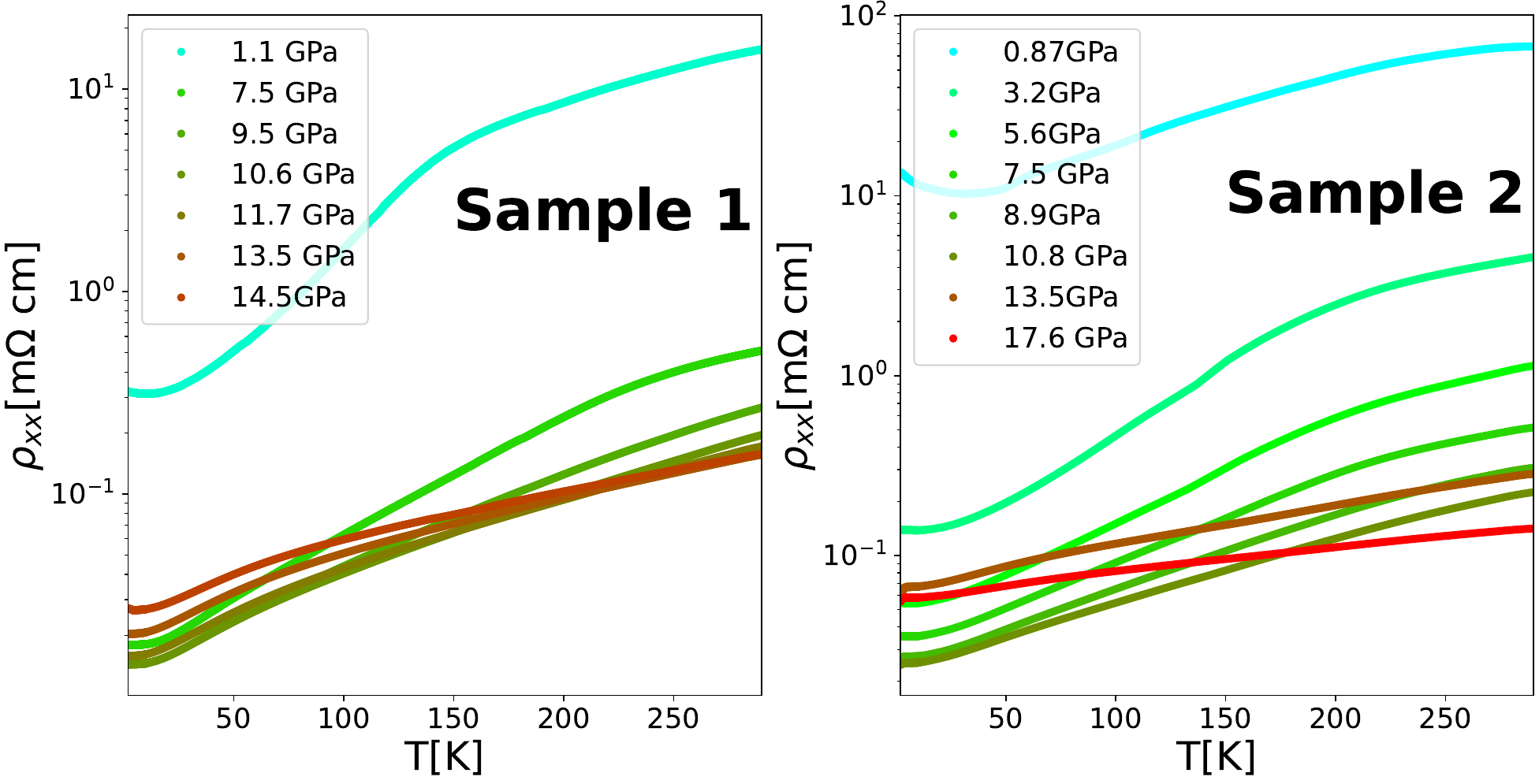}
  
   \caption{The longitudinal resistivity as a function of temperature at different pressures presented in log-scale, on the left in the first sample and on the right in the second sample.}
  
  \vspace{-0.5em}
\end{figure}

\vspace{-0.5cm}

\begin{figure}[ht]
  \centering
  \includegraphics[width=\columnwidth]{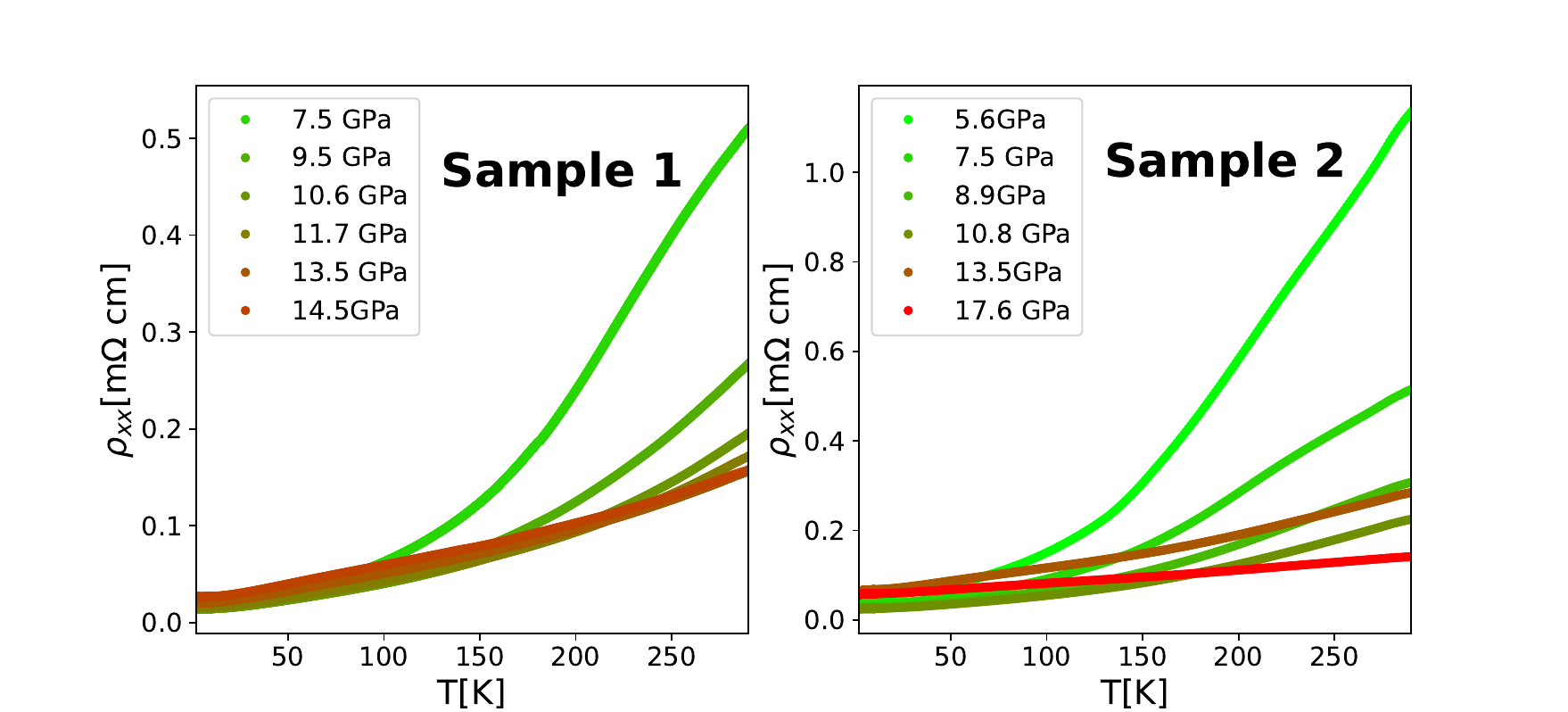}
  
   \caption{\label{rhoxx(T) graphs}The longitudinal resistivity as a function of temperature at different pressures in the metallic state, on the left in the first sample and on the right in the second sample.}
  
  \vspace{-0.5em}
\end{figure}

\vspace{-0.5cm}

\newpage

\begin{figure}[ht]
  \centering
  \includegraphics[width=0.65\columnwidth]{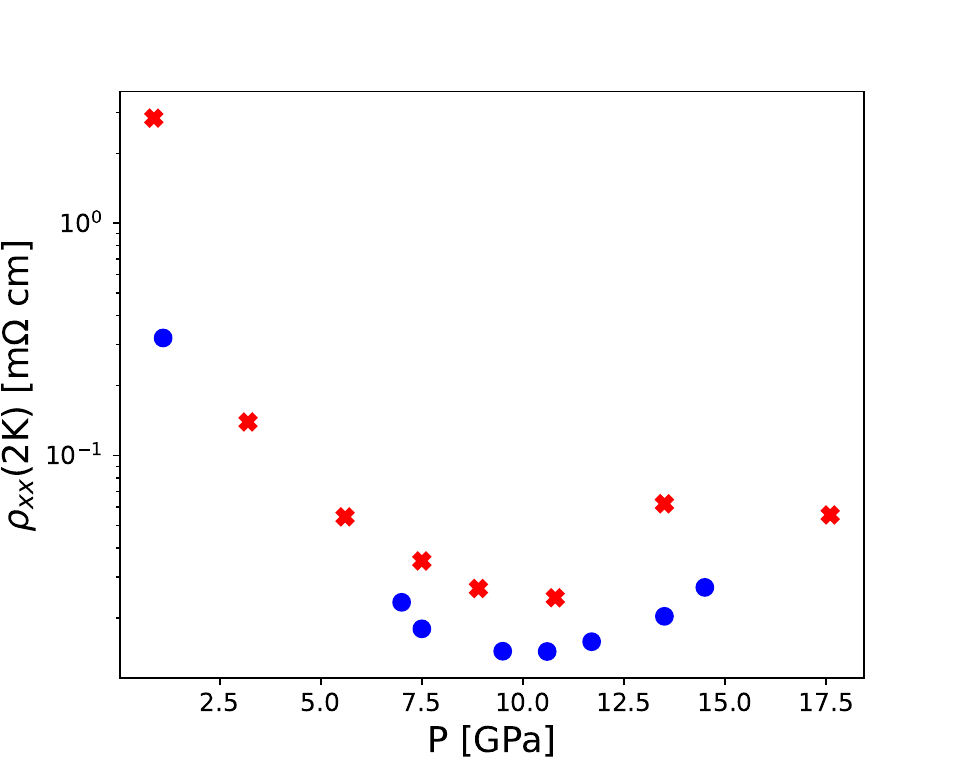}
  
   \caption{\label{rhoxx(P) at 2K}The longitudinal resistivity at low temperature (2\,K) as a function of the pressure in log-scale. The Blue and the red points are from the first and second cells, respectively. Their resistivity values are scaled by a single geometric factor of order unity which was used throughout the manuscript for each longitudinal measurement.}
  
  %\vspace{11cm}
\end{figure}

\newpage

\section{S4 - The Hall slopes measured at different pressures and temperatures\label{RH section}}\label{section S4}
\vspace{-1em}

Here, we present the Hall slopes as a function of temperature, measured at different pressures. 
In most measurements, the Hall slope is positive, meaning that although there is a mix of electrons and holes in all pressures, in most of the pressures, we can treat the transport as of hole-like charge carriers. However, at $\mathrm{3.2\, GPa}$ and $\mathrm{14.5\, GPa}$, there is a change in the sign of the Hall slope, indicating that at these pressures, both holes and electrons contribute to transport where their contributions are temperature dependent, which can also be seen in Figure 1 (b) in the text. At these pressures, we cannot treat the transport as dominated by a single charge carrier. The fact that the Hall slope changes sign as a function of the temperature at those pressures but not before might indicate changes in the band structure of the \CGT, which may result in a change in the integrated Berry curvature.

\vspace{-0.4cm}

\begin{figure}[ht]
  \centering
  \includegraphics[width=0.65\columnwidth]{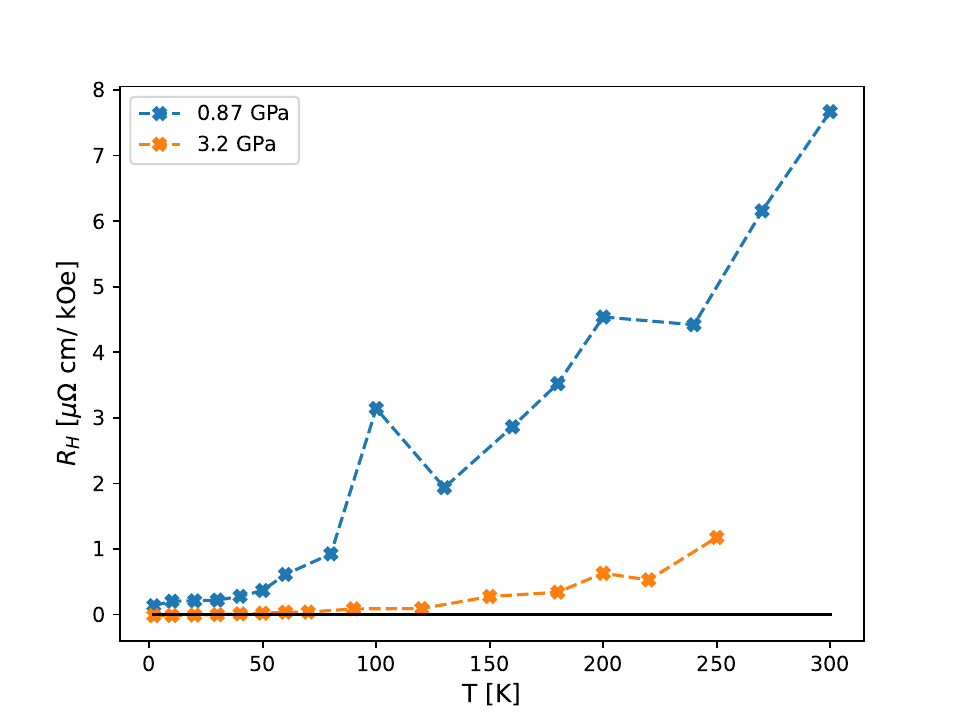}
  
  \caption{The Hall slopes as a function of temperature, measured at $\mathrm{0.87\, GPa}$ and at $\mathrm{3.2\, GPa}$.}
  
  \vspace{-0.5em}
\end{figure}

\vspace{-0.5cm}

\begin{figure}[ht]
  \centering
  \includegraphics[width=0.65\columnwidth]{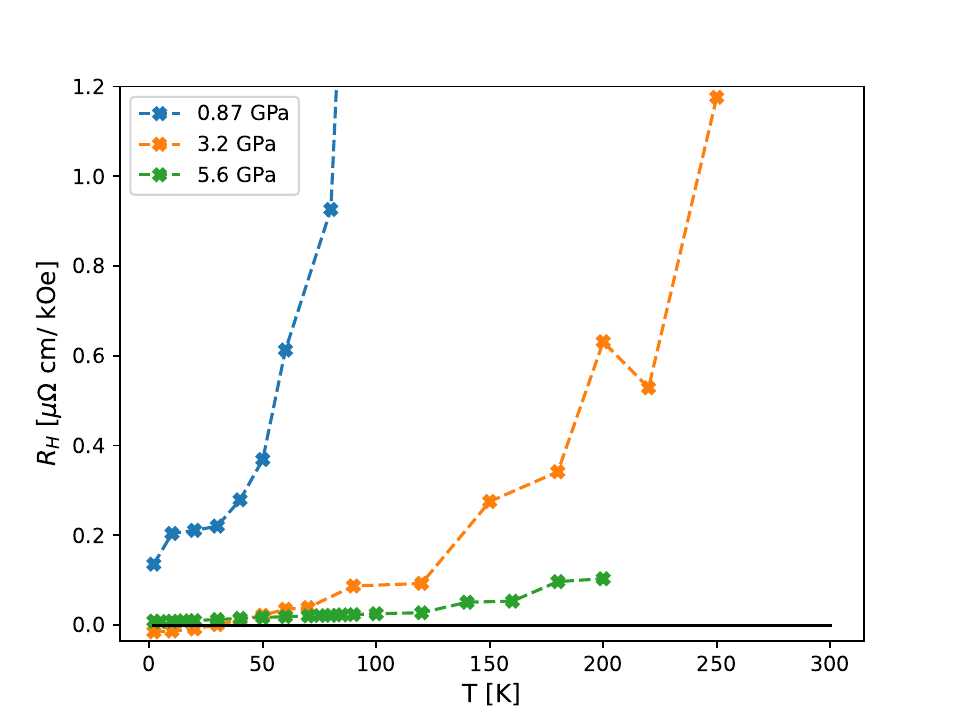}
  
  \caption{The Hall slopes as a function of temperature, measured at $\mathrm{0.87\, GPa}$, $\mathrm{3.2\, GPa}$, and $\mathrm{5.6\, GPa}$.}
  
  \vspace{-0.5em}
\end{figure}

\begin{figure}[ht]
  \centering
  \includegraphics[width=0.65\columnwidth]{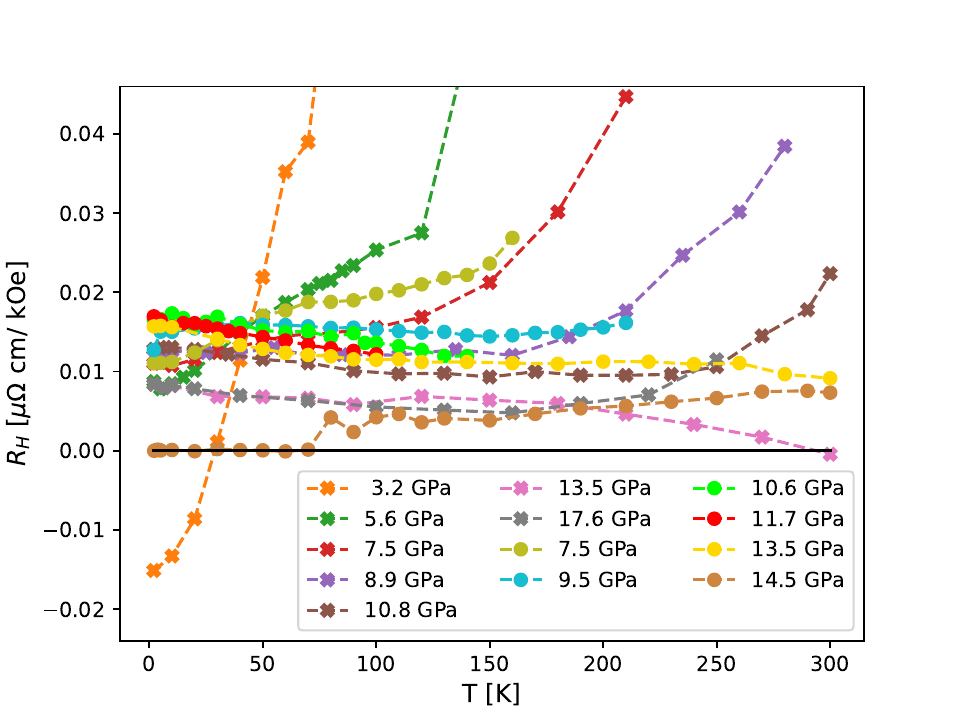}
  
  \caption{The Hall slopes as a function of temperature, measured at different pressures. The dots represent data from the first cell, and the Xs denote measurements from the second cell.}
  
  \vspace{-0.5em}
\end{figure}

\clearpage

\section{S5 - The extraction of \rhoAHE~from the measurements}\label{section S5}
\begin{figure}[ht]
  \centering
  \includegraphics[width=1\columnwidth]{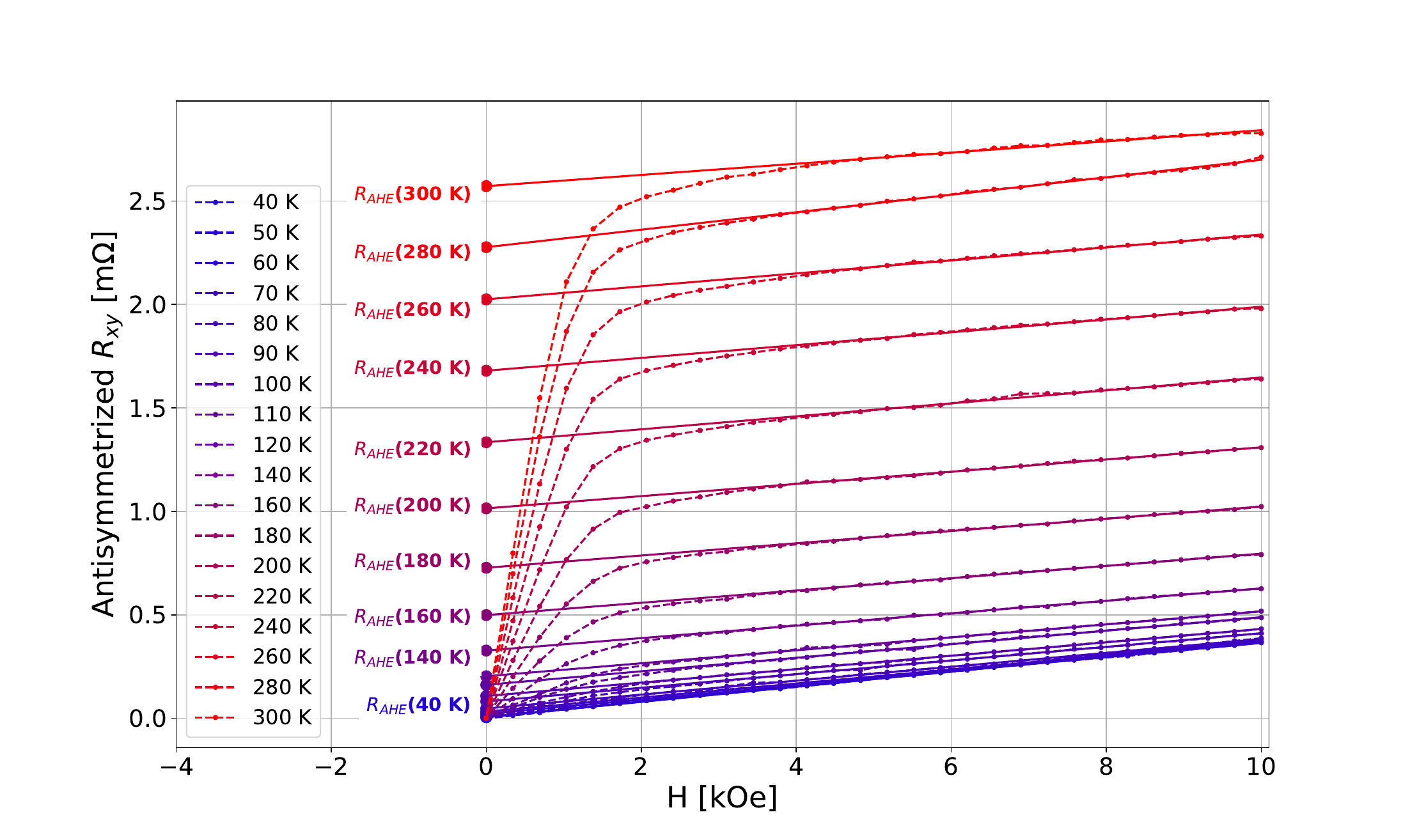}
  
  \caption{Here we show an example of how we extracted the AHE resistance ($\mathrm{R_{AHE}}$) for each pressure at different temperatures. The figure displays the antisymmetrization of the raw data of $\mathrm{R_{xy}}$~as a function of the applied field H, measured at several different temperatures for the first sample in the metallic state at a pressure of $\mathrm{13.5\, GPa}$. The solid lines are the linear fit for the high-fields regime ($\mathrm{4\, kOe<H}$) at each temperature, and the big dots represent the intersection of each fit with the y-axis. The intersection of each fit is $\mathrm{R_{AHE}}$ measured at each temperature. \label{extracting R_AHE}}
  
  \vspace{-0.5em}
\end{figure}

\begin{figure}[ht]
  \centering
  \includegraphics[width=0.7\columnwidth]{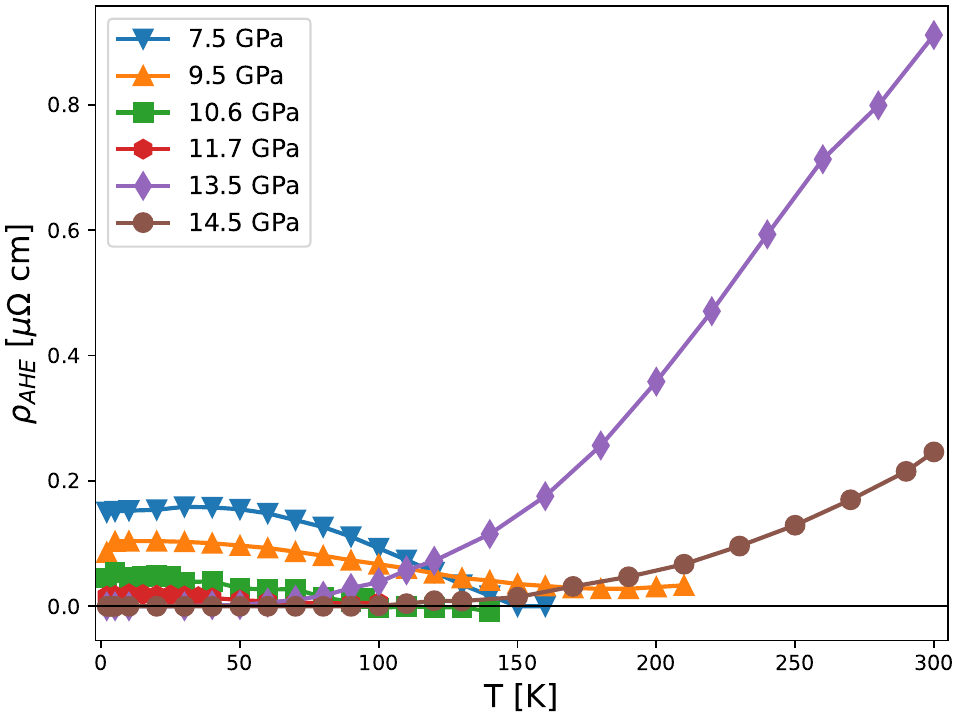}
  
  \caption{\label{rhoAH(T) 1st sample}\rhoAHE~as a function of temperature for the various pressures measured for sample 1.}
  
  \vspace{-0.5em}
\end{figure}

\vspace{-1em}
\rhoAHE~in a specific temperature and pressure, is extracted from the measurements by measuring $\mathrm{R_{xy}(H)}$ and antisymmetrize it. This results in graphs as shown in section \hyperref[raw data + AS]{S6}. Then we do a linear fit to the high-field regime ($\mathrm{4\, kOe < H}$), and the intersection of the fit with the y-axis is the AHE resistance ($\mathrm{R_{AHE}}$) (see Fig.\ref{extracting R_AHE}). Finally, by multiplying $\mathrm{R_{AHE}}$ with the width of the sample, we get \rhoAHE.

\newpage

\section{S6 - THE AHE measured in the first cell}\label{section S6}
\vspace{-1em}
Here we present our measurements of \rhoAHE~as a function of temperature for the various pressures measured in the first sample. As was also observed in the second sample ( in the main text), At pressures below $\mathrm{13\, GPa}$, $\rho_{AHE}\neq0$ at low temperatures and decays smoothly as the temperature increases. In contrast, for $\mathrm{P > 13\, GPa}$, at low temperatures $\rho_{AHE}=0$, and increases as the temperature increases.  

\newpage

\section{S7 - Raw Data measurements of $\mathrm{R_{xy}}$~and the resulted anti-symmetric plots for all pressures and cells \label{raw data + AS}}
\vspace{-1em}

Here we present measurements of $\mathrm{R_{xy}}$~as a function of the applied field H at various pressures and temperatures, both in their raw form and after undergoing antisymmetrization. When there is significant mixing of $\mathrm{R_{xx}}$~and $\mathrm{R_{xy}}$~in the measurements, it is reflected in the raw data, which appears neither symmetric nor antisymmetric. This effect has been observed multiple times, particularly in the low-pressure regime before the metal-insulator transition.

  %\vspace{-0.5em}

\begin{figure}[ht]
  \centering
  \includegraphics[width=0.835\columnwidth]{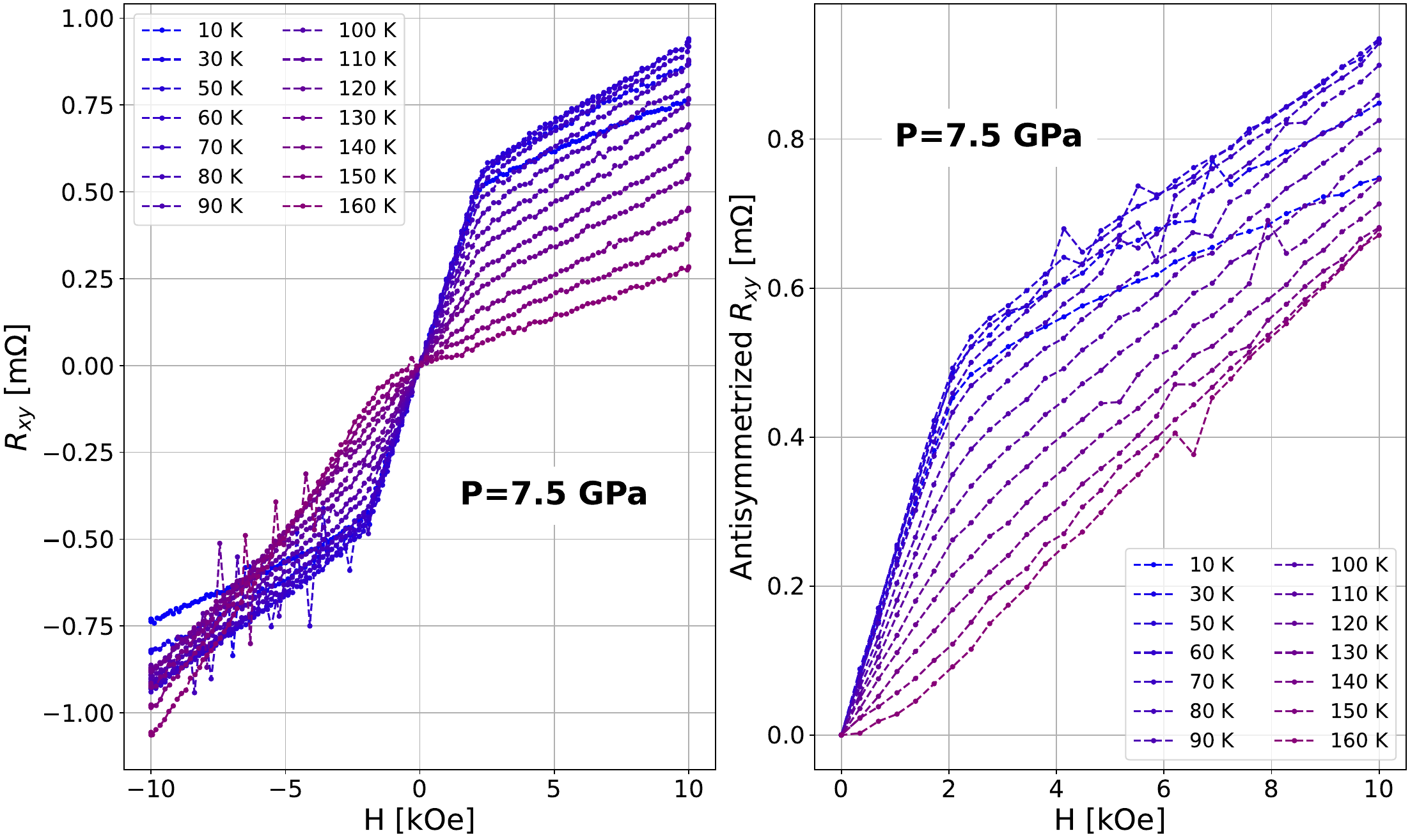}
  
  \caption{The left panel displays the raw data of $\mathrm{R_{xy}}$~as a function of the applied field H for the first sample in the metallic state at a pressure of $\mathrm{7.5\, GPa}$. The right panel shows the same data after antisymmetrization.}
  
  \vspace{-1.5em}
  %\vspace{-0.5em}
\end{figure}

\begin{figure}[ht]
  \centering
  \includegraphics[width=0.835\columnwidth]{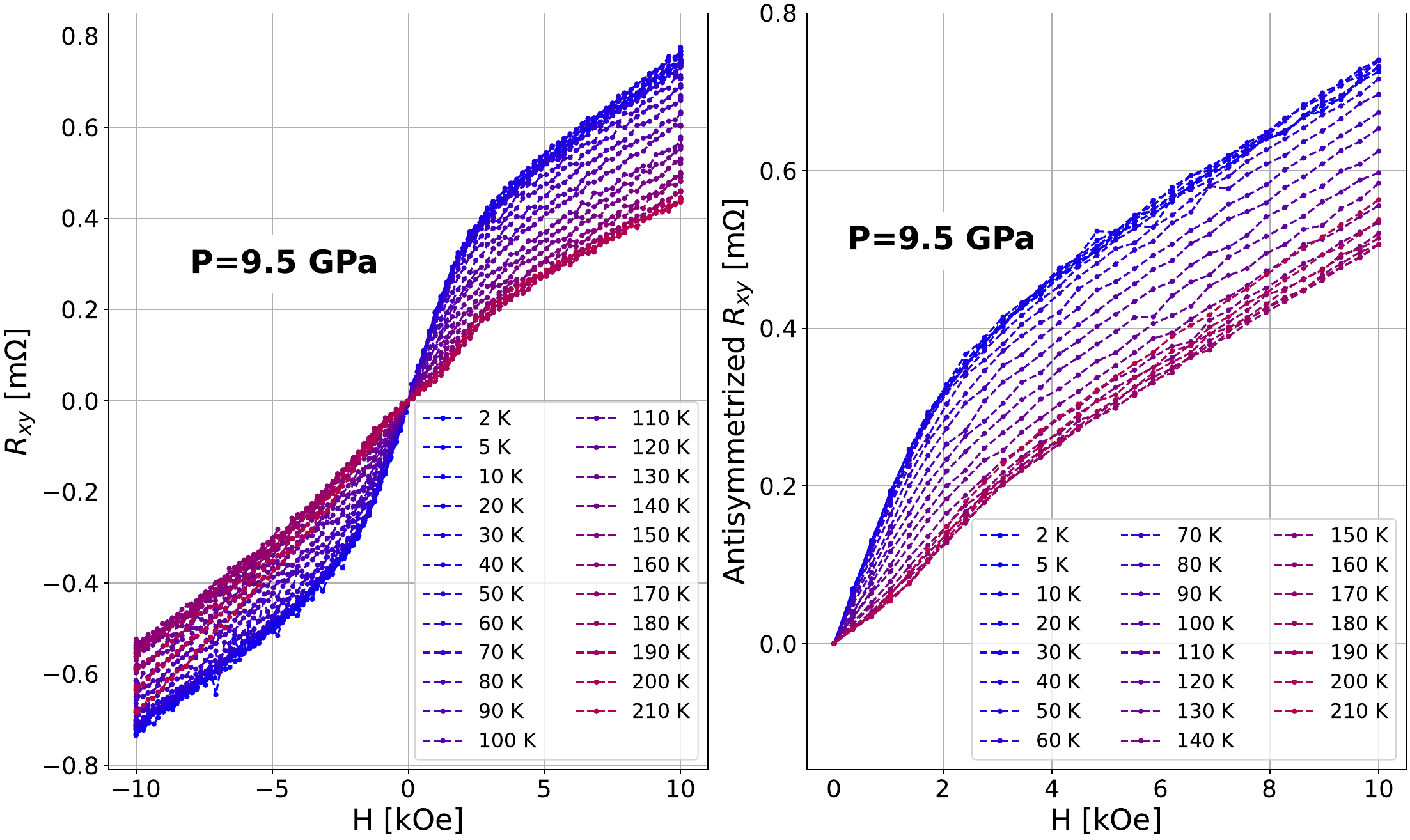}
  
  \caption{The left panel displays the raw data of $\mathrm{R_{xy}}$~as a function of the applied field H for the first sample in the metallic state at a pressure of $\mathrm{9.5\, GPa}$. The right panel shows the same data after antisymmetrization.}
  
  \vspace{-0.5em}
\end{figure}

\begin{figure}[ht]
  \centering
  \includegraphics[width=0.95\columnwidth]{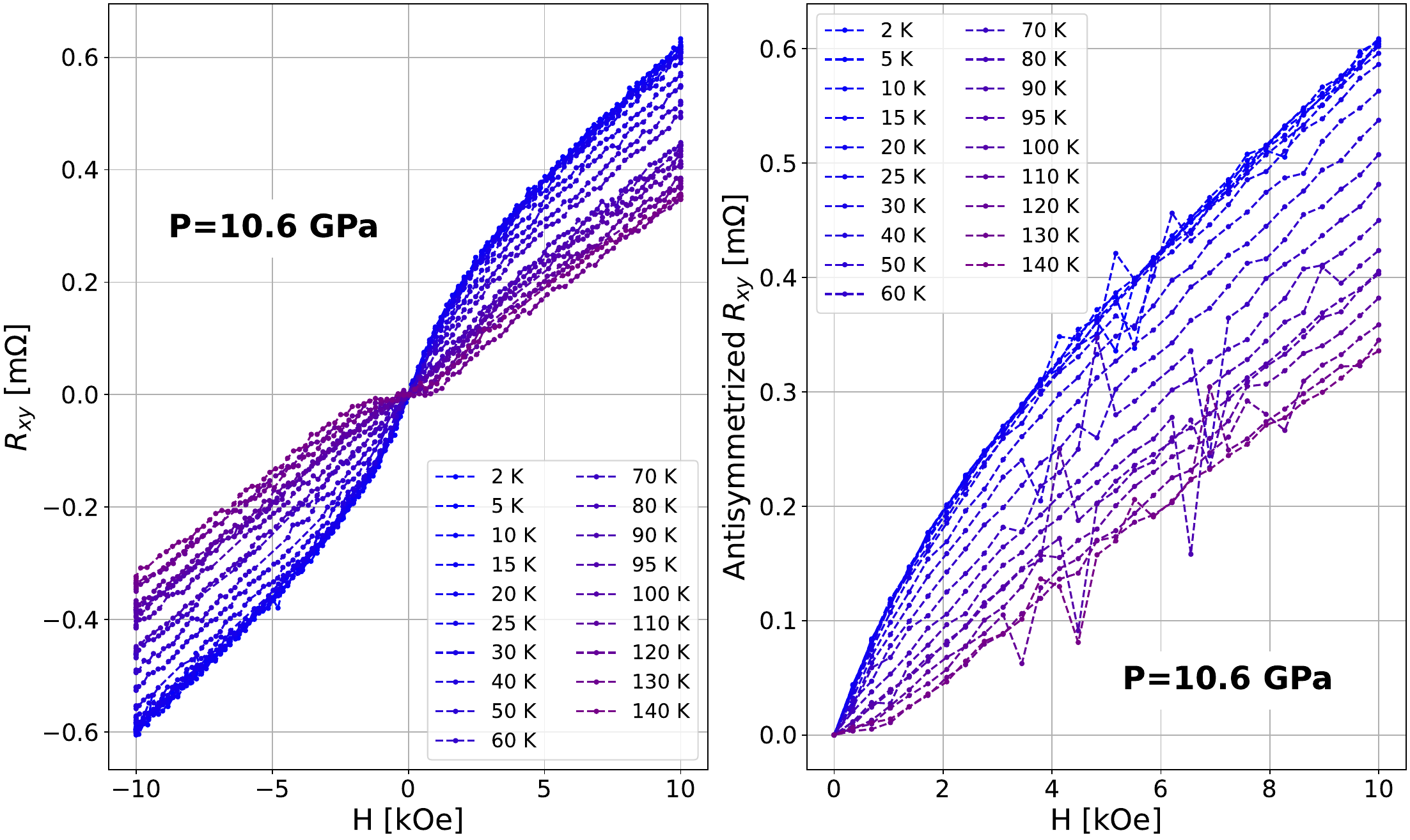}
  
  \caption{The left panel displays the raw data of $\mathrm{R_{xy}}$~as a function of the applied field H for the first sample in the metallic state at a pressure of $\mathrm{10.6\, GPa}$. The right panel shows the same data after antisymmetrization.}
  
  \vspace{-0.5em}
\end{figure}

\begin{figure}[ht]
  \centering
  \includegraphics[width=0.95\columnwidth]{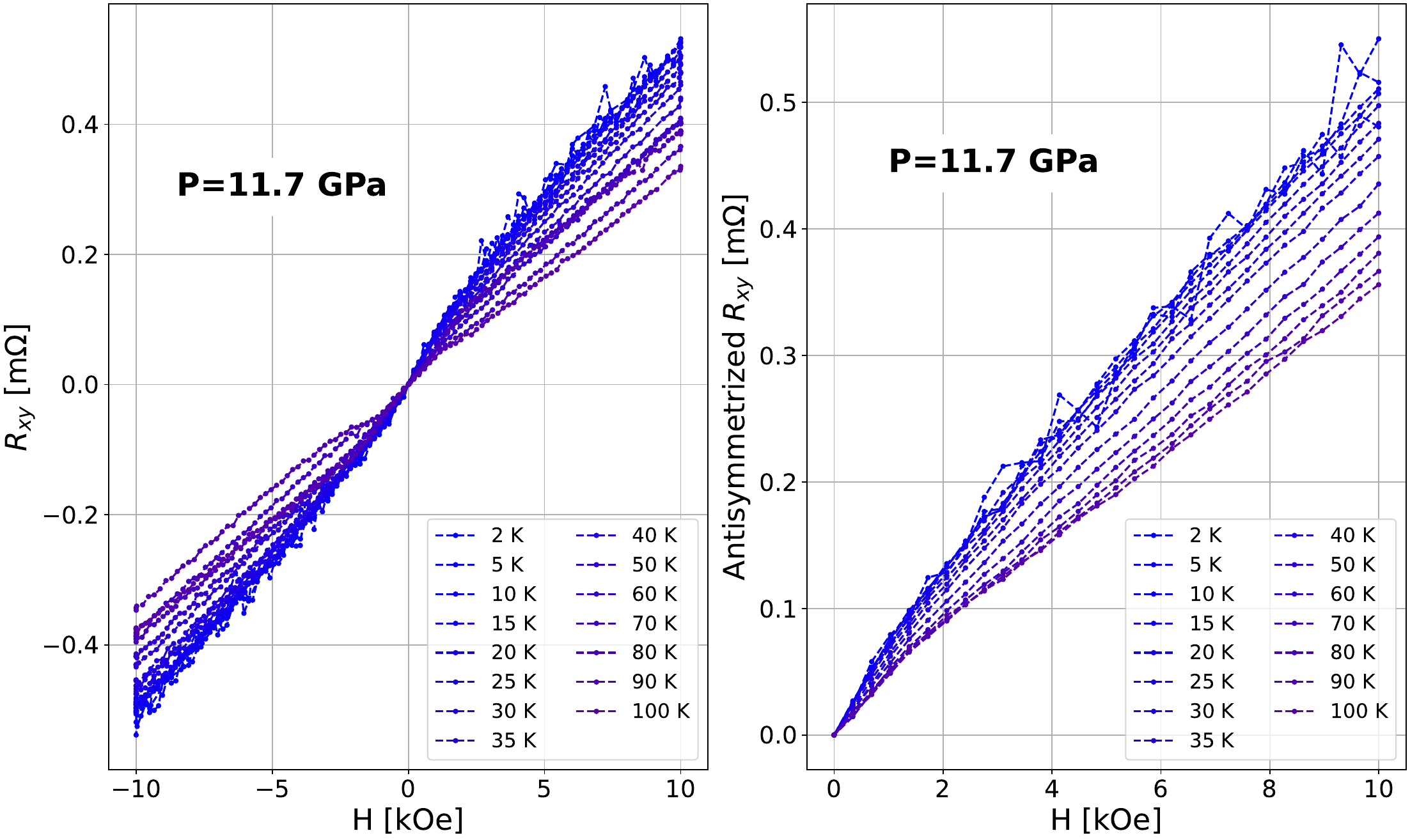}
  
  \caption{The left panel displays the raw data of $\mathrm{R_{xy}}$~as a function of the applied field H for the first sample in the metallic state at a pressure of $\mathrm{11.7\, GPa}$. The right panel shows the same data after antisymmetrization.}
  
  \vspace{-0.5em}
\end{figure}

\begin{figure}[ht]
  \centering
  \includegraphics[width=0.965\columnwidth]{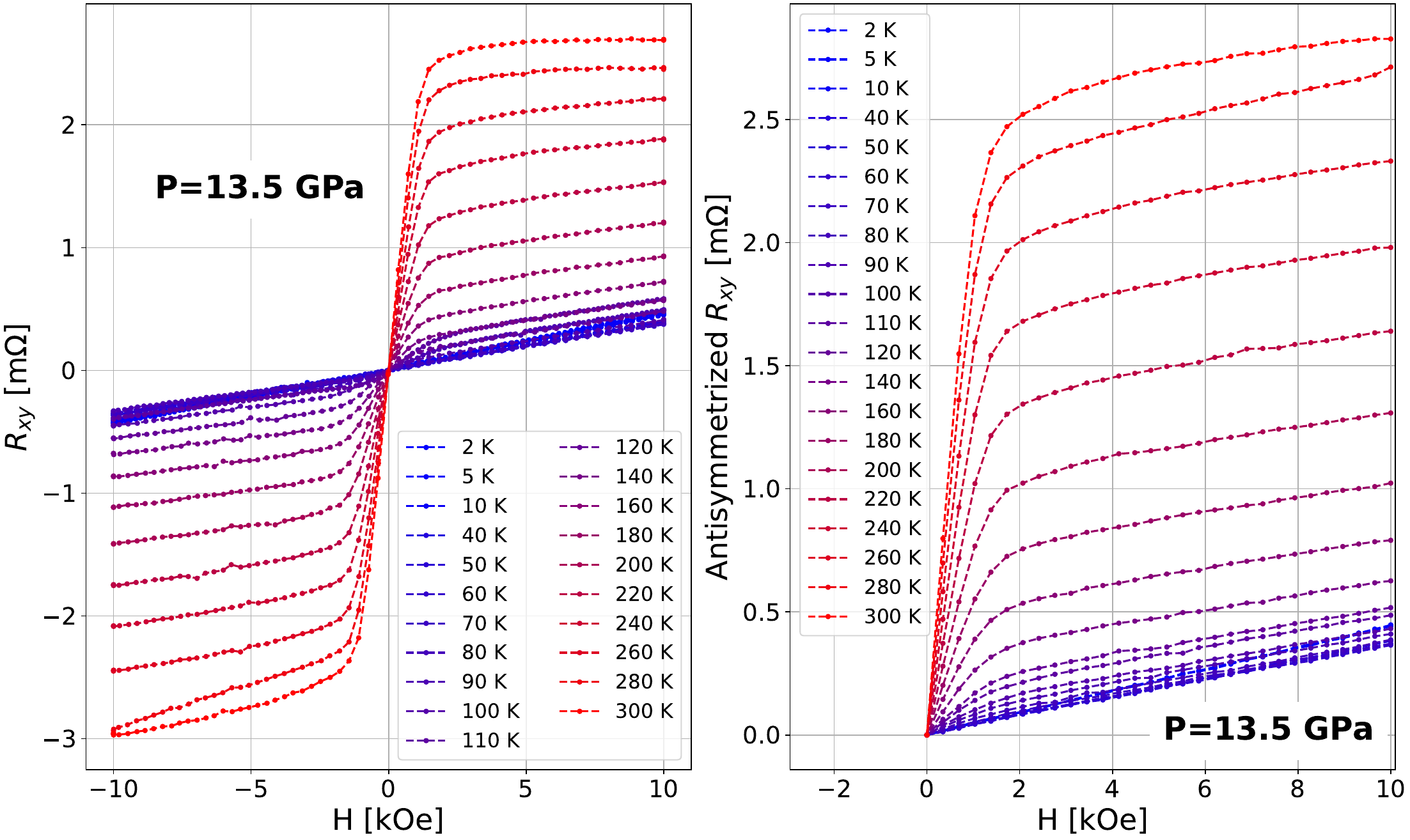}
  
  \caption{The left panel displays the raw data of $\mathrm{R_{xy}}$~as a function of the applied field H for the first sample in the metallic state at a pressure of $\mathrm{13.5\, GPa}$. The right panel shows the same data after antisymmetrization.}
  
  \vspace{-0.5em}
\end{figure}

\begin{figure}[ht]
  \centering
  \includegraphics[width=0.965\columnwidth]{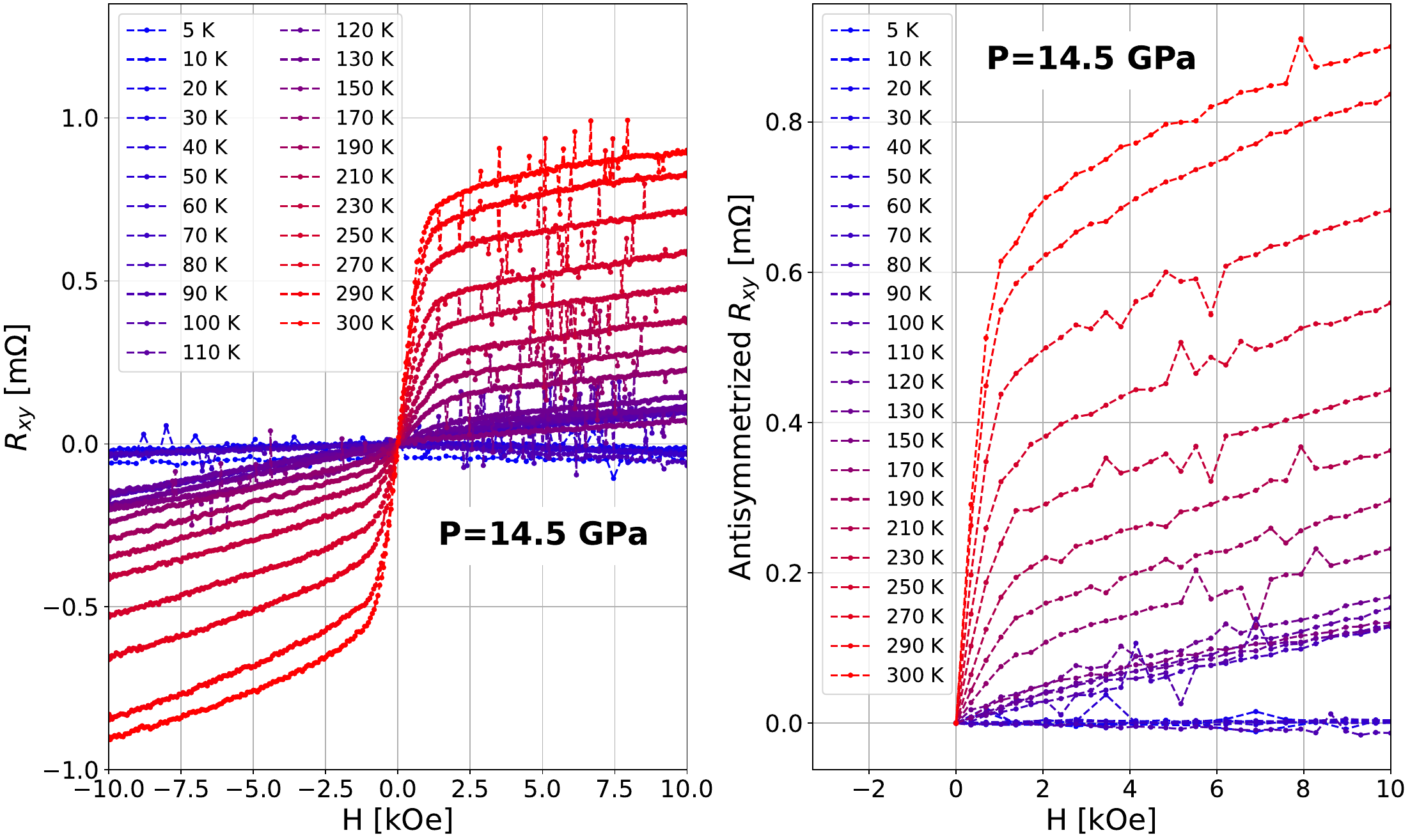}
  
  \caption{The left panel displays the raw data of $\mathrm{R_{xy}}$~as a function of the applied field H for the first sample in the metallic state at a pressure of $\mathrm{14.5\, GPa}$. The right panel shows the same data after antisymmetrization.}
  
  \vspace{-0.5em}
\end{figure}

\begin{figure}[ht]
  \centering
  \includegraphics[width=0.91\columnwidth]{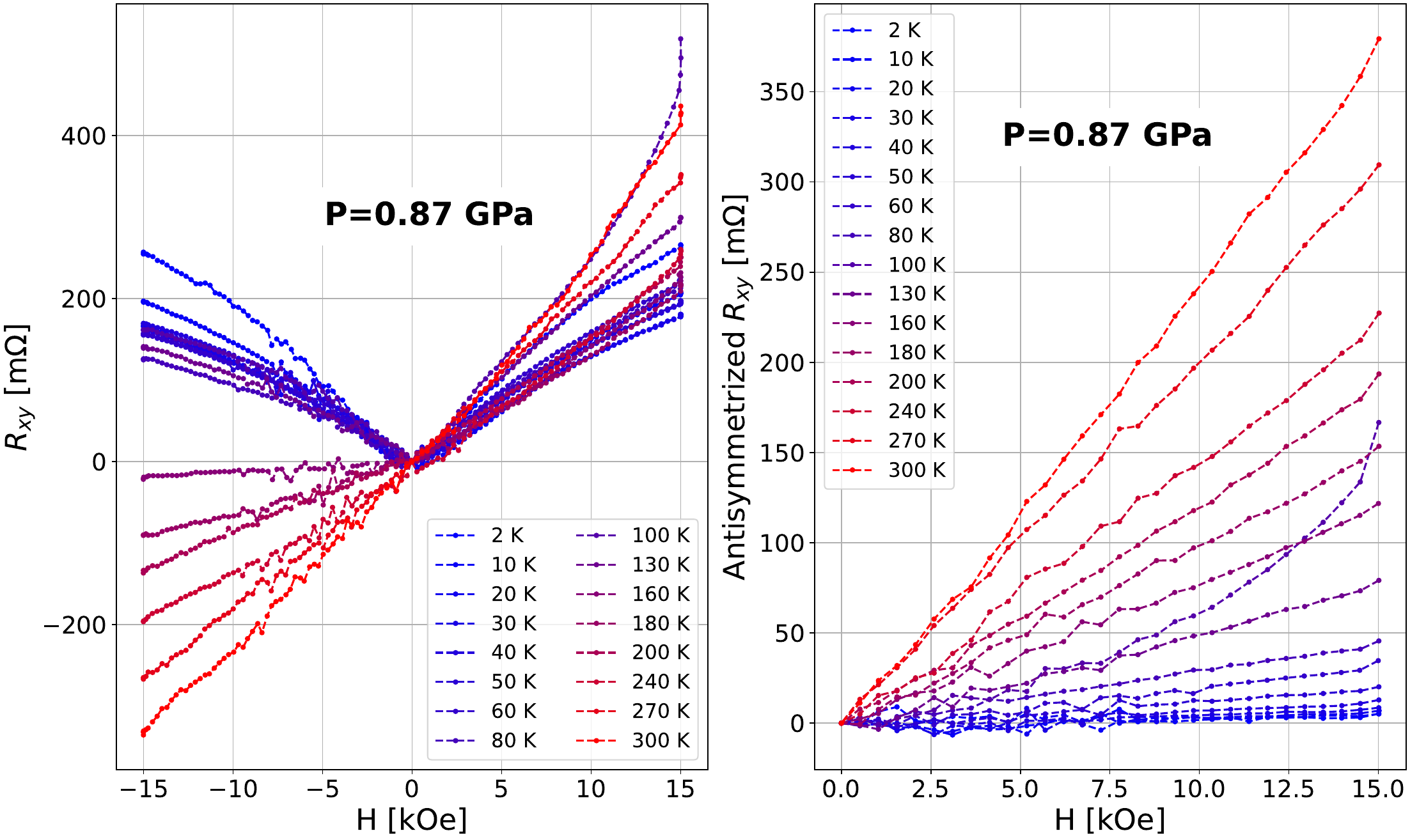}
  
  \caption{The left panel displays raw data of $\mathrm{R_{xy}}$~as a function of the applied field H for the second sample in the insulating state at a pressure of $\mathrm{0.87\, GPa}$. The right panel presents the same data after antisymmetrization. The presence of significant intermixing between $\mathrm{R_{xx}}$~and $\mathrm{R_{xy}}$~can be easily identified by the absence of symmetry or antisymmetry in the raw data.}
  
  \vspace{-0.5em}
\end{figure}

\begin{figure}[ht]
  \centering
  \includegraphics[width=0.91\columnwidth]{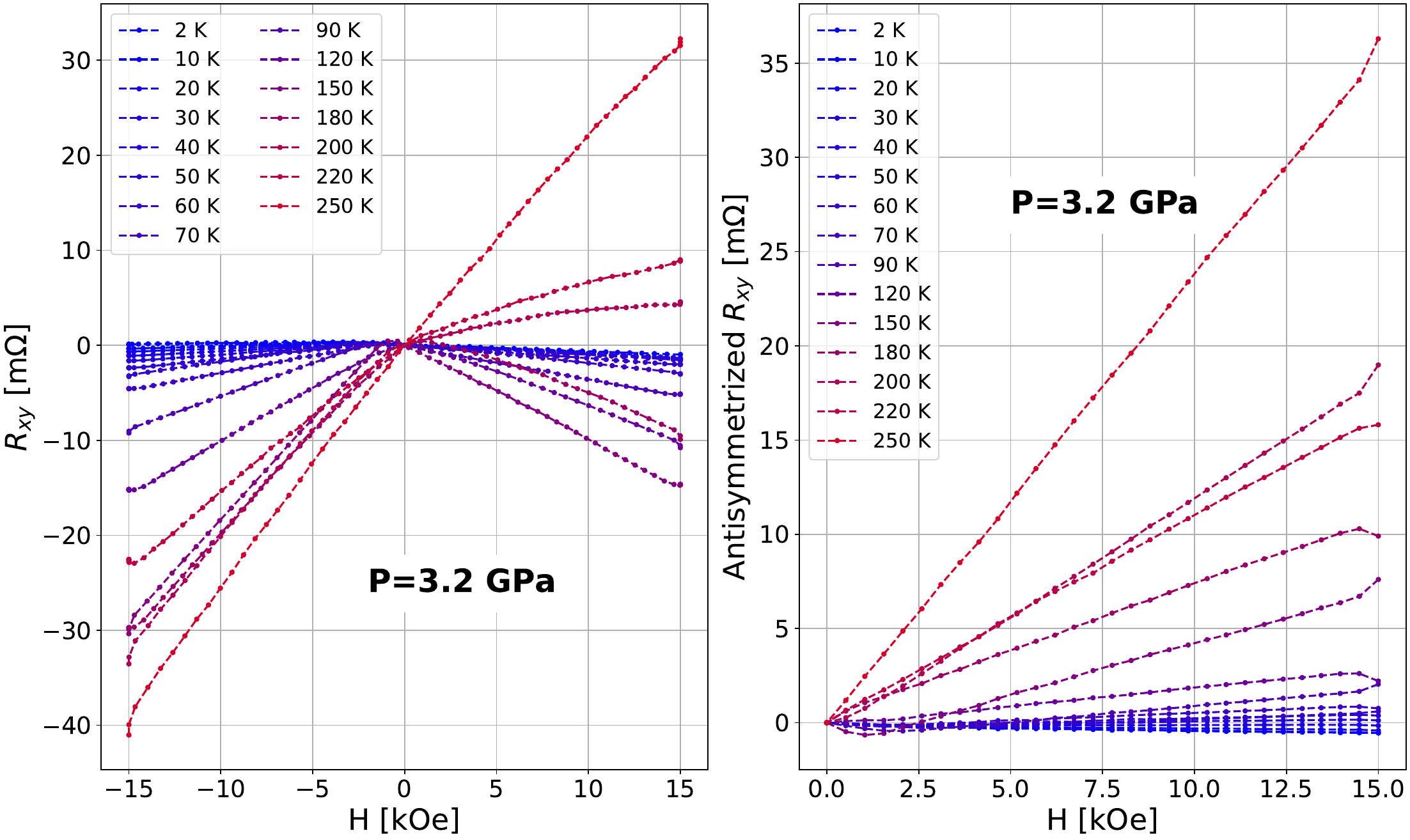}
  
  \caption{The left panel displays raw data of $\mathrm{R_{xy}}$~as a function of the applied field H for the second sample in the insulating state at a pressure of $\mathrm{3.2\, GPa}$. The right panel presents the same data after antisymmetrization. The presence of significant intermixing between $\mathrm{R_{xx}}$~and $\mathrm{R_{xy}}$~can be easily identified by the absence of symmetry or antisymmetry in the raw data.}
  
  \vspace{-0.5em}
\end{figure}

\begin{figure}[ht]
  \centering
  \includegraphics[width=0.985\columnwidth]{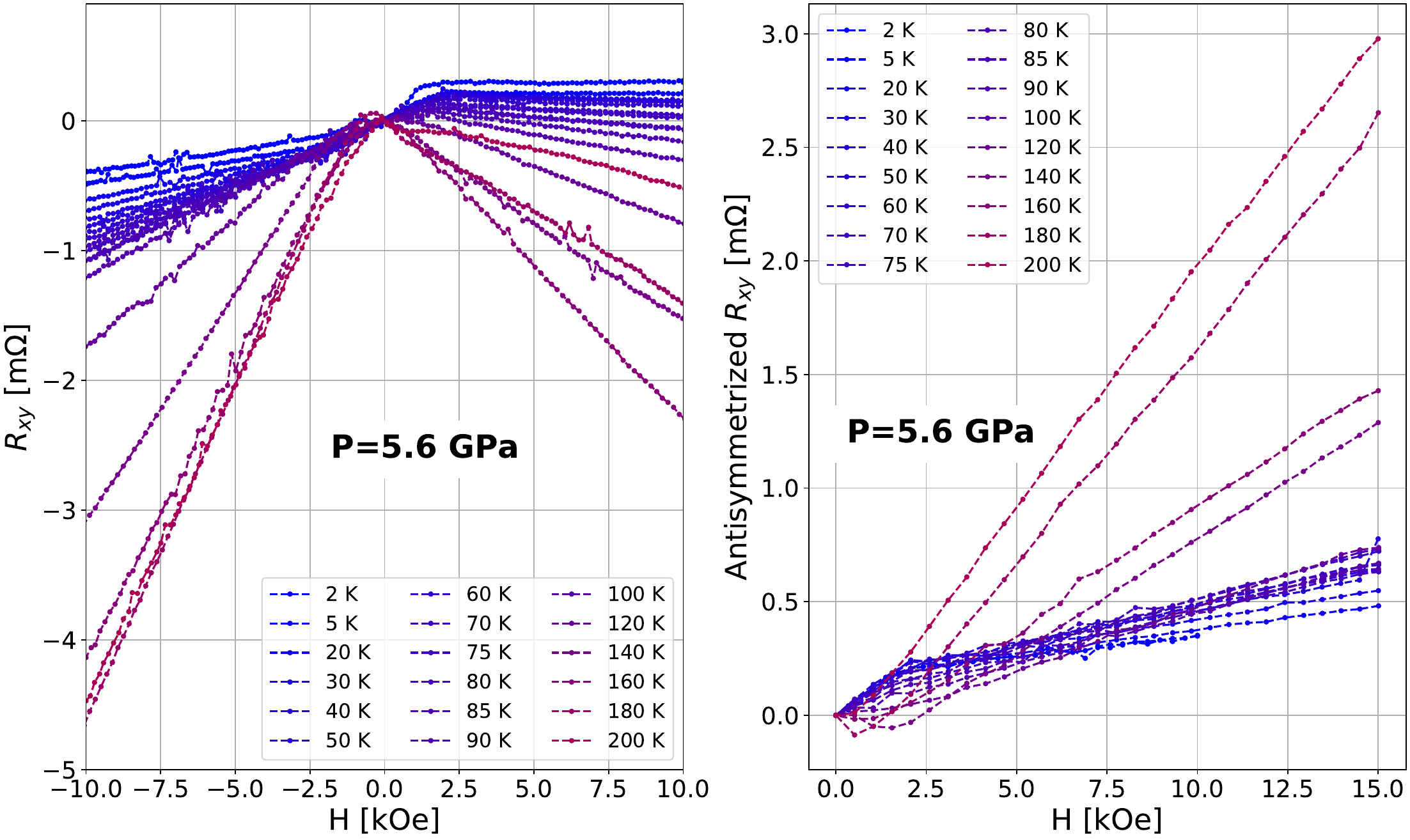}
  
  \caption{The left panel displays the raw data of $\mathrm{R_{xy}}$~as a function of the applied field H for the second sample in the metallic state at a pressure of $\mathrm{5.6\, GPa}$. The right panel shows the same data after antisymmetrization.}
  
  \vspace{-0.5em}
\end{figure}

\begin{figure}[ht]
  \centering
  \includegraphics[width=0.985\columnwidth]{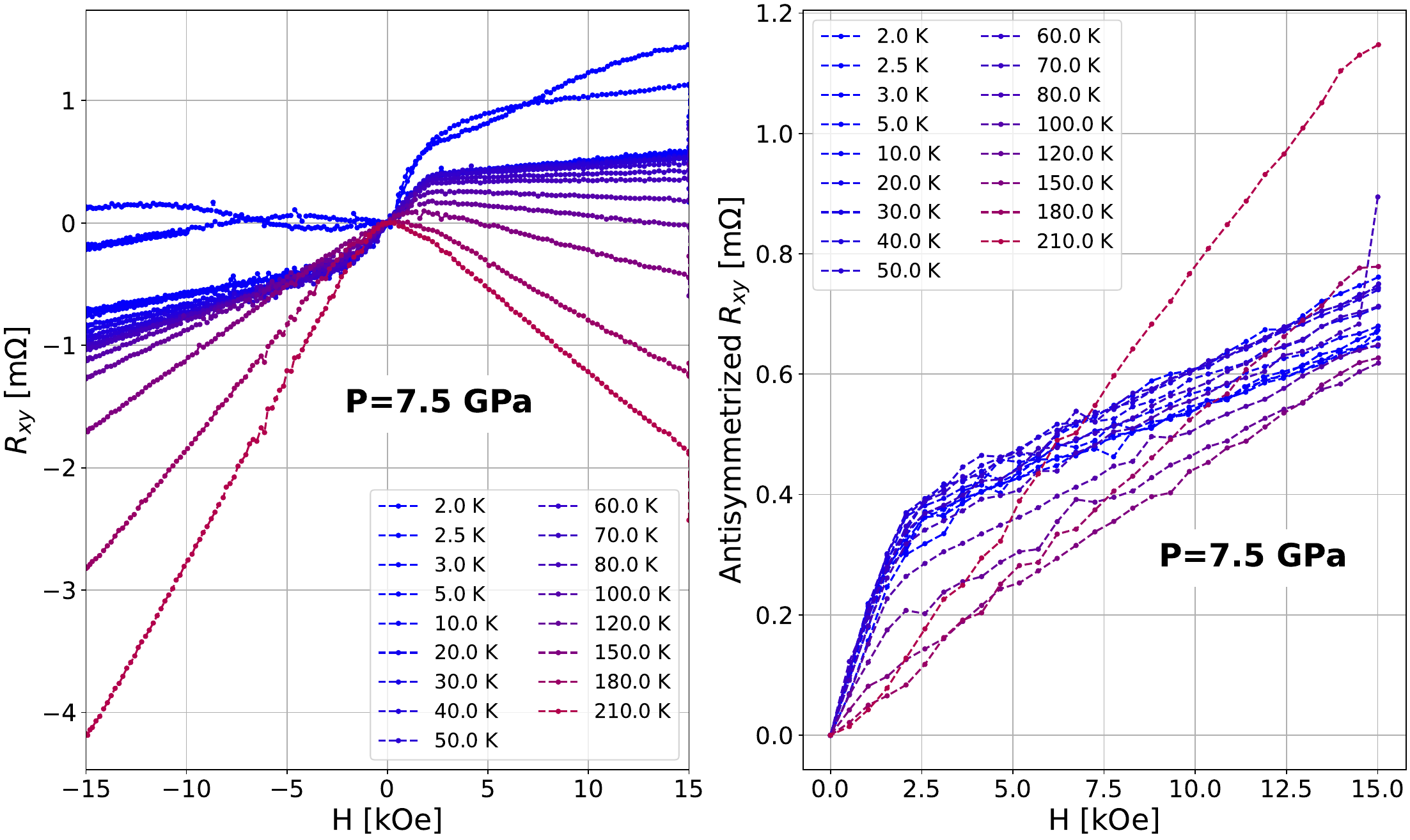}
  
  \caption{The left panel displays the raw data of $\mathrm{R_{xy}}$~as a function of the applied field H for the second sample in the metallic state at a pressure of $\mathrm{7.5\, GPa}$. The right panel shows the same data after antisymmetrization.}
  
  \vspace{-0.5em}
\end{figure}

\begin{figure}[ht]
  \centering
  \includegraphics[width=0.985\columnwidth]{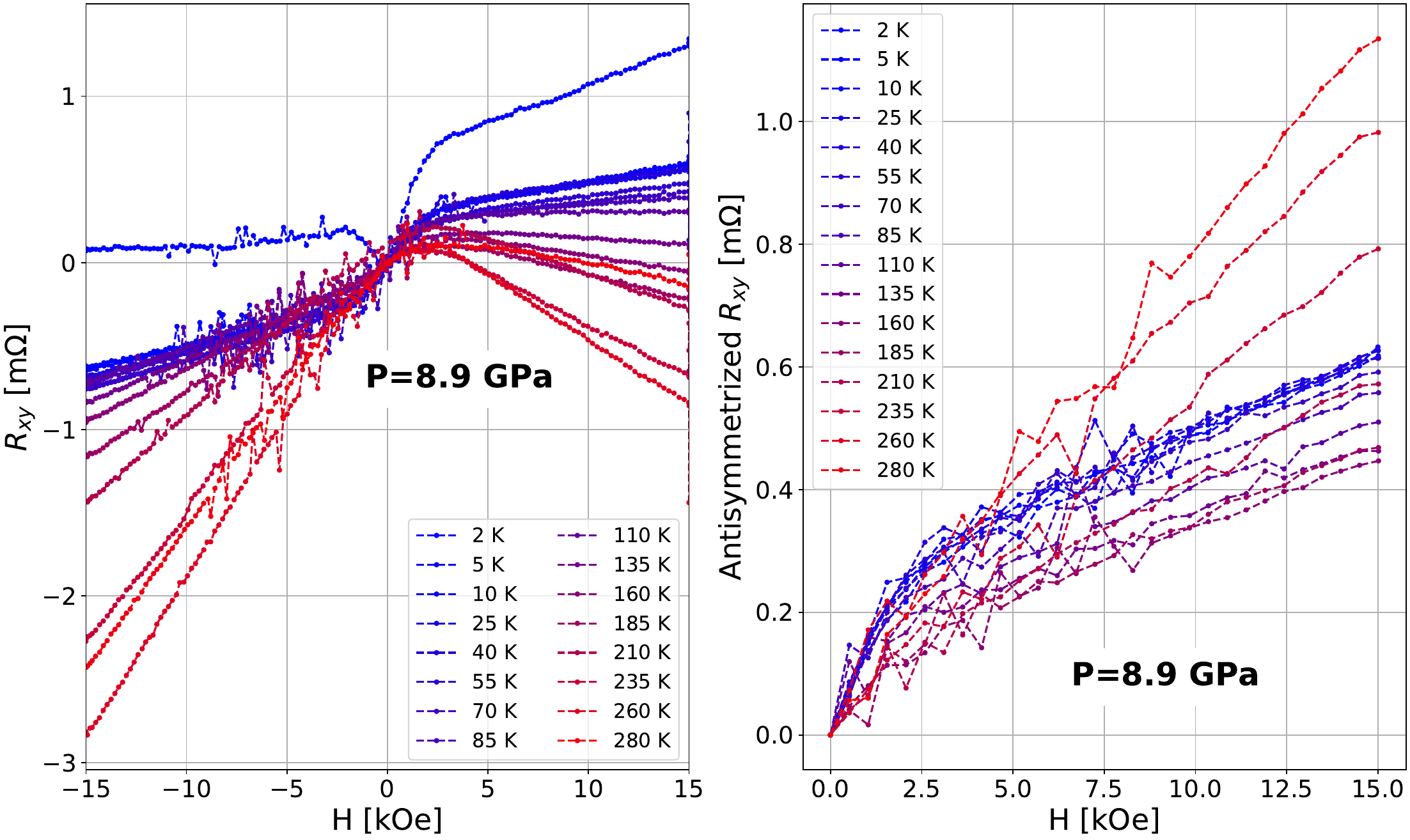}
  
  \caption{The left panel displays the raw data of $\mathrm{R_{xy}}$~as a function of the applied field H for the second sample in the metallic state at a pressure of $\mathrm{8.9\, GPa}$. The right panel shows the same data after antisymmetrization.}
  
  \vspace{-0.5em}
\end{figure}

\begin{figure}[ht]
  \centering
  \includegraphics[width=0.985\columnwidth]{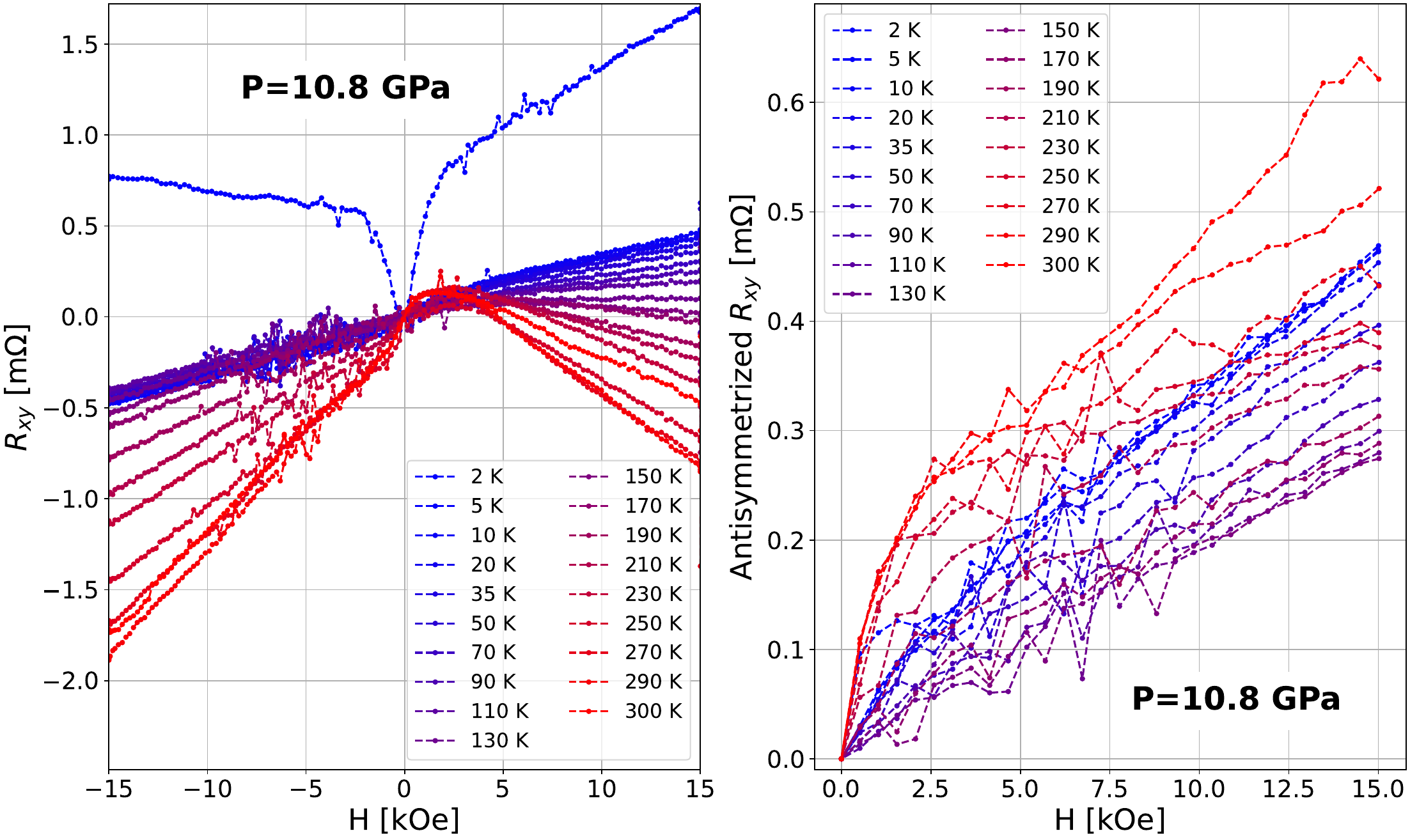}
  
  \caption{The left panel displays the raw data of $\mathrm{R_{xy}}$~as a function of the applied field H for the second sample in the metallic state at a pressure of $\mathrm{10.8\, GPa}$. The right panel shows the same data after antisymmetrization.}
  
  \vspace{-0.5em}
\end{figure}

\begin{figure}[ht]
  \centering
  \includegraphics[width=0.985\columnwidth]{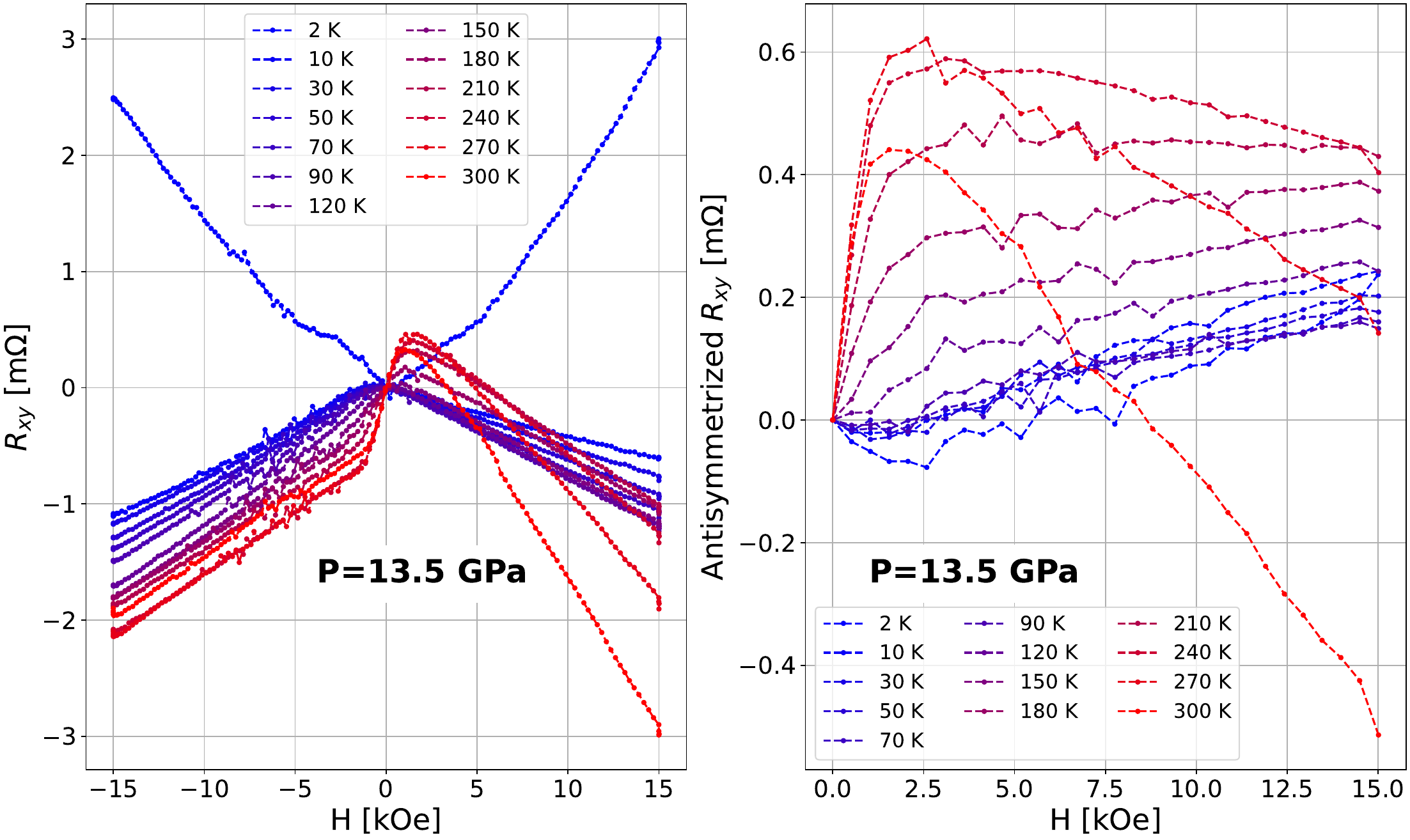}
  
  \caption{The left panel displays the raw data of $\mathrm{R_{xy}}$~as a function of the applied field H for the second sample in the metallic state at a pressure of $\mathrm{13.5\, GPa}$. The right panel shows the same data after antisymmetrization.}
  
  \vspace{-0.5em}
\end{figure}

\begin{figure}[ht]
  \centering
  \includegraphics[width=0.985\columnwidth]{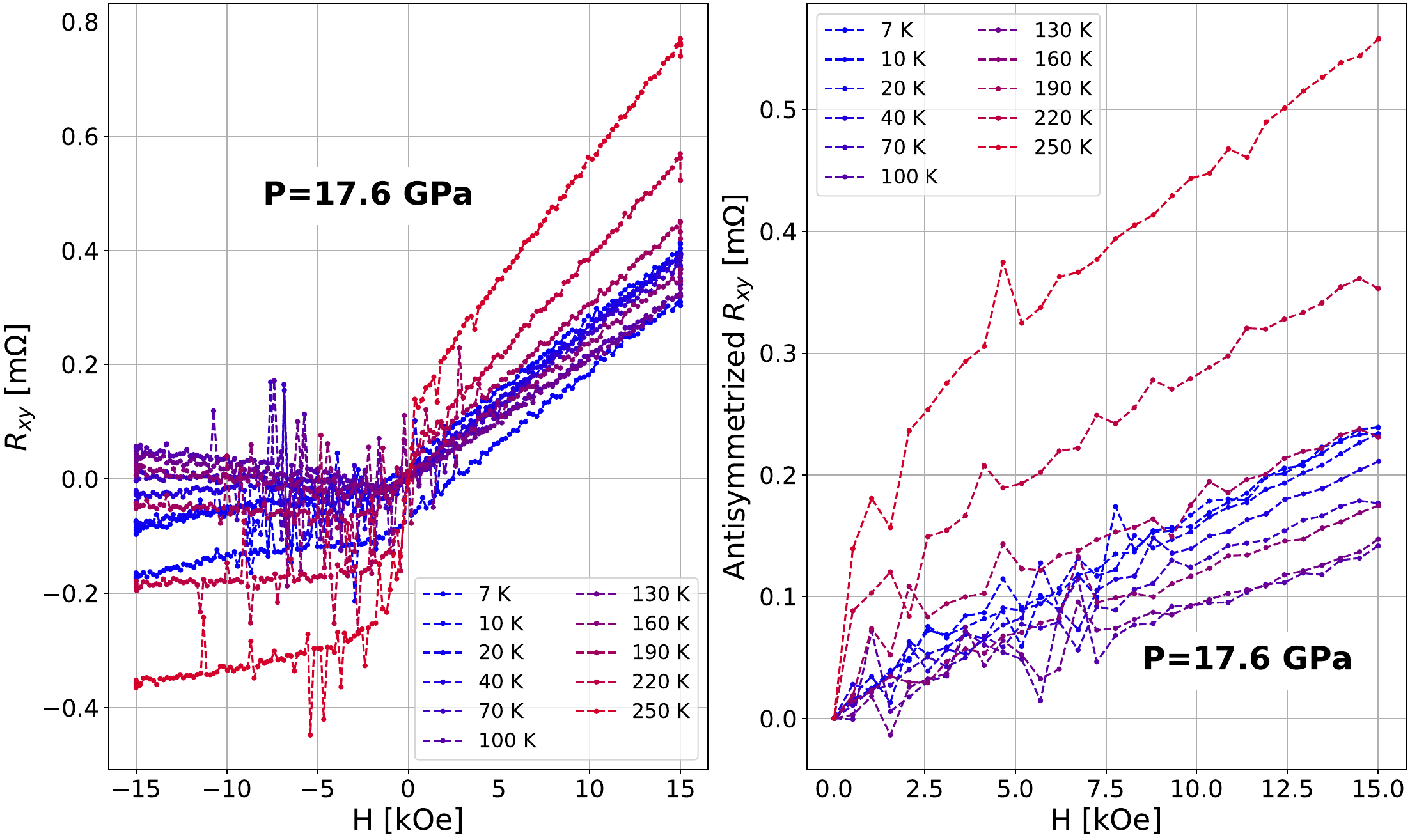}
  
  \caption{The left panel displays the raw data of $\mathrm{R_{xy}}$~as a function of the applied field H for the second sample in the metallic state at a pressure of $\mathrm{17.6\, GPa}$. The right panel shows the same data after antisymmetrization.}
  
  \vspace{-0.5em}
\end{figure}

\clearpage

\section{S8 - Effect of magnetic quantization axis rotation on Berry curvature}\label{section S8}
We have performed additional DFT calculations, where we rotated the quantization axis (equivalent to the direction of magnetic field) from the $k_z$-direction towards the $k_x$-direction. For these additional calculations we evaluated the Berry curvature for $5 \cdot 10^5$ $k$-points for each pressure value and rotation angle in each of the ten independent Monte Carlo calculations.

\begin{figure}[ht]
  \centering
  \includegraphics[width=0.65\columnwidth]{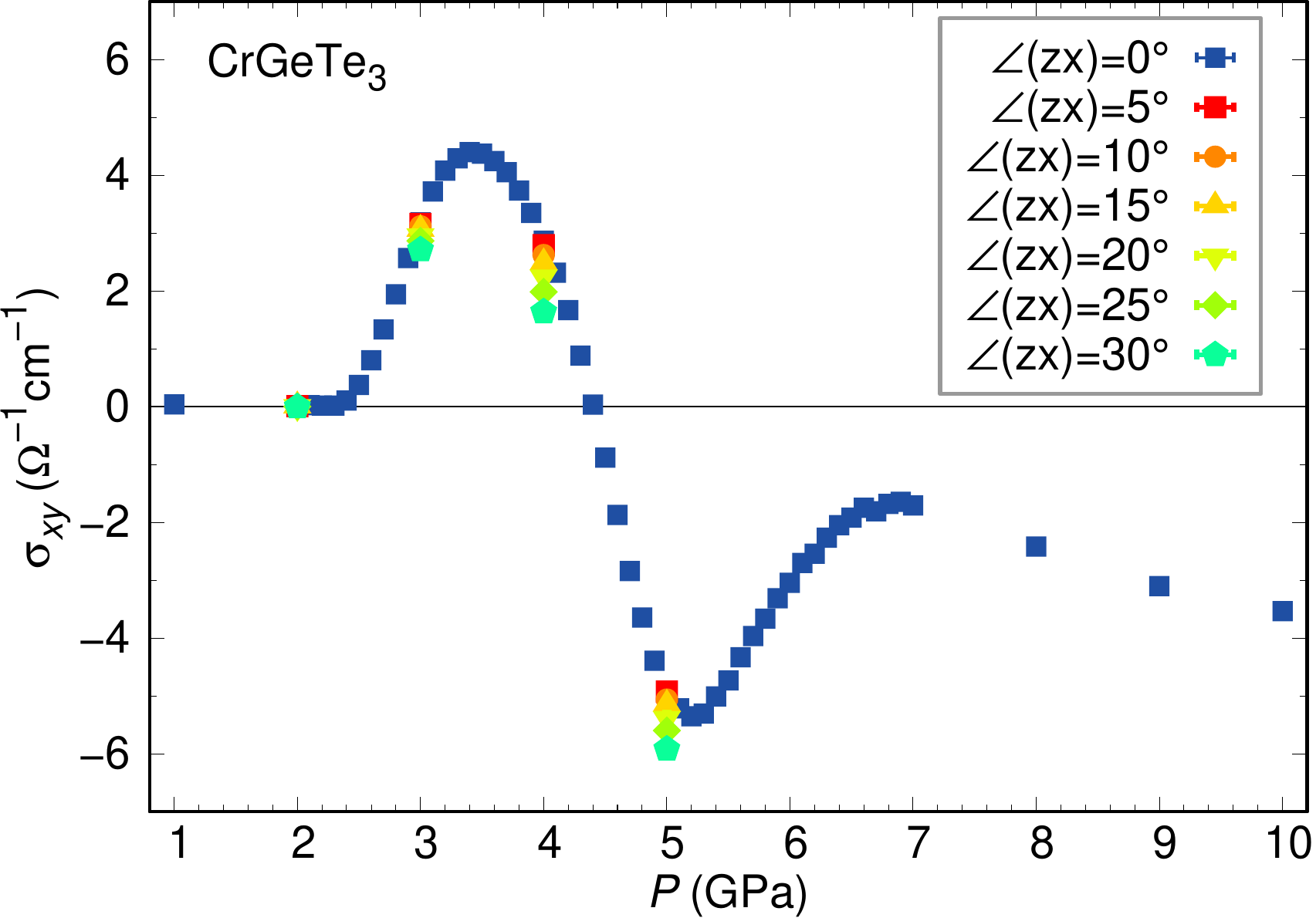}
  
  \caption{Sensitivity of anomalous Hall conductivity on the direction of the quantization axis. The angle denotes the rotation from the $k_x$-direction towards the $k_z$-direction.}
  \label{fig:dftquantaxis}
  
  \vspace{-0.5em}
\end{figure}

The rotation of the quantization axis leads to a small but systematic effect on the anomalous Hall conductivity, which is shown in Fig.~\ref{fig:dftquantaxis}.

% \end{document}


\begin{thebibliography}{38}%
\makeatletter
\providecommand \@ifxundefined [1]{%
 \@ifx{#1\undefined}
}%
\providecommand \@ifnum [1]{%
 \ifnum #1\expandafter \@firstoftwo
 \else \expandafter \@secondoftwo
 \fi
}%
\providecommand \@ifx [1]{%
 \ifx #1\expandafter \@firstoftwo
 \else \expandafter \@secondoftwo
 \fi
}%
\providecommand \natexlab [1]{#1}%
\providecommand \enquote  [1]{``#1''}%
\providecommand \bibnamefont  [1]{#1}%
\providecommand \bibfnamefont [1]{#1}%
\providecommand \citenamefont [1]{#1}%
\providecommand \href@noop [0]{\@secondoftwo}%
\providecommand \href [0]{\begingroup \@sanitize@url \@href}%
\providecommand \@href[1]{\@@startlink{#1}\@@href}%
\providecommand \@@href[1]{\endgroup#1\@@endlink}%
\providecommand \@sanitize@url [0]{\catcode `\\12\catcode `\$12\catcode `\&12\catcode `\#12\catcode `\^12\catcode `\_12\catcode `\%12\relax}%
\providecommand \@@startlink[1]{}%
\providecommand \@@endlink[0]{}%
\providecommand \url  [0]{\begingroup\@sanitize@url \@url }%
\providecommand \@url [1]{\endgroup\@href {#1}{\urlprefix }}%
\providecommand \urlprefix  [0]{URL }%
\providecommand \Eprint [0]{\href }%
\providecommand \doibase [0]{https://doi.org/}%
\providecommand \selectlanguage [0]{\@gobble}%
\providecommand \bibinfo  [0]{\@secondoftwo}%
\providecommand \bibfield  [0]{\@secondoftwo}%
\providecommand \translation [1]{[#1]}%
\providecommand \BibitemOpen [0]{}%
\providecommand \bibitemStop [0]{}%
\providecommand \bibitemNoStop [0]{.\EOS\space}%
\providecommand \EOS [0]{\spacefactor3000\relax}%
\providecommand \BibitemShut  [1]{\csname bibitem#1\endcsname}%
\let\auto@bib@innerbib\@empty
%</preamble>
\bibitem [{\citenamefont {Xiao}\ \emph {et~al.}(2010)\citenamefont {Xiao}, \citenamefont {Chang},\ and\ \citenamefont {Niu}}]{RevModPhys.82.1959}%
  \BibitemOpen
  \bibfield  {author} {\bibinfo {author} {\bibfnamefont {D.}~\bibnamefont {Xiao}}, \bibinfo {author} {\bibfnamefont {M.-C.}\ \bibnamefont {Chang}},\ and\ \bibinfo {author} {\bibfnamefont {Q.}~\bibnamefont {Niu}},\ }\bibfield  {title} {\bibinfo {title} {Berry phase effects on electronic properties},\ }\href {https://doi.org/10.1103/RevModPhys.82.1959} {\bibfield  {journal} {\bibinfo  {journal} {Rev. Mod. Phys.}\ }\textbf {\bibinfo {volume} {82}},\ \bibinfo {pages} {1959} (\bibinfo {year} {2010})}\BibitemShut {NoStop}%
\bibitem [{\citenamefont {Bhoi}\ \emph {et~al.}(2021)\citenamefont {Bhoi}, \citenamefont {Gouchi}, \citenamefont {Hiraoka}, \citenamefont {Zhang}, \citenamefont {Ogita}, \citenamefont {Hasegawa}, \citenamefont {Kitagawa}, \citenamefont {Takagi}, \citenamefont {Kim},\ and\ \citenamefont {Uwatoko}}]{bhoi2021nearly}%
  \BibitemOpen
  \bibfield  {author} {\bibinfo {author} {\bibfnamefont {D.}~\bibnamefont {Bhoi}}, \bibinfo {author} {\bibfnamefont {J.}~\bibnamefont {Gouchi}}, \bibinfo {author} {\bibfnamefont {N.}~\bibnamefont {Hiraoka}}, \bibinfo {author} {\bibfnamefont {Y.}~\bibnamefont {Zhang}}, \bibinfo {author} {\bibfnamefont {N.}~\bibnamefont {Ogita}}, \bibinfo {author} {\bibfnamefont {T.}~\bibnamefont {Hasegawa}}, \bibinfo {author} {\bibfnamefont {K.}~\bibnamefont {Kitagawa}}, \bibinfo {author} {\bibfnamefont {H.}~\bibnamefont {Takagi}}, \bibinfo {author} {\bibfnamefont {K.~H.}\ \bibnamefont {Kim}},\ and\ \bibinfo {author} {\bibfnamefont {Y.}~\bibnamefont {Uwatoko}},\ }\bibfield  {title} {\bibinfo {title} {Nearly room-temperature ferromagnetism in a pressure-induced correlated metallic state of the van der {Waals} insulator {${\mathrm{CrGeTe}}_{3}$}},\ }\href {https://doi.org/10.1103/PhysRevLett.127.217203} {\bibfield  {journal} {\bibinfo  {journal} {Phys. Rev. Lett.}\ }\textbf {\bibinfo {volume} {127}},\ \bibinfo {pages} {217203}
  (\bibinfo {year} {2021})}\BibitemShut {NoStop}%
\bibitem [{\citenamefont {Nagaosa}\ \emph {et~al.}(2010)\citenamefont {Nagaosa}, \citenamefont {Sinova}, \citenamefont {Onoda}, \citenamefont {MacDonald},\ and\ \citenamefont {Ong}}]{nagaosa2010anomalous}%
  \BibitemOpen
  \bibfield  {author} {\bibinfo {author} {\bibfnamefont {N.}~\bibnamefont {Nagaosa}}, \bibinfo {author} {\bibfnamefont {J.}~\bibnamefont {Sinova}}, \bibinfo {author} {\bibfnamefont {S.}~\bibnamefont {Onoda}}, \bibinfo {author} {\bibfnamefont {A.~H.}\ \bibnamefont {MacDonald}},\ and\ \bibinfo {author} {\bibfnamefont {N.~P.}\ \bibnamefont {Ong}},\ }\bibfield  {title} {\bibinfo {title} {Anomalous {Hall} effect},\ }\href {https://doi.org/10.1103/RevModPhys.82.1539} {\bibfield  {journal} {\bibinfo  {journal} {Rev. Mod. Phys.}\ }\textbf {\bibinfo {volume} {82}},\ \bibinfo {pages} {1539} (\bibinfo {year} {2010})}\BibitemShut {NoStop}%
\bibitem [{\citenamefont {Carteaux}\ \emph {et~al.}(1995)\citenamefont {Carteaux}, \citenamefont {Brunet}, \citenamefont {Ouvrard},\ and\ \citenamefont {Andre}}]{carteaux1995crystallographic}%
  \BibitemOpen
  \bibfield  {author} {\bibinfo {author} {\bibfnamefont {V.}~\bibnamefont {Carteaux}}, \bibinfo {author} {\bibfnamefont {D.}~\bibnamefont {Brunet}}, \bibinfo {author} {\bibfnamefont {G.}~\bibnamefont {Ouvrard}},\ and\ \bibinfo {author} {\bibfnamefont {G.}~\bibnamefont {Andre}},\ }\bibfield  {title} {\bibinfo {title} {Crystallographic, magnetic and electronic structures of a new layered ferromagnetic compound {Cr$_2$Ge$_2$Te$_6$}},\ }\href {https://doi.org/10.1088/0953-8984/7/1/008} {\bibfield  {journal} {\bibinfo  {journal} {J. Phys. Condens. Matter}\ }\textbf {\bibinfo {volume} {7}},\ \bibinfo {pages} {69} (\bibinfo {year} {1995})}\BibitemShut {NoStop}%
\bibitem [{\citenamefont {Zhu}\ \emph {et~al.}(2021)\citenamefont {Zhu}, \citenamefont {Zhang}, \citenamefont {Wang}, \citenamefont {dos Santos}, \citenamefont {Song}, \citenamefont {Mueller}, \citenamefont {Schmalzl}, \citenamefont {Schmidt}, \citenamefont {Ivanov}, \citenamefont {Park}, \citenamefont {Xu}, \citenamefont {Ma}, \citenamefont {Lounis}, \citenamefont {Bl\"ugel}, \citenamefont {Su},\ and\ \citenamefont {Br\"uckel}}]{zhu2021topological}%
  \BibitemOpen
  \bibfield  {author} {\bibinfo {author} {\bibfnamefont {F.}~\bibnamefont {Zhu}}, \bibinfo {author} {\bibfnamefont {L.}~\bibnamefont {Zhang}}, \bibinfo {author} {\bibfnamefont {X.}~\bibnamefont {Wang}}, \bibinfo {author} {\bibfnamefont {F.~J.}\ \bibnamefont {dos Santos}}, \bibinfo {author} {\bibfnamefont {J.}~\bibnamefont {Song}}, \bibinfo {author} {\bibfnamefont {T.}~\bibnamefont {Mueller}}, \bibinfo {author} {\bibfnamefont {K.}~\bibnamefont {Schmalzl}}, \bibinfo {author} {\bibfnamefont {W.~F.}\ \bibnamefont {Schmidt}}, \bibinfo {author} {\bibfnamefont {A.}~\bibnamefont {Ivanov}}, \bibinfo {author} {\bibfnamefont {J.~T.}\ \bibnamefont {Park}}, \bibinfo {author} {\bibfnamefont {J.}~\bibnamefont {Xu}}, \bibinfo {author} {\bibfnamefont {J.}~\bibnamefont {Ma}}, \bibinfo {author} {\bibfnamefont {S.}~\bibnamefont {Lounis}}, \bibinfo {author} {\bibfnamefont {S.~Y.}\ \bibnamefont {Bl\"ugel}, \bibfnamefont {Mokrousov}}, \bibinfo {author} {\bibfnamefont {Y.}~\bibnamefont {Su}},\ and\ \bibinfo {author} {\bibfnamefont
  {T.}~\bibnamefont {Br\"uckel}},\ }\bibfield  {title} {\bibinfo {title} {Topological magnon insulators in two-dimensional van der {Waals} ferromagnets {CrSiTe$_3$} and {CrGeTe$_3$}: Toward intrinsic gap-tunability},\ }\href {https://doi.org/10.1126/sciadv.abi7532} {\bibfield  {journal} {\bibinfo  {journal} {Sci. Adv.}\ }\textbf {\bibinfo {volume} {7}},\ \bibinfo {pages} {eabi7532} (\bibinfo {year} {2021})}\BibitemShut {NoStop}%
\bibitem [{\citenamefont {Li}\ and\ \citenamefont {Yang}(2014)}]{monolayercalculationCGT}%
  \BibitemOpen
  \bibfield  {author} {\bibinfo {author} {\bibfnamefont {X.}~\bibnamefont {Li}}\ and\ \bibinfo {author} {\bibfnamefont {J.}~\bibnamefont {Yang}},\ }\bibfield  {title} {\bibinfo {title} {{CrXTe$_3$ (X = Si{,} Ge)} nanosheets: two dimensional intrinsic ferromagnetic semiconductors},\ }\href {https://doi.org/10.1039/C4TC01193G} {\bibfield  {journal} {\bibinfo  {journal} {J. Mater. Chem. C}\ }\textbf {\bibinfo {volume} {2}},\ \bibinfo {pages} {7071} (\bibinfo {year} {2014})}\BibitemShut {NoStop}%
\bibitem [{\citenamefont {Xu}\ \emph {et~al.}(2018)\citenamefont {Xu}, \citenamefont {Feng}, \citenamefont {Xiang},\ and\ \citenamefont {Bellaiche}}]{xu2018interplay}%
  \BibitemOpen
  \bibfield  {author} {\bibinfo {author} {\bibfnamefont {C.}~\bibnamefont {Xu}}, \bibinfo {author} {\bibfnamefont {J.}~\bibnamefont {Feng}}, \bibinfo {author} {\bibfnamefont {H.}~\bibnamefont {Xiang}},\ and\ \bibinfo {author} {\bibfnamefont {L.}~\bibnamefont {Bellaiche}},\ }\bibfield  {title} {\bibinfo {title} {Interplay between kitaev interaction and single ion anisotropy in ferromagnetic {CrI$_3$} and {CrGeTe$_3$} monolayers},\ }\href {https://doi.org/10.1038/s41524-018-0115-6} {\bibfield  {journal} {\bibinfo  {journal} {npj Comput. Mater.}\ }\textbf {\bibinfo {volume} {4}},\ \bibinfo {pages} {57} (\bibinfo {year} {2018})}\BibitemShut {NoStop}%
\bibitem [{\citenamefont {Zhuang}\ \emph {et~al.}(2015)\citenamefont {Zhuang}, \citenamefont {Xie}, \citenamefont {Kent},\ and\ \citenamefont {Ganesh}}]{zhuang2015computational}%
  \BibitemOpen
  \bibfield  {author} {\bibinfo {author} {\bibfnamefont {H.~L.}\ \bibnamefont {Zhuang}}, \bibinfo {author} {\bibfnamefont {Y.}~\bibnamefont {Xie}}, \bibinfo {author} {\bibfnamefont {P.~R.~C.}\ \bibnamefont {Kent}},\ and\ \bibinfo {author} {\bibfnamefont {P.}~\bibnamefont {Ganesh}},\ }\bibfield  {title} {\bibinfo {title} {Computational discovery of ferromagnetic semiconducting single-layer {${\mathrm{CrSnTe}}_{3}$}},\ }\href {https://doi.org/10.1103/PhysRevB.92.035407} {\bibfield  {journal} {\bibinfo  {journal} {Phys. Rev. B}\ }\textbf {\bibinfo {volume} {92}},\ \bibinfo {pages} {035407} (\bibinfo {year} {2015})}\BibitemShut {NoStop}%
\bibitem [{\citenamefont {Chen}\ \emph {et~al.}(2022)\citenamefont {Chen}, \citenamefont {Mao}, \citenamefont {Chung}, \citenamefont {Stone}, \citenamefont {Kolesnikov}, \citenamefont {Wang}, \citenamefont {Murai}, \citenamefont {Gao}, \citenamefont {Delaire},\ and\ \citenamefont {Dai}}]{chen2022anisotropic}%
  \BibitemOpen
  \bibfield  {author} {\bibinfo {author} {\bibfnamefont {L.}~\bibnamefont {Chen}}, \bibinfo {author} {\bibfnamefont {C.}~\bibnamefont {Mao}}, \bibinfo {author} {\bibfnamefont {J.-H.}\ \bibnamefont {Chung}}, \bibinfo {author} {\bibfnamefont {M.~B.}\ \bibnamefont {Stone}}, \bibinfo {author} {\bibfnamefont {A.~I.}\ \bibnamefont {Kolesnikov}}, \bibinfo {author} {\bibfnamefont {X.}~\bibnamefont {Wang}}, \bibinfo {author} {\bibfnamefont {N.}~\bibnamefont {Murai}}, \bibinfo {author} {\bibfnamefont {B.}~\bibnamefont {Gao}}, \bibinfo {author} {\bibfnamefont {O.}~\bibnamefont {Delaire}},\ and\ \bibinfo {author} {\bibfnamefont {P.}~\bibnamefont {Dai}},\ }\bibfield  {title} {\bibinfo {title} {Anisotropic magnon damping by zero-temperature quantum fluctuations in ferromagnetic {CrGeTe$_3$}},\ }\href {https://doi.org/10.1038/s41467-022-31612-w} {\bibfield  {journal} {\bibinfo  {journal} {Nat. Commun.}\ }\textbf {\bibinfo {volume} {13}},\ \bibinfo {pages} {4037} (\bibinfo {year} {2022})}\BibitemShut {NoStop}%
\bibitem [{\citenamefont {Lin}\ \emph {et~al.}(2017)\citenamefont {Lin}, \citenamefont {Zhuang}, \citenamefont {Luo}, \citenamefont {Liu}, \citenamefont {Chen}, \citenamefont {Yan}, \citenamefont {Sun}, \citenamefont {Zhou}, \citenamefont {Lu}, \citenamefont {Tong}, \citenamefont {Sheng}, \citenamefont {Qu}, \citenamefont {Song}, \citenamefont {Zhu},\ and\ \citenamefont {Sun}}]{lin2017tricritical}%
  \BibitemOpen
  \bibfield  {author} {\bibinfo {author} {\bibfnamefont {G.~T.}\ \bibnamefont {Lin}}, \bibinfo {author} {\bibfnamefont {H.~L.}\ \bibnamefont {Zhuang}}, \bibinfo {author} {\bibfnamefont {X.}~\bibnamefont {Luo}}, \bibinfo {author} {\bibfnamefont {B.~J.}\ \bibnamefont {Liu}}, \bibinfo {author} {\bibfnamefont {F.~C.}\ \bibnamefont {Chen}}, \bibinfo {author} {\bibfnamefont {J.}~\bibnamefont {Yan}}, \bibinfo {author} {\bibfnamefont {Y.}~\bibnamefont {Sun}}, \bibinfo {author} {\bibfnamefont {J.}~\bibnamefont {Zhou}}, \bibinfo {author} {\bibfnamefont {W.~J.}\ \bibnamefont {Lu}}, \bibinfo {author} {\bibfnamefont {P.}~\bibnamefont {Tong}}, \bibinfo {author} {\bibfnamefont {Z.~G.}\ \bibnamefont {Sheng}}, \bibinfo {author} {\bibfnamefont {Z.}~\bibnamefont {Qu}}, \bibinfo {author} {\bibfnamefont {W.~H.}\ \bibnamefont {Song}}, \bibinfo {author} {\bibfnamefont {X.~B.}\ \bibnamefont {Zhu}},\ and\ \bibinfo {author} {\bibfnamefont {Y.~P.}\ \bibnamefont {Sun}},\ }\bibfield  {title} {\bibinfo {title} {Tricritical behavior of the
  two-dimensional intrinsically ferromagnetic semiconductor {CrGeTe$_3$}},\ }\href {https://doi.org/10.1103/PhysRevB.95.245212} {\bibfield  {journal} {\bibinfo  {journal} {Phys. Rev. B}\ }\textbf {\bibinfo {volume} {95}},\ \bibinfo {pages} {245212} (\bibinfo {year} {2017})}\BibitemShut {NoStop}%
\bibitem [{\citenamefont {Ron}\ \emph {et~al.}(2019)\citenamefont {Ron}, \citenamefont {Zoghlin}, \citenamefont {Balents}, \citenamefont {Wilson},\ and\ \citenamefont {Hsieh}}]{ron2019dimensional}%
  \BibitemOpen
  \bibfield  {author} {\bibinfo {author} {\bibfnamefont {A.}~\bibnamefont {Ron}}, \bibinfo {author} {\bibfnamefont {E.}~\bibnamefont {Zoghlin}}, \bibinfo {author} {\bibfnamefont {L.}~\bibnamefont {Balents}}, \bibinfo {author} {\bibfnamefont {S.~D.}\ \bibnamefont {Wilson}},\ and\ \bibinfo {author} {\bibfnamefont {D.}~\bibnamefont {Hsieh}},\ }\bibfield  {title} {\bibinfo {title} {Dimensional crossover in a layered ferromagnet detected by spin correlation driven distortions},\ }\href {https://doi.org/10.1038/s41467-019-09663-3} {\bibfield  {journal} {\bibinfo  {journal} {Nat. Commun.}\ }\textbf {\bibinfo {volume} {10}},\ \bibinfo {pages} {1654} (\bibinfo {year} {2019})}\BibitemShut {NoStop}%
\bibitem [{\citenamefont {Ron}\ \emph {et~al.}(2020)\citenamefont {Ron}, \citenamefont {Chaudhary}, \citenamefont {Zhang}, \citenamefont {Ning}, \citenamefont {Zoghlin}, \citenamefont {Wilson}, \citenamefont {Averitt}, \citenamefont {Refael},\ and\ \citenamefont {Hsieh}}]{ron2020ultrafast}%
  \BibitemOpen
  \bibfield  {author} {\bibinfo {author} {\bibfnamefont {A.}~\bibnamefont {Ron}}, \bibinfo {author} {\bibfnamefont {S.}~\bibnamefont {Chaudhary}}, \bibinfo {author} {\bibfnamefont {G.}~\bibnamefont {Zhang}}, \bibinfo {author} {\bibfnamefont {H.}~\bibnamefont {Ning}}, \bibinfo {author} {\bibfnamefont {E.}~\bibnamefont {Zoghlin}}, \bibinfo {author} {\bibfnamefont {S.~D.}\ \bibnamefont {Wilson}}, \bibinfo {author} {\bibfnamefont {R.~D.}\ \bibnamefont {Averitt}}, \bibinfo {author} {\bibfnamefont {G.}~\bibnamefont {Refael}},\ and\ \bibinfo {author} {\bibfnamefont {D.}~\bibnamefont {Hsieh}},\ }\bibfield  {title} {\bibinfo {title} {Ultrafast enhancement of ferromagnetic spin exchange induced by ligand-to-metal charge transfer},\ }\href {https://doi.org/10.1103/PhysRevLett.125.197203} {\bibfield  {journal} {\bibinfo  {journal} {Phys. Rev. Lett.}\ }\textbf {\bibinfo {volume} {125}},\ \bibinfo {pages} {197203} (\bibinfo {year} {2020})}\BibitemShut {NoStop}%
\bibitem [{\citenamefont {Williams}\ \emph {et~al.}(2015)\citenamefont {Williams}, \citenamefont {Aczel}, \citenamefont {Lumsden}, \citenamefont {Nagler}, \citenamefont {Stone}, \citenamefont {Yan},\ and\ \citenamefont {Mandrus}}]{williams2015magnetic}%
  \BibitemOpen
  \bibfield  {author} {\bibinfo {author} {\bibfnamefont {T.~J.}\ \bibnamefont {Williams}}, \bibinfo {author} {\bibfnamefont {A.~A.}\ \bibnamefont {Aczel}}, \bibinfo {author} {\bibfnamefont {M.~D.}\ \bibnamefont {Lumsden}}, \bibinfo {author} {\bibfnamefont {S.~E.}\ \bibnamefont {Nagler}}, \bibinfo {author} {\bibfnamefont {M.~B.}\ \bibnamefont {Stone}}, \bibinfo {author} {\bibfnamefont {J.-Q.}\ \bibnamefont {Yan}},\ and\ \bibinfo {author} {\bibfnamefont {D.}~\bibnamefont {Mandrus}},\ }\bibfield  {title} {\bibinfo {title} {Magnetic correlations in the quasi-two-dimensional semiconducting ferromagnet {${\text{CrSiTe}}_{3}$}},\ }\href {https://doi.org/10.1103/PhysRevB.92.144404} {\bibfield  {journal} {\bibinfo  {journal} {Phys. Rev. B}\ }\textbf {\bibinfo {volume} {92}},\ \bibinfo {pages} {144404} (\bibinfo {year} {2015})}\BibitemShut {NoStop}%
\bibitem [{\citenamefont {Sakurai}\ \emph {et~al.}(2021)\citenamefont {Sakurai}, \citenamefont {Rubrecht}, \citenamefont {Corredor}, \citenamefont {Takehara}, \citenamefont {Yasutani}, \citenamefont {Zeisner}, \citenamefont {Alfonsov}, \citenamefont {Selter}, \citenamefont {Aswartham}, \citenamefont {Wolter}, \citenamefont {Büchner}, \citenamefont {Ohta},\ and\ \citenamefont {Kataev}}]{sakurai2021pressure}%
  \BibitemOpen
  \bibfield  {author} {\bibinfo {author} {\bibfnamefont {T.}~\bibnamefont {Sakurai}}, \bibinfo {author} {\bibfnamefont {B.}~\bibnamefont {Rubrecht}}, \bibinfo {author} {\bibfnamefont {L.~T.}\ \bibnamefont {Corredor}}, \bibinfo {author} {\bibfnamefont {R.}~\bibnamefont {Takehara}}, \bibinfo {author} {\bibfnamefont {M.}~\bibnamefont {Yasutani}}, \bibinfo {author} {\bibfnamefont {J.}~\bibnamefont {Zeisner}}, \bibinfo {author} {\bibfnamefont {A.}~\bibnamefont {Alfonsov}}, \bibinfo {author} {\bibfnamefont {S.}~\bibnamefont {Selter}}, \bibinfo {author} {\bibfnamefont {S.}~\bibnamefont {Aswartham}}, \bibinfo {author} {\bibfnamefont {A.~U.~B.}\ \bibnamefont {Wolter}}, \bibinfo {author} {\bibfnamefont {B.}~\bibnamefont {Büchner}}, \bibinfo {author} {\bibfnamefont {H.}~\bibnamefont {Ohta}},\ and\ \bibinfo {author} {\bibfnamefont {V.}~\bibnamefont {Kataev}},\ }\bibfield  {title} {\bibinfo {title} {Pressure control of the magnetic anisotropy of the quasi-two-dimensional van der {Waals} ferromagnet
  {Cr$_{2}$Ge$_{2}$Te$_{6}$}},\ }\href {https://doi.org/10.1103/physrevb.103.024404} {\bibfield  {journal} {\bibinfo  {journal} {Phys. Rev. B}\ }\textbf {\bibinfo {volume} {103}},\ \bibinfo {pages} {024404} (\bibinfo {year} {2021})}\BibitemShut {NoStop}%
\bibitem [{\citenamefont {Ebad-Allah}\ \emph {et~al.}()\citenamefont {Ebad-Allah}, \citenamefont {Guterding}, \citenamefont {Varma}, \citenamefont {Diware}, \citenamefont {Ganorkar}, \citenamefont {Jeschke},\ and\ \citenamefont {Kuntscher}}]{EbadAllah2024}%
  \BibitemOpen
  \bibfield  {author} {\bibinfo {author} {\bibfnamefont {J.}~\bibnamefont {Ebad-Allah}}, \bibinfo {author} {\bibfnamefont {D.}~\bibnamefont {Guterding}}, \bibinfo {author} {\bibfnamefont {M.}~\bibnamefont {Varma}}, \bibinfo {author} {\bibfnamefont {M.}~\bibnamefont {Diware}}, \bibinfo {author} {\bibfnamefont {S.}~\bibnamefont {Ganorkar}}, \bibinfo {author} {\bibfnamefont {H.~O.}\ \bibnamefont {Jeschke}},\ and\ \bibinfo {author} {\bibfnamefont {C.~A.}\ \bibnamefont {Kuntscher}},\ }\href@noop {} {\bibinfo {title} {Near room-temperature ferromagnetism from double-exchange in van der {Waals} material {CrGeTe$_3$}: evidence from optical conductivity under pressure}},\ \Eprint {https://arxiv.org/abs/2410.02522} {arXiv:2410.02522 [cond-mat.str-el]} \BibitemShut {NoStop}%
\bibitem [{\citenamefont {Zhang}\ \emph {et~al.}(2024)\citenamefont {Zhang}, \citenamefont {Harii}, \citenamefont {Yokouchi}, \citenamefont {Okayasu},\ and\ \citenamefont {Shiomi}}]{Zhang2023}%
  \BibitemOpen
  \bibfield  {author} {\bibinfo {author} {\bibfnamefont {S.}~\bibnamefont {Zhang}}, \bibinfo {author} {\bibfnamefont {K.}~\bibnamefont {Harii}}, \bibinfo {author} {\bibfnamefont {T.}~\bibnamefont {Yokouchi}}, \bibinfo {author} {\bibfnamefont {S.}~\bibnamefont {Okayasu}},\ and\ \bibinfo {author} {\bibfnamefont {Y.}~\bibnamefont {Shiomi}},\ }\bibfield  {title} {\bibinfo {title} {Amorphous ferromagnetic metal in van der {Waals} materials},\ }\href {https://doi.org/10.1002/aelm.202300609} {\bibfield  {journal} {\bibinfo  {journal} {Adv. Electron. Mater.}\ }\textbf {\bibinfo {volume} {10}},\ \bibinfo {pages} {2300609} (\bibinfo {year} {2024})}\BibitemShut {NoStop}%
\bibitem [{\citenamefont {Sterer}\ \emph {et~al.}(1990)\citenamefont {Sterer}, \citenamefont {Pasternak},\ and\ \citenamefont {Taylor}}]{Diamond_anvil_cell}%
  \BibitemOpen
  \bibfield  {author} {\bibinfo {author} {\bibfnamefont {E.}~\bibnamefont {Sterer}}, \bibinfo {author} {\bibfnamefont {M.~P.}\ \bibnamefont {Pasternak}},\ and\ \bibinfo {author} {\bibfnamefont {R.~D.}\ \bibnamefont {Taylor}},\ }\bibfield  {title} {\bibinfo {title} {{A multipurpose miniature diamond anvil cell}},\ }\href {https://doi.org/10.1063/1.1141433} {\bibfield  {journal} {\bibinfo  {journal} {Rev. Sci. Instrum.}\ }\textbf {\bibinfo {volume} {61}},\ \bibinfo {pages} {1117} (\bibinfo {year} {1990})}\BibitemShut {NoStop}%
\bibitem [{\citenamefont {Dewaele}\ \emph {et~al.}(2008)\citenamefont {Dewaele}, \citenamefont {Torrent}, \citenamefont {Loubeyre},\ and\ \citenamefont {Mezouar}}]{PhysRevB.78.104102}%
  \BibitemOpen
  \bibfield  {author} {\bibinfo {author} {\bibfnamefont {A.}~\bibnamefont {Dewaele}}, \bibinfo {author} {\bibfnamefont {M.}~\bibnamefont {Torrent}}, \bibinfo {author} {\bibfnamefont {P.}~\bibnamefont {Loubeyre}},\ and\ \bibinfo {author} {\bibfnamefont {M.}~\bibnamefont {Mezouar}},\ }\bibfield  {title} {\bibinfo {title} {Compression curves of transition metals in the mbar range: Experiments and projector augmented-wave calculations},\ }\href {https://doi.org/10.1103/PhysRevB.78.104102} {\bibfield  {journal} {\bibinfo  {journal} {Phys. Rev. B}\ }\textbf {\bibinfo {volume} {78}},\ \bibinfo {pages} {104102} (\bibinfo {year} {2008})}\BibitemShut {NoStop}%
\bibitem [{\citenamefont {Koepernik}\ and\ \citenamefont {Eschrig}(1999)}]{Koepernik1999}%
  \BibitemOpen
  \bibfield  {author} {\bibinfo {author} {\bibfnamefont {K.}~\bibnamefont {Koepernik}}\ and\ \bibinfo {author} {\bibfnamefont {H.}~\bibnamefont {Eschrig}},\ }\bibfield  {title} {\bibinfo {title} {Full-potential nonorthogonal local-orbital minimum-basis band-structure scheme},\ }\href {https://doi.org/10.1103/PhysRevB.59.1743} {\bibfield  {journal} {\bibinfo  {journal} {Phys. Rev. B}\ }\textbf {\bibinfo {volume} {59}},\ \bibinfo {pages} {1743} (\bibinfo {year} {1999})}\BibitemShut {NoStop}%
\bibitem [{\citenamefont {Perdew}\ \emph {et~al.}(1996)\citenamefont {Perdew}, \citenamefont {Burke},\ and\ \citenamefont {Ernzerhof}}]{Perdew1996}%
  \BibitemOpen
  \bibfield  {author} {\bibinfo {author} {\bibfnamefont {J.~P.}\ \bibnamefont {Perdew}}, \bibinfo {author} {\bibfnamefont {K.}~\bibnamefont {Burke}},\ and\ \bibinfo {author} {\bibfnamefont {M.}~\bibnamefont {Ernzerhof}},\ }\bibfield  {title} {\bibinfo {title} {Generalized gradient approximation made simple},\ }\href {https://doi.org/10.1103/PhysRevLett.77.3865} {\bibfield  {journal} {\bibinfo  {journal} {Phys. Rev. Lett.}\ }\textbf {\bibinfo {volume} {77}},\ \bibinfo {pages} {3865} (\bibinfo {year} {1996})}\BibitemShut {NoStop}%
\bibitem [{\citenamefont {Yu}\ \emph {et~al.}(2019)\citenamefont {Yu}, \citenamefont {Xia}, \citenamefont {Xu}, \citenamefont {Xu}, \citenamefont {Wang}, \citenamefont {Wang}, \citenamefont {Yu}, \citenamefont {Zou}, \citenamefont {Zhao}, \citenamefont {Wang}, \citenamefont {Miao},\ and\ \citenamefont {Guo}}]{yu2019pressure}%
  \BibitemOpen
  \bibfield  {author} {\bibinfo {author} {\bibfnamefont {Z.}~\bibnamefont {Yu}}, \bibinfo {author} {\bibfnamefont {W.}~\bibnamefont {Xia}}, \bibinfo {author} {\bibfnamefont {K.}~\bibnamefont {Xu}}, \bibinfo {author} {\bibfnamefont {M.}~\bibnamefont {Xu}}, \bibinfo {author} {\bibfnamefont {H.}~\bibnamefont {Wang}}, \bibinfo {author} {\bibfnamefont {X.}~\bibnamefont {Wang}}, \bibinfo {author} {\bibfnamefont {N.}~\bibnamefont {Yu}}, \bibinfo {author} {\bibfnamefont {Z.}~\bibnamefont {Zou}}, \bibinfo {author} {\bibfnamefont {J.}~\bibnamefont {Zhao}}, \bibinfo {author} {\bibfnamefont {L.}~\bibnamefont {Wang}}, \bibinfo {author} {\bibfnamefont {X.}~\bibnamefont {Miao}},\ and\ \bibinfo {author} {\bibfnamefont {Y.}~\bibnamefont {Guo}},\ }\bibfield  {title} {\bibinfo {title} {Pressure-induced structural phase transition and a special amorphization phase of two-dimensional ferromagnetic semiconductor {Cr$_2$Ge$_2$Te$_6$}},\ }\href {https://doi.org/10.1021/acs.jpcc.9b02415} {\bibfield  {journal} {\bibinfo  {journal} {J.
  Phys. Chem. C}\ }\textbf {\bibinfo {volume} {123}},\ \bibinfo {pages} {13885} (\bibinfo {year} {2019})}\BibitemShut {NoStop}%
\bibitem [{\citenamefont {Xu}\ \emph {et~al.}(2023)\citenamefont {Xu}, \citenamefont {Shimizu}, \citenamefont {Guterding}, \citenamefont {Otsuki},\ and\ \citenamefont {Jeschke}}]{Xu2023}%
  \BibitemOpen
  \bibfield  {author} {\bibinfo {author} {\bibfnamefont {H.-X.}\ \bibnamefont {Xu}}, \bibinfo {author} {\bibfnamefont {M.}~\bibnamefont {Shimizu}}, \bibinfo {author} {\bibfnamefont {D.}~\bibnamefont {Guterding}}, \bibinfo {author} {\bibfnamefont {J.}~\bibnamefont {Otsuki}},\ and\ \bibinfo {author} {\bibfnamefont {H.~O.}\ \bibnamefont {Jeschke}},\ }\bibfield  {title} {\bibinfo {title} {Pressure evolution of electronic structure and magnetism in the layered van der {Waals} ferromagnet {${\mathrm{CrGeTe}}_{3}$}},\ }\href {https://doi.org/10.1103/PhysRevB.108.125142} {\bibfield  {journal} {\bibinfo  {journal} {Phys. Rev. B}\ }\textbf {\bibinfo {volume} {108}},\ \bibinfo {pages} {125142} (\bibinfo {year} {2023})}\BibitemShut {NoStop}%
\bibitem [{\citenamefont {Wang}\ \emph {et~al.}(2006)\citenamefont {Wang}, \citenamefont {Yates}, \citenamefont {Souza},\ and\ \citenamefont {Vanderbilt}}]{Wang2006}%
  \BibitemOpen
  \bibfield  {author} {\bibinfo {author} {\bibfnamefont {X.}~\bibnamefont {Wang}}, \bibinfo {author} {\bibfnamefont {J.~R.}\ \bibnamefont {Yates}}, \bibinfo {author} {\bibfnamefont {I.}~\bibnamefont {Souza}},\ and\ \bibinfo {author} {\bibfnamefont {D.}~\bibnamefont {Vanderbilt}},\ }\bibfield  {title} {\bibinfo {title} {Ab initio calculation of the anomalous {Hall} conductivity by {Wannier} interpolation},\ }\href {https://doi.org/10.1103/PhysRevB.74.195118} {\bibfield  {journal} {\bibinfo  {journal} {Phys. Rev. B}\ }\textbf {\bibinfo {volume} {74}},\ \bibinfo {pages} {195118} (\bibinfo {year} {2006})}\BibitemShut {NoStop}%
\bibitem [{\citenamefont {Lepage}(1978)}]{Lepage1978}%
  \BibitemOpen
  \bibfield  {author} {\bibinfo {author} {\bibfnamefont {G.~P.}\ \bibnamefont {Lepage}},\ }\bibfield  {title} {\bibinfo {title} {A new algorithm for adaptive multidimensional integration},\ }\href {https://doi.org/10.1016/0021-9991(78)90004-9} {\bibfield  {journal} {\bibinfo  {journal} {J. Comput. Phys.}\ }\textbf {\bibinfo {volume} {27}},\ \bibinfo {pages} {192} (\bibinfo {year} {1978})}\BibitemShut {NoStop}%
\bibitem [{\citenamefont {Lepage}(2021)}]{Lepage2021}%
  \BibitemOpen
  \bibfield  {author} {\bibinfo {author} {\bibfnamefont {G.~P.}\ \bibnamefont {Lepage}},\ }\bibfield  {title} {\bibinfo {title} {Adaptive multidimensional integration: vegas enhanced},\ }\href {https://doi.org/10.1016/j.jcp.2021.110386} {\bibfield  {journal} {\bibinfo  {journal} {J. Comput. Phys.}\ }\textbf {\bibinfo {volume} {439}},\ \bibinfo {pages} {110386} (\bibinfo {year} {2021})}\BibitemShut {NoStop}%
\bibitem [{\citenamefont {Wang}\ \emph {et~al.}(2007)\citenamefont {Wang}, \citenamefont {Vanderbilt}, \citenamefont {Yates},\ and\ \citenamefont {Souza}}]{Wang2007}%
  \BibitemOpen
  \bibfield  {author} {\bibinfo {author} {\bibfnamefont {X.}~\bibnamefont {Wang}}, \bibinfo {author} {\bibfnamefont {D.}~\bibnamefont {Vanderbilt}}, \bibinfo {author} {\bibfnamefont {J.~R.}\ \bibnamefont {Yates}},\ and\ \bibinfo {author} {\bibfnamefont {I.}~\bibnamefont {Souza}},\ }\bibfield  {title} {\bibinfo {title} {Fermi-surface calculation of the anomalous {Hall} conductivity},\ }\href {https://doi.org/10.1103/PhysRevB.76.195109} {\bibfield  {journal} {\bibinfo  {journal} {Phys. Rev. B}\ }\textbf {\bibinfo {volume} {76}},\ \bibinfo {pages} {195109} (\bibinfo {year} {2007})}\BibitemShut {NoStop}%
\bibitem [{\citenamefont {Maryenko}\ \emph {et~al.}(2017)\citenamefont {Maryenko}, \citenamefont {Mishchenko}, \citenamefont {Bahramy}, \citenamefont {Ernst}, \citenamefont {Falson}, \citenamefont {Kozuka}, \citenamefont {Tsukazaki}, \citenamefont {Nagaosa},\ and\ \citenamefont {Kawasaki}}]{maryenko2017observation}%
  \BibitemOpen
  \bibfield  {author} {\bibinfo {author} {\bibfnamefont {D.}~\bibnamefont {Maryenko}}, \bibinfo {author} {\bibfnamefont {A.~S.}\ \bibnamefont {Mishchenko}}, \bibinfo {author} {\bibfnamefont {M.~S.}\ \bibnamefont {Bahramy}}, \bibinfo {author} {\bibfnamefont {A.}~\bibnamefont {Ernst}}, \bibinfo {author} {\bibfnamefont {J.}~\bibnamefont {Falson}}, \bibinfo {author} {\bibfnamefont {Y.}~\bibnamefont {Kozuka}}, \bibinfo {author} {\bibfnamefont {A.}~\bibnamefont {Tsukazaki}}, \bibinfo {author} {\bibfnamefont {N.}~\bibnamefont {Nagaosa}},\ and\ \bibinfo {author} {\bibfnamefont {M.}~\bibnamefont {Kawasaki}},\ }\bibfield  {title} {\bibinfo {title} {Observation of anomalous {Hall} effect in a non-magnetic two-dimensional electron system},\ }\href {https://doi.org/10.1038/ncomms14777} {\bibfield  {journal} {\bibinfo  {journal} {Nat. Commun.}\ }\textbf {\bibinfo {volume} {8}},\ \bibinfo {pages} {14777} (\bibinfo {year} {2017})}\BibitemShut {NoStop}%
\bibitem [{\citenamefont {Culcer}\ \emph {et~al.}(2003)\citenamefont {Culcer}, \citenamefont {MacDonald},\ and\ \citenamefont {Niu}}]{PhysRevB.68.045327}%
  \BibitemOpen
  \bibfield  {author} {\bibinfo {author} {\bibfnamefont {D.}~\bibnamefont {Culcer}}, \bibinfo {author} {\bibfnamefont {A.}~\bibnamefont {MacDonald}},\ and\ \bibinfo {author} {\bibfnamefont {Q.}~\bibnamefont {Niu}},\ }\bibfield  {title} {\bibinfo {title} {Anomalous {Hall} effect in paramagnetic two-dimensional systems},\ }\href {https://doi.org/10.1103/PhysRevB.68.045327} {\bibfield  {journal} {\bibinfo  {journal} {Phys. Rev. B}\ }\textbf {\bibinfo {volume} {68}},\ \bibinfo {pages} {045327} (\bibinfo {year} {2003})}\BibitemShut {NoStop}%
\bibitem [{\citenamefont {Hou}\ \emph {et~al.}(2015)\citenamefont {Hou}, \citenamefont {Su}, \citenamefont {Tian}, \citenamefont {Jin}, \citenamefont {Yang},\ and\ \citenamefont {Niu}}]{hou2015multivariable}%
  \BibitemOpen
  \bibfield  {author} {\bibinfo {author} {\bibfnamefont {D.}~\bibnamefont {Hou}}, \bibinfo {author} {\bibfnamefont {G.}~\bibnamefont {Su}}, \bibinfo {author} {\bibfnamefont {Y.}~\bibnamefont {Tian}}, \bibinfo {author} {\bibfnamefont {X.}~\bibnamefont {Jin}}, \bibinfo {author} {\bibfnamefont {S.~A.}\ \bibnamefont {Yang}},\ and\ \bibinfo {author} {\bibfnamefont {Q.}~\bibnamefont {Niu}},\ }\bibfield  {title} {\bibinfo {title} {Multivariable scaling for the anomalous {Hall} effect},\ }\href {https://doi.org/10.1103/PhysRevLett.114.217203} {\bibfield  {journal} {\bibinfo  {journal} {Phys. Rev. Lett.}\ }\textbf {\bibinfo {volume} {114}},\ \bibinfo {pages} {217203} (\bibinfo {year} {2015})}\BibitemShut {NoStop}%
\bibitem [{\citenamefont {Liu}\ \emph {et~al.}(2018)\citenamefont {Liu}, \citenamefont {Sun}, \citenamefont {Kumar}, \citenamefont {Muechler}, \citenamefont {Sun}, \citenamefont {Jiao}, \citenamefont {Yang}, \citenamefont {Liu}, \citenamefont {Liang}, \citenamefont {Xu}, \citenamefont {Kroder}, \citenamefont {S{\"u}{\ss}}, \citenamefont {Borrmann}, \citenamefont {Shekhar}, \citenamefont {Wang}, \citenamefont {Xi}, \citenamefont {Wang}, \citenamefont {Schnelle}, \citenamefont {Wirth}, \citenamefont {Chen}, \citenamefont {Goennenwein},\ and\ \citenamefont {Felser}}]{liu2018giant}%
  \BibitemOpen
  \bibfield  {author} {\bibinfo {author} {\bibfnamefont {E.}~\bibnamefont {Liu}}, \bibinfo {author} {\bibfnamefont {Y.}~\bibnamefont {Sun}}, \bibinfo {author} {\bibfnamefont {N.}~\bibnamefont {Kumar}}, \bibinfo {author} {\bibfnamefont {L.}~\bibnamefont {Muechler}}, \bibinfo {author} {\bibfnamefont {A.}~\bibnamefont {Sun}}, \bibinfo {author} {\bibfnamefont {L.}~\bibnamefont {Jiao}}, \bibinfo {author} {\bibfnamefont {S.-Y.}\ \bibnamefont {Yang}}, \bibinfo {author} {\bibfnamefont {D.}~\bibnamefont {Liu}}, \bibinfo {author} {\bibfnamefont {A.}~\bibnamefont {Liang}}, \bibinfo {author} {\bibfnamefont {Q.}~\bibnamefont {Xu}}, \bibinfo {author} {\bibfnamefont {J.}~\bibnamefont {Kroder}}, \bibinfo {author} {\bibfnamefont {V.}~\bibnamefont {S{\"u}{\ss}}}, \bibinfo {author} {\bibfnamefont {H.}~\bibnamefont {Borrmann}}, \bibinfo {author} {\bibfnamefont {C.}~\bibnamefont {Shekhar}}, \bibinfo {author} {\bibfnamefont {Z.}~\bibnamefont {Wang}}, \bibinfo {author} {\bibfnamefont {C.}~\bibnamefont {Xi}}, \bibinfo {author}
  {\bibfnamefont {W.}~\bibnamefont {Wang}}, \bibinfo {author} {\bibfnamefont {W.}~\bibnamefont {Schnelle}}, \bibinfo {author} {\bibfnamefont {S.}~\bibnamefont {Wirth}}, \bibinfo {author} {\bibfnamefont {Y.}~\bibnamefont {Chen}}, \bibinfo {author} {\bibfnamefont {S.~T.~B.}\ \bibnamefont {Goennenwein}},\ and\ \bibinfo {author} {\bibfnamefont {C.}~\bibnamefont {Felser}},\ }\bibfield  {title} {\bibinfo {title} {Giant anomalous {Hall} effect in a ferromagnetic kagome-lattice semimetal},\ }\href {https://doi.org/10.1038/s41567-018-0234-5} {\bibfield  {journal} {\bibinfo  {journal} {Nat. Phys.}\ }\textbf {\bibinfo {volume} {14}},\ \bibinfo {pages} {1125} (\bibinfo {year} {2018})}\BibitemShut {NoStop}%
\bibitem [{\citenamefont {Fang}\ \emph {et~al.}(2003)\citenamefont {Fang}, \citenamefont {Nagaosa}, \citenamefont {Takahashi}, \citenamefont {Asamitsu}, \citenamefont {Mathieu}, \citenamefont {Ogasawara}, \citenamefont {Yamada}, \citenamefont {Kawasaki}, \citenamefont {Tokura},\ and\ \citenamefont {Terakura}}]{fang2003anomalous}%
  \BibitemOpen
  \bibfield  {author} {\bibinfo {author} {\bibfnamefont {Z.}~\bibnamefont {Fang}}, \bibinfo {author} {\bibfnamefont {N.}~\bibnamefont {Nagaosa}}, \bibinfo {author} {\bibfnamefont {K.~S.}\ \bibnamefont {Takahashi}}, \bibinfo {author} {\bibfnamefont {A.}~\bibnamefont {Asamitsu}}, \bibinfo {author} {\bibfnamefont {R.}~\bibnamefont {Mathieu}}, \bibinfo {author} {\bibfnamefont {T.}~\bibnamefont {Ogasawara}}, \bibinfo {author} {\bibfnamefont {H.}~\bibnamefont {Yamada}}, \bibinfo {author} {\bibfnamefont {M.}~\bibnamefont {Kawasaki}}, \bibinfo {author} {\bibfnamefont {Y.}~\bibnamefont {Tokura}},\ and\ \bibinfo {author} {\bibfnamefont {K.}~\bibnamefont {Terakura}},\ }\bibfield  {title} {\bibinfo {title} {The anomalous {Hall} effect and magnetic monopoles in momentum space},\ }\href {https://doi.org/10.1126/science.1089408} {\bibfield  {journal} {\bibinfo  {journal} {Science}\ }\textbf {\bibinfo {volume} {302}},\ \bibinfo {pages} {92} (\bibinfo {year} {2003})}\BibitemShut {NoStop}%
\bibitem [{\citenamefont {Pureur}\ \emph {et~al.}(2004)\citenamefont {Pureur}, \citenamefont {Fabris}, \citenamefont {Schaf},\ and\ \citenamefont {Campbell}}]{P.Pureur2004}%
  \BibitemOpen
  \bibfield  {author} {\bibinfo {author} {\bibfnamefont {P.}~\bibnamefont {Pureur}}, \bibinfo {author} {\bibfnamefont {F.~W.}\ \bibnamefont {Fabris}}, \bibinfo {author} {\bibfnamefont {J.}~\bibnamefont {Schaf}},\ and\ \bibinfo {author} {\bibfnamefont {I.~A.}\ \bibnamefont {Campbell}},\ }\bibfield  {title} {\bibinfo {title} {Chiral susceptibility in canonical spin glass and re-entrant alloys from {Hall} effect measurements},\ }\href {https://doi.org/10.1209/epl/i2004-10042-8} {\bibfield  {journal} {\bibinfo  {journal} {EPL}\ }\textbf {\bibinfo {volume} {67}},\ \bibinfo {pages} {123} (\bibinfo {year} {2004})}\BibitemShut {NoStop}%
\bibitem [{\citenamefont {Zeng}\ \emph {et~al.}(2006)\citenamefont {Zeng}, \citenamefont {Yao}, \citenamefont {Niu},\ and\ \citenamefont {Weitering}}]{PhysRevLett.96.037204}%
  \BibitemOpen
  \bibfield  {author} {\bibinfo {author} {\bibfnamefont {C.}~\bibnamefont {Zeng}}, \bibinfo {author} {\bibfnamefont {Y.}~\bibnamefont {Yao}}, \bibinfo {author} {\bibfnamefont {Q.}~\bibnamefont {Niu}},\ and\ \bibinfo {author} {\bibfnamefont {H.~H.}\ \bibnamefont {Weitering}},\ }\bibfield  {title} {\bibinfo {title} {Linear magnetization dependence of the intrinsic anomalous {Hall} effect},\ }\href {https://doi.org/10.1103/PhysRevLett.96.037204} {\bibfield  {journal} {\bibinfo  {journal} {Phys. Rev. Lett.}\ }\textbf {\bibinfo {volume} {96}},\ \bibinfo {pages} {037204} (\bibinfo {year} {2006})}\BibitemShut {NoStop}%
\bibitem [{\citenamefont {Haham}\ \emph {et~al.}(2011)\citenamefont {Haham}, \citenamefont {Shperber}, \citenamefont {Schultz}, \citenamefont {Naftalis}, \citenamefont {Shimshoni}, \citenamefont {Reiner},\ and\ \citenamefont {Klein}}]{haham2011scaling}%
  \BibitemOpen
  \bibfield  {author} {\bibinfo {author} {\bibfnamefont {N.}~\bibnamefont {Haham}}, \bibinfo {author} {\bibfnamefont {Y.}~\bibnamefont {Shperber}}, \bibinfo {author} {\bibfnamefont {M.}~\bibnamefont {Schultz}}, \bibinfo {author} {\bibfnamefont {N.}~\bibnamefont {Naftalis}}, \bibinfo {author} {\bibfnamefont {E.}~\bibnamefont {Shimshoni}}, \bibinfo {author} {\bibfnamefont {J.~W.}\ \bibnamefont {Reiner}},\ and\ \bibinfo {author} {\bibfnamefont {L.}~\bibnamefont {Klein}},\ }\bibfield  {title} {\bibinfo {title} {Scaling of the anomalous {Hall} effect in {SrRuO${}_{3}$}},\ }\href {https://doi.org/10.1103/PhysRevB.84.174439} {\bibfield  {journal} {\bibinfo  {journal} {Phys. Rev. B}\ }\textbf {\bibinfo {volume} {84}},\ \bibinfo {pages} {174439} (\bibinfo {year} {2011})}\BibitemShut {NoStop}%
\bibitem [{\citenamefont {Piva}\ \emph {et~al.}(2023)\citenamefont {Piva}, \citenamefont {Souza}, \citenamefont {Brousseau-Couture}, \citenamefont {Sorn}, \citenamefont {Pakuszewski}, \citenamefont {John}, \citenamefont {Adriano}, \citenamefont {C\^ot\'e}, \citenamefont {Pagliuso}, \citenamefont {Paramekanti},\ and\ \citenamefont {Nicklas}}]{PhysRevResearch.5.013068}%
  \BibitemOpen
  \bibfield  {author} {\bibinfo {author} {\bibfnamefont {M.~M.}\ \bibnamefont {Piva}}, \bibinfo {author} {\bibfnamefont {J.~C.}\ \bibnamefont {Souza}}, \bibinfo {author} {\bibfnamefont {V.}~\bibnamefont {Brousseau-Couture}}, \bibinfo {author} {\bibfnamefont {S.}~\bibnamefont {Sorn}}, \bibinfo {author} {\bibfnamefont {K.~R.}\ \bibnamefont {Pakuszewski}}, \bibinfo {author} {\bibfnamefont {J.~K.}\ \bibnamefont {John}}, \bibinfo {author} {\bibfnamefont {C.}~\bibnamefont {Adriano}}, \bibinfo {author} {\bibfnamefont {M.}~\bibnamefont {C\^ot\'e}}, \bibinfo {author} {\bibfnamefont {P.~G.}\ \bibnamefont {Pagliuso}}, \bibinfo {author} {\bibfnamefont {A.}~\bibnamefont {Paramekanti}},\ and\ \bibinfo {author} {\bibfnamefont {M.}~\bibnamefont {Nicklas}},\ }\bibfield  {title} {\bibinfo {title} {Topological features in the ferromagnetic {Weyl} semimetal {CeAlSi}: Role of domain walls},\ }\href {https://doi.org/10.1103/PhysRevResearch.5.013068} {\bibfield  {journal} {\bibinfo  {journal} {Phys. Rev. Res.}\ }\textbf {\bibinfo
  {volume} {5}},\ \bibinfo {pages} {013068} (\bibinfo {year} {2023})}\BibitemShut {NoStop}%
\bibitem [{\citenamefont {Liu}\ \emph {et~al.}(2020)\citenamefont {Liu}, \citenamefont {Zhang}, \citenamefont {Xu}, \citenamefont {Yang}, \citenamefont {Wang}, \citenamefont {Lei}, \citenamefont {Sui}, \citenamefont {Uwatoko}, \citenamefont {Wang}, \citenamefont {Weng}, \citenamefont {Sun},\ and\ \citenamefont {Cheng}}]{liu2020pressure}%
  \BibitemOpen
  \bibfield  {author} {\bibinfo {author} {\bibfnamefont {Z.~Y.}\ \bibnamefont {Liu}}, \bibinfo {author} {\bibfnamefont {T.}~\bibnamefont {Zhang}}, \bibinfo {author} {\bibfnamefont {S.~X.}\ \bibnamefont {Xu}}, \bibinfo {author} {\bibfnamefont {P.~T.}\ \bibnamefont {Yang}}, \bibinfo {author} {\bibfnamefont {Q.}~\bibnamefont {Wang}}, \bibinfo {author} {\bibfnamefont {H.~C.}\ \bibnamefont {Lei}}, \bibinfo {author} {\bibfnamefont {Y.}~\bibnamefont {Sui}}, \bibinfo {author} {\bibfnamefont {Y.}~\bibnamefont {Uwatoko}}, \bibinfo {author} {\bibfnamefont {B.~S.}\ \bibnamefont {Wang}}, \bibinfo {author} {\bibfnamefont {H.~M.}\ \bibnamefont {Weng}}, \bibinfo {author} {\bibfnamefont {J.~P.}\ \bibnamefont {Sun}},\ and\ \bibinfo {author} {\bibfnamefont {J.-G.}\ \bibnamefont {Cheng}},\ }\bibfield  {title} {\bibinfo {title} {Pressure effect on the anomalous {Hall} effect of ferromagnetic {Weyl} semimetal {Co$_3$Sn$_2$S$_2$}},\ }\href {https://doi.org/10.1103/PhysRevMaterials.4.044203} {\bibfield  {journal} {\bibinfo
  {journal} {Phys. Rev. Mater.}\ }\textbf {\bibinfo {volume} {4}},\ \bibinfo {pages} {044203} (\bibinfo {year} {2020})}\BibitemShut {NoStop}%
\bibitem [{\citenamefont {Shen}\ \emph {et~al.}(2023)\citenamefont {Shen}, \citenamefont {Gao}, \citenamefont {Yi}, \citenamefont {Li}, \citenamefont {Zhang}, \citenamefont {Yang}, \citenamefont {Wang}, \citenamefont {Zhou}, \citenamefont {Huang}, \citenamefont {Wei}, \citenamefont {Yang}, \citenamefont {Shi}, \citenamefont {Xu}, \citenamefont {Gao}, \citenamefont {Shen}, \citenamefont {Li}, \citenamefont {Wang},\ and\ \citenamefont {Liu}}]{Shen2023}%
  \BibitemOpen
  \bibfield  {author} {\bibinfo {author} {\bibfnamefont {J.}~\bibnamefont {Shen}}, \bibinfo {author} {\bibfnamefont {J.}~\bibnamefont {Gao}}, \bibinfo {author} {\bibfnamefont {C.}~\bibnamefont {Yi}}, \bibinfo {author} {\bibfnamefont {M.}~\bibnamefont {Li}}, \bibinfo {author} {\bibfnamefont {S.}~\bibnamefont {Zhang}}, \bibinfo {author} {\bibfnamefont {J.}~\bibnamefont {Yang}}, \bibinfo {author} {\bibfnamefont {B.}~\bibnamefont {Wang}}, \bibinfo {author} {\bibfnamefont {M.}~\bibnamefont {Zhou}}, \bibinfo {author} {\bibfnamefont {R.}~\bibnamefont {Huang}}, \bibinfo {author} {\bibfnamefont {H.}~\bibnamefont {Wei}}, \bibinfo {author} {\bibfnamefont {H.}~\bibnamefont {Yang}}, \bibinfo {author} {\bibfnamefont {Y.}~\bibnamefont {Shi}}, \bibinfo {author} {\bibfnamefont {X.}~\bibnamefont {Xu}}, \bibinfo {author} {\bibfnamefont {H.-J.}\ \bibnamefont {Gao}}, \bibinfo {author} {\bibfnamefont {B.}~\bibnamefont {Shen}}, \bibinfo {author} {\bibfnamefont {G.}~\bibnamefont {Li}}, \bibinfo {author} {\bibfnamefont
  {Z.}~\bibnamefont {Wang}},\ and\ \bibinfo {author} {\bibfnamefont {E.}~\bibnamefont {Liu}},\ }\bibfield  {title} {\bibinfo {title} {Magnetic-field modulation of topological electronic state and emergent magneto-transport in a magnetic {Weyl} semimetal},\ }\href {https://doi.org/10.1016/j.xinn.2023.100399} {\bibfield  {journal} {\bibinfo  {journal} {The Innovation}\ }\textbf {\bibinfo {volume} {4}},\ \bibinfo {pages} {100399} (\bibinfo {year} {2023})}\BibitemShut {NoStop}%
\end{thebibliography}
\end{document}